\documentstyle[psfig]{mn}

\def\gs{\mathrel{\raise0.35ex\hbox{$\scriptstyle >$}\kern-0.6em
\lower0.40ex\hbox{{$\scriptstyle \sim$}}}}
\def\ls{\mathrel{\raise0.35ex\hbox{$\scriptstyle <$}\kern-0.6em
\lower0.40ex\hbox{{$\scriptstyle \sim$}}}}

\title[CO observations of cooling-flow clusters]
      {The detection of molecular gas in the central galaxies of cooling-flow clusters}

\author[A.\,C.\ Edge ]
       {A.\,C.\ Edge,
        \vspace*{1mm}\\
        Department of Physics, University of Durham, South Road,
        Durham DH1 3LE\\}

\date{Accepted ... ; Received ... ; in original form 11th January 2001}

\pagerange{000--000}

\begin{document}

\maketitle

\begin{abstract}

We present the detections of CO line emission in the
central galaxy of  sixteen extreme cooling flow clusters
using  the {IRAM} 30m and the {JCMT} 15m.
These detections of CO(1-0), CO(2-1),
CO(3-2) and CO(4-3) are 
consistent with the presence of a substantial
mass of warm molecular gas ($10^{9-11.5}M_\odot$)
within 50~kpc radius of the central galaxy.
We present limits on  thirteen other galaxies
in similarly extreme cooling flow clusters. 
These results are consistent with the presence of a 
massive starburst in the central galaxy which warms
a population of cold gas clouds producing both optical
and NIR emission lines and significant CO line emission. 
Curiously, our CO detections are restricted to the
lower radio power central galaxies. These are the first
detections of molecular gas in a cooling flow other
than NGC\,1275 in the Perseus cluster. 
As four of our targets have firm limits
on their dust mass from {SCUBA} and the rest have crude limits
from {IRAS}, we can 
calculate gas-to-dust ratios. Simple
analysis indicates that the best secondary indicator
of molecular gas is optical line luminosity. 
We review the implications of these results and
the prospects for observations in the near future.

\end{abstract}

\begin{keywords}
    galaxies: active 
--- galaxies: starburst
--- galaxies: cooling flow
--- galaxies: individual: A11, A262, A291, A478, A646, A1068, A1664, A1795, A1835, A2146, A2204, A2390, A2597, RXJ0338+09, 
RXJ0352+19, RXJ0439+05, RXJ0747-19, RXJ0821+07, RXJ1347$-$11, RXJ1532+30, Zw2089, Zw3146, Zw3916, Zw7160, Zw8193, Zw8197, Zw8276
4C$+$55.16
--- X-ray: cooling flow
\end{keywords}

\section{Introduction}

The existence of cooling flows and the ultimate fate of this cooling gas has been the subject of
an extensive and strongly contested debate for several decades (see Fabian 1994).
The gas in the cores of massive, relaxed clusters of galaxies can  
cool and recombine through X-ray emission
initiating a cooling flow (Fabian \& Nulsen 1977; Cowie \& Binney 1977).  
The resulting reservoir of cold gas
has not been detected in molecular form and so is inferred either to reside in a
phase with $T_{\rm gas} << 100$\,{\sc k} (Ferland, Fabian
\& Johnstone 1994) or to indicate that cooling flows deposit much less gas
if at all (O'Dea et al.\ 1994; Braine et al.\ 1995, Voit \& Donahue 1995).
The only cooling flow known to contain 
molecular gas is that around NGC~1275 in Perseus (Gear et al.\ 1985;
Lester et al.\ 1995, Bridges \& Irwin 1998), 
although the interpretation in this source is
complicated by the strongly varying nuclear component.  Moreover, the
presence of the molecular gas may be related to an apparently on-going merger in
this system, which has been the subject of a long-running debate (Van
den Bergh 1977; Hu et al.\ 1983; Pedlar et al.\ 1990; Holtzman et
al.\ 1992; Norgaard-Nielsen et al.\ 1993).

Recent results 
from optical emission-line ratios (Hansen, Jorgensen \&
Norgaard-Nielsen 1995; Allen 1995), {\it Hubble Space Telescope}
({\it HST}) imaging (McNamara et al.\ 1996; Pinkney et al.\ 1996)
and sub-mm dust emission (Edge et al.\ 1999) all indicate that
dust is present the cores of massive cooling flows.  As dust is
rarely seen in the Universe without some accompanying molecular
gas, it is timely to return to the limits on molecular gas
in the massive cooling flows selected from the {\it ROSAT} All Sky
Survey (e.g. Zw3146 Edge et al. 1994 and 
RXJ1347$-$11 Allen 1996). This field has been dominated
by non-detections (Grabelsky \& Ulmer 1990;
McNamara \& Jaffe 1992; O'Dea et al. 1994;
Braine \& Dupraz 1994; Fujita et al. 2000) but improvements
in receiver technology combined with an increased pool of 
extreme cooling flows selected from {\it ROSAT} samples
offer new opportunities. Cooling flows with mass deposition
rates of $>$100~M$_\odot$~yr$^{-1}$ will accumulate
molecular gas masses of $>10^{11}$~M$_\odot$ in just 10$^9$~yr
which is now detectable out to $z=0.3$.
Selecting more distant cooling flows allows more of the 
cooled gas to fall in the telescope beam and to observe
lines in more favourable frequencies (i.e. CO(3-2) in the
A-band of {JCMT} for $z>0.25$).
With these factors in mind we obtained 2 shifts of {JCMT}
observations to search for CO(3-2) in the three most
distant, massive cooling flows known at the time.

Throughout we assume $\Omega_0=1$ and $H_0 = 50$\,km\,s$^{-1}$\,Mpc$^{-1}$.

\section{Observations}

The initial observations presented in this paper were obtained on 26th and 27th December 1998
using the recently commissioned A3i receiver on the James Clark Maxwell Telescope (JCMT)
on Mauna Kea in good conditions ($\tau_{CSO}=0.05$--0.12). Three targets were
observed,  Zw3146 ($z=$0.2906, Edge et al. 1994), RXJ1347$-$11 ($z=0.4503$, Schindler et al. 1995)
and RXJ1532+30 ($z=$0.3615, Ebeling et al. 1998) searching for the CO(3-2) transition
redshifted into the A-band (210--276 GHz) and the exposures are summarised in
Table~1. This project benefits greatly from the significant
improvement in the performance of A3i over the previous A2 receiver in
the 260--276~GHz range. However,
due to the unexpectedly poor performance of the A3i receiver in the frequency range
of 245--255~GHz, the observation of RXJ1532+30 is much less sensitive than
the other two ($T_{sys}$ of 800~K as opposed to 280--340~K) hence the limit
set on this system is much weaker.
Standard spectra of  IRC+1021 were taken and flux calibration was made with Mars.

These observations were augmented by additional A3i observations on 12$^{\rm th}$ March,
1$^{\rm st}$--3$^{\rm rd}$ and 15$^{\rm th}$ May 
1999 of Zw3146 using a wider bandwidth and slightly lower central frequency
to compensate for the $-150$~km~s$^{-1}$ shift seen in the December data. The 
February observations were taken in poorer conditions ($\tau_{CSO}>0.15$
and low elevation)
so do not achieve sufficient sensitivity to detect the line but the May data are
better and confirm the initial detection (Fig. 1). 
During the May observations, we
also obtained service time to observe $^{12}$CO in A1835 and Zw7160 
(both $z\approx0.25$, Allen et al. 1992) and $^{13}$CO in A1835. 
All these {JCMT} observations
were made in beamswitching mode with a throw of 60 or 90$''$.

Given the detections of CO(3-2) our next step was to obtain
CO(1-0) observations with the {IRAM} 30m telescope. We were awarded time
contingent on a detection of Zw3146 in service time on May 7th 1999.
This data recovered a line of similar width and redshift to
the initial {JCMT} detection so we were allocated a total of 68 hours
in August 1999. The data from this campaign were taken in 
good conditions for daytime, summer observing (2--8mm atmospheric water vapour)
so rapid progress was made. All the {IRAM} observations were made in
beam-switching mode with a throw of 210$''$. For our strongest
$^{12}$CO(1-0) detection (A1068), we also obtained $^{13}$CO(1-0) data
which sets a strong limit on the isotopic ratio. For marginal detections
we obtained, where possible, further data shifted in frequency by 150--400~MHz to eliminate
possible non-linear baseline effects. These data confirmed the detections
in most cases {\it but not all}. These are discussed individually in Section 4.

\begin{table*}
\caption{Log of {JCMT} Observations.}
\begin{tabular}{lclclcccc}
\noalign{\medskip \hrule \medskip}
Cluster & Date & Instrument & Frequency & Bandwidth & Exposure & Conditions     & T$_{sys}$ & Beam Size \\
        &      &            & (GHz)     & (MHz)     & (s)      & ($\tau_{CSO}$) & (K)       & ($''$)\\
 & & & & & & & & \\
$^{12}$CO(3-2) & & & & & & & & \\
Zw3146     &  26/12/98 & A3i & 267.934 & 1000 & 6000 & 0.08 & 354 &   18.0 \\
RXJ1347$-$11 &  26/12/98 & A3i & 238.434 & 1000 & 7800 & 0.08 & 346 &   20.9 \\
RXJ1532+30 &  26/12/98 & A3i & 253.982 & 1000 & 6600 & 0.08 & 825 &   19.0 \\
Zw3146     &  27/12/98 & A3i & 267.934 & 1000 & 7200 & 0.05 & 292 &   18.0 \\
RXJ1347$-$11 &  27/12/98 & A3i & 238.434 & 1000 & 6600 & 0.05 & 279 &   20.9 \\
RXJ1532+30 &  27/12/98 & A3i & 253.982 & 1000 & 6000 & 0.05 & 748 &   19.0 \\
Zw3146     &  12/03/99 & A3i & 268.140 & 1800 & 5400 & 0.18 & 450 &   18.0 \\
Zw3146     &  01/05/99 & A3i & 268.140 & 1800 & 5400 & 0.12 & 413 &   18.0 \\
A1835      &  2-3/05/99 & A3i & 276.130 & 1800 & 5700 & 0.13 & 452 &  17.5 \\
A1835      &  15/05/99 & A3i & 276.130 & 1800 & 1800 & 0.12 & 496 & 17.5 \\
Zw7160     &  15/05/99 & A3i & 274.920 & 1800 & 1800 & 0.12 & 542 & 17.6 \\
Zw7160     &  24/05/99 & A3i & 274.920 & 1800 & 2700 & 0.14 & 486 & 17.6 \\
A1835      &  31/05/99 & A3i & 276.130 & 1800 & 6600 & 0.15 & 480 & 17.5 \\
Zw7160     &  02/06/99 & A3i & 274.920 & 1800 & 6000 & 0.20 & 577 & 17.6 \\
Zw7160     &  19/06/99 & A3i & 274.920 & 1800 & 3600 & 0.10 & 389 & 17.6 \\
 & & & & & & & & \\
$^{13}$CO(3-2) & & & & & & & & \\
A1835      & 14-15/07/99 &  A3i & 263.980 & 1800 & 7200 & 0.06 & 317 & 18.3 \\
 & & & & & & & & \\
$^{12}$CO(4-3) & & & & & & & & \\
Zw3146     &  28-29/01/00 &  B3 & 357.410 & 1800 & 10200 & 0.07 & 410 & 13.6 \\
RXJ1532+30 &  28-29/01/00 &  B3 & 338.488 & 1800 & 11400 & 0.05 & 310 & 14.3 \\
Zw7160     &  \ \  20/07/97 &  B3 & 366.600 &  900 &  4200 & 0.05 & 603 & 13.2 \\
\noalign{\smallskip \hrule}
\end{tabular}
\end{table*}

\begin{table*}
\caption{Log of {IRAM} Observations from 1999.}
\begin{tabular}{lclclcccc}
\noalign{\medskip \hrule \medskip}
Cluster & Date & Instrument & Frequency & Bandwidth & Exposure & Conditions & T$_{sys}$ & Beam Size \\
        &      &            &  (GHz)    &   (MHz)   &  (s)     &  (mm H$_2$O) & (K) & ($''$) \\
 & & & & & & & & \\
$^{12}$CO(1-0) & & & & & & & & \\
Zw3146    & 6-7/05/99 & A100 & \ \ 89.316 & 500 & 4920 & 4 & 125 & 26.9 \\
A1068     & 31/07/99 & A100 & 101.239 & 500 & 2640 & 4 & 137 & 23.8 \\
A1835     & 31/07/99 & A100 & \ \  92.048 & 500 & 3840 & 4 & 130 &  26.1\\
Zw3146    & 12/08/99 & A100 & \ \ 89.316 & 500 & 3600 & 2 & 115 & 26.9 \\
RXJ1532+30 & 12/08/99 & A100 & \ \ 84.665 & 500 & 3300 & 3 & 113 & 28.4 \\
A2204     & 12/08/99 & A100 & 100.114 & 500 & 3600 & 2 & 121 & 24.0 \\
Zw3916    & 13/08/99 & A100 & \ \ 95.740 & 500 & 5640 & 7 & 141 & 25.1 \\
Zw8193    & 13/08/99 & A100 & \ \ 97.448 & 500 & 6900 & 6 & 131 & 24.7 \\
A2390     & 13/08/99 & A100 & \ \ 93.542 & 500 & 6000 & 9 & 165 & 25.7 \\
RXJ0821+07 & 14/08/99 & A100 & 103.848 & 500 & 3300 & 9 & 168 & 23.2 \\
A2146     & 14/08/99 & A100 & \ \ 93.390 & 500 & 5700 & 9 & 165 & 25.8 \\
Zw8193    & 14/08/99 & A100 & \ \ 97.210 & 500 & 2400 & 7 & 136 & 24.7 \\
Zw7160    & 14/08/99 & A100 & \ \ 91.645 & 500 & 6000 & 6 & 144 & 26.2 \\
Zw8197    & 14/08/99 & A100 & 103.475 & 500 & 5400 & 3 & 128 &  23.2\\
RXJ0352+19 & 15/08/99 & A100 & 103.942 & 500 & \ \ 900 & 9 & 211 & 23.1  \\
A646      & 15/08/99 & A100 & 102.300 & 500 & 6120 & 8 & 154 & 23.5 \\
A2204     & 15/08/99 & A100 & 100.114 & 500 & 8400 & 9 & 148 &  24.0\\
A2390     & 15/08/99 & A100 & \ \ 93.585 & 500 & 5220 & 7 & 147 & 25.7 \\
A1068     & 16/08/99 & A100 & 101.239    & 500 &  \ \ 900 & 2 & 117 & 23.8 \\
RXJ1532+30 & 16/08/99 & A100 & \ \ 84.600 & 500 & 4800 & 4 & 119 & 28.4 \\
Zw7160    & 16/08/99 & A100 & \ \ 91.740 & 500 & 4950 & 2 & 132 & 26.2 \\
Zw8197    & 16/08/99 & A100 & 103.540    & 500 & 5700 & 1 & 117 & 23.2 \\
 & & & & & & & \\ 
$^{13}$CO(1-0) & & & & & & & \\ 
A1068     & 01/08/99 & A100 & \ \ 96.788 & 500 & 7680  & 5  & 124 & 24.8 \\
 & & & & & & & \\ 
$^{12}$CO(2-1) & & & & & & & \\ 
A1068     & 16/08/99 & A230 & 202.487 & 500 &  \ \ 900 & 2 & 301 & 11.9 \\
\noalign{\smallskip \hrule}
\end{tabular}
\end{table*}

\begin{table*}
\caption{Log of {IRAM} Observations from 2000. The exposures marked with an asterisk are
those where only one receiver (A100) was available due to a fault in B100.}
\begin{tabular}{lclclcccc}
\noalign{\medskip \hrule \medskip}
Cluster & Date & Instrument & Frequency & Bandwidth & Exposure & Conditions & T$_{sys}$ & Beam Size \\
        &      &            &  (GHz)    & (MHz)     & (s)      & (mm H$_2$O) & (K) & ($''$) \\
 & & & & & & & & \\
$^{12}$CO(1-0) & & & & & & & & \\
RXJ0821+07 & 19/04/00 &  A100 & 103.760 & 500 & 1500 & 5  & 133 & 23.2 \\
A1068     & 19/04/00 &  A100 & 101.245 & 1000 & 2100 & 5 & 107 & 23.8 \\
A1664     & 19/04/00 &  A100 & 102.227 & 500 & 3600 & 5 & 160 & 23.5  \\
A2390     & 20/04/00 &  A100 & \ \ 93.585 & 500 & 3600 & 3 & 113 & 25.7 \\
A2597     & 20/04/00 &  A100 & 106.221 & 500 & 3900 & 4 & 138 & 22.6 \\
RXJ0352+19 & 20/04/00 &  A100 & 103.942 & 500 & 3600 & 3 & 132 & 23.1  \\
RXJ0439+05 & 20/04/00 &  A100 & \ \ 95.423 & 500 & 3600$^*$ & 2 & 114 & 25.2 \\
RXJ0338+09 & 21/04/00 &  A100 & 111.502 & 500 & 2700$^*$ & 2 & 148 & 21.6 \\
RXJ0338+09 & 21/04/00 &  A100 & 111.440 & 500 & 1800$^*$ & 2 & 150 & 21.6 \\
RXJ0439+05 & 21/04/00 &  A100 & \ \ 93.580 & 500 & 4500 & 2 & 114 & 25.2 \\ 
A2390     & 23/04/00 &  A100 & \ \ 93.635 & 500 & 2400 & 1 & 103 & 25.7 \\
A2597     & 23/04/00 &  A100 & 106.150 & 500 & 3600 & 1 & 117 & 22.7 \\
A262      & 23/04/00 &  A100 & 113.435 & 500 & 1800 & 1 & 148 & 21.2 \\
RXJ0338+09B& 23/04/00 &  A100 & 111.440 & 500 & 1950 & 2 & 137 & 21.6 \\
RXJ0352+19 & 23/04/00 &  A100 & 104.000 & 500 & 1800 & 1 & 105 & 23.1  \\
RXJ0747-19 & 23/04/00 &  A100 & 104.054 & 500 & 3600 & 1 & 124 & 23.1  \\
4C+55.16  & 23/04/00 &  A100 & \ \ 92.811 & 500 & 1800 & 1 & 104 & 25.9  \\
4C+55.16  & 23/04/00 &  A100 & \ \ 92.717 & 500 & 2400 & 1 & 104 & 25.9  \\
RXJ0338+09B& 24/04/00 &  A100 & 111.300 & 500 & 1800$^*$ & 2 & 122 & 21.6 \\
A262      & 24/04/00 &  A100 & 113.435 & 500 & 1800$^*$ & 2 & 137 & 21.2 \\
A478      & 24/04/00 &  A100 & 106.143 & 500 & 1800 & 1 & 109 & 22.7 \\
RXJ0747-19 & 24/04/00 &  A100 & 103.950 & 500 & 1800 & 2 & 124 & 23.1  \\
Hydra-A   & 24/04/00 &  A100 & 109.382 & 500 & 1500 & 2 & 137 & 22.0  \\
A646      & 24/04/00 &  A100 & 102.240 & 500 & 3600 & 2 & 107 & 23.5  \\
A2390     & 25/04/00 &  A100 & \ \ 93.750 & 500 & 2250 & 1 & 109 & 25.6 \\
Zw2089    & 16/06/00 &  A100 & \ \ 93.284 & 500 & 3900 & 7 & 130 & 25.7 \\
A1664     & 16/06/00 &  A100 & 102.280 & 500 & 4320 & 5 & 139 & 23.5  \\
Zw2089    & 17/06/00 &  A100 & \ \ 93.284 & 500 & 4500 & 5 & 131 & 25.7 \\
A1795     & 17/06/00 &  A100 & 108.419 & 500 & 4320 & 5 & 133 & 22.2  \\
Zw8276    & 25/07/00 &  A100 & 107.179 & 500 & 4050 & 6 & 136 & 22.4  \\
A2390     & 25/07/00 &  A100 & \ \ 93.750 & 500 & 3600 & 7 & 117 & 25.6 \\
A291      & 25/07/00 &  A100 & \ \ 96.381 & 500 & 4200 & 6 & 124 & 24.9 \\
A11       & 02/08/00 &  A100 & 100.34 & 500 & 2850 & 12 & 182 & 24.0 \\
A11       & 05/08/00 &  A100 & 100.25 & 500 & 3300 & 11 & 184 & 24.0 \\
 & & & & & & & \\ 
$^{12}$CO(2-1)  & & & & & & & \\ 
RXJ0821+07 &  19/04/00 &  A230 & 207.516 & 500 & 3600 & 5 & 352 & 11.6 \\
A1068     &  19/04/00 &  A230 & 202.486 & 1000 & 2100 & 5 & 331 & 11.9 \\
A2597     &  21/04/00 &  A230 & 212.336 & 1000 & 4800 & 2 & 306 & 11.3 \\
RXJ0352+19 &  23/04/00 &  A230 & 207.879 & 1000 & 1800 & 1 & 212 & 11.6  \\
RXJ0747-19 &  23/04/00 &  A230 & 208.104 & 1000 & 900 & 1 & 305 & 11.6  \\
RXJ0338+09B&  24/04/00 &  A230 & 222.877 & 1000 & 1800 & 2 & 222 & 10.8 \\
A262      &  24/04/00 &  A230 & 226.866 & 1000 & 1800 & 2 & 249 & 10.6 \\
A478      &  24/04/00 &  A230 & 212.282 & 1000 & 1800 & 1 & 228 & 11.3 \\
\noalign{\smallskip \hrule}
\end{tabular}
\end{table*}

\section{Analysis}

The {JCMT} data were analysed using the standard {\it STARLINK} reduction package {\it SPECX}.
After merging the separate sections of the DAS spectra and removing a linear
baseline from the spectra (excluding the  800 km~s$^{-1}$ around the line), the
spectra were co-added and binned to 12.5MHz and Hanning smoothed. 
Figs~1-3 present the spectra for CO(3-2), CO(4-3) and $^{13}$CO(3-2) for all our {JCMT} observations.

Fortuitously our observations on the 27$^{\rm th}$ December
were made in very dry and stable sky conditions which are
rarely used for low frequency observations.
The observations in this shift
were exceptionally stable in baseline shifts (rms of 0.7~mK
compared to 5.7~mK on the 26$^{\rm th}$). This stability results
in spectra far superior to all those in other sessions where
the water vapour level was substantially higher and/or the baseline
stability poorer. Therefore we have treated these spectra individually
and used data taken in other sessions for confirmation of these results.

\begin{table*}
\caption{Summary of {JCMT} results. All velocity shifts are quoted relative
to the published optical redshift and are not relative to the observed frequency.}
\begin{tabular}{lcclclcc}
\noalign{\medskip \hrule \medskip}
Cluster & Date & Line & Noise & Area & Peak & Width & velocity shift \\
        &      &      & (mK)  & (K km~s$^{-1}$) & (mK) & (km~s$^{-1}$) & (km~s$^{-1}$) \\
 & & & & & & & \\
Zw3146     & 26/12/98    & CO(3-2) & 1.7 & 0.90$\pm$0.33 & 3.2$\pm$1.3& 264$\pm$120 & $-$162$\pm$103 \\
           & 27/12/98    & CO(3-2) & 1.1 & 1.32$\pm$0.23 & 4.3$\pm$1.0 & 289$\pm$89 & $-$194$\pm$45 \\
           & 12/03/99    & CO(3-2) & 3.6 & 0.77$\pm$0.36 & 2.4$\pm$1.1 & 303$\pm$120 & $-$204$\pm$93 \\ 
           & 01/05/99    & CO(3-2) & 1.6 &  1.11$\pm$0.28 & 3.2$\pm$0.9 & 320$\pm$95 & $-$218$\pm$52 \\ 
           & 28-29/01/00 & CO(4-3) & 1.8 & 2.02$\pm$0.33 & 5.5$\pm$1.3& 345$\pm$90 & $-$280$\pm$44 \\ 
& & & & & & &  \\ 
RXJ1347$-$11 & 26-27/12/98 & CO(3-2) & 0.7 & $<$0.15 & $<$0.5 & 300Fixed & -300--300 \\
 & & & & & & &  \\
RXJ1532+30 & 26-27/12/98 & CO(3-2) & 2.6 & 0.74$\pm$0.25 & 4.3$\pm$2.1 & 173$\pm$103 & $-$88$\pm$35  \\
           & 28-29/01/00 & CO(4-3) & 1.9 & 1.00$\pm$0.20 & 2.3$\pm$0.9 & 457$\pm$103 & $-$189$\pm$35  \\ 
 & & & & & & &   \\
A1835      & 2-3/05/99   & CO(3-2) & 2.2 & 0.77$\pm$0.23 & 6.3$\pm$2.3 & 129$\pm$91 & $-$108$\pm$55 \\
           & 15/05/99    & CO(3-2) & 3.7 & 0.98$\pm$0.45 & 6.7$\pm$2.9 & 149$\pm$78  & $-$119$\pm$87  \\
           & 31/05/99    & CO(3-2) & 1.7 & 1.03$\pm$0.25 & 5.7$\pm$2.2 & 168$\pm$58  & $-$77$\pm$69  \\
           & 14-15/07/99 & $^{13}$CO(3-2) & 1.3 & $<$0.2 & $<$1.2 & 168Fixed & $-$77Fixed \\
 & & & & & & &  \\
Zw7160     & all data    & CO(3-2) & 2.1 & $<$0.25 & $<0.7$ & 380Fixed & $-$233Fixed \\
           & 20/07/97 & CO(4-3) & 3.7 & $<1.42$ & $<3.5$ & 380Fixed & $-$233Fixed \\
\noalign{\smallskip \hrule}
\end{tabular}
\end{table*}

The {IRAM} 30m data were analysed using the {\it CLASS} package. The
data were combined for both 500MHz backends and both autocorrelators
for the A100 and B100 receivers. The data had a linear baseline subtracted
and were Hanning smoothed to 8~MHz resolution. 

The weather through both the runs (largely day time observations) was variable
but only a few periods were poor (precipitable water above 8mm) although
fourteen hours were lost to high winds. Pointing observations were
made every hour and temperature calibrations every 30 minutes.
The pointing was always better than 5$''$ throughout all observations.

Table 7 gives the continuum value where one is detected.

\begin{table*}
\caption{Summary of {IRAM} results for 1999 data. The noise values are the rms in 8MHz bins.
Again all velocity shifts are quoted relative to the published optical redshift.}
\begin{tabular}{lcclclcc}
\noalign{\medskip \hrule \medskip}
Cluster & Date & Line & Noise & Area & Peak & Width & velocity shift \\
        &      &      & (mK)  & (K km~s$^{-1}$) & (mK) & (km~s$^{-1}$) & (km~s$^{-1}$) \\
 & & & & & & & \\
Zw3146     & 07/05/99    & CO(1-0) & 0.6 &  0.58$\pm$0.09 & 1.5$\pm$0.2 & 355$\pm$69 & $-$179$\pm$25\\
           & 12/08/99    & CO(1-0) & 0.6 &  0.74$\pm$0.13 & 1.7$\pm$0.3 & 412$\pm$92 & $-$178$\pm$39 \\
 & & & & & & &  \\ 
A1068      & 31/07/99    & CO(1-0) & 1.0 &  1.82$\pm$0.13 & 8.3$\pm$0.6 & 207$\pm$18 &  $-$16$\pm$7  \\
           & 01/08/99    & $^{13}$CO(1-0)  & 0.5 & $<0.08$ & $<$0.4 & 207Fixed & $-$16Fixed \\
           & 16/08/99    & CO(2-1) & 3.6 &  3.73$\pm$0.07 & 13.9$\pm$0.3 & 251$\pm$21 & $-$42$\pm$13 \\
           & 16/08/99    & CO(1-0) & 1.6 &  2.60$\pm$0.35 & 7.7$\pm$1.0 & 319$\pm$58 & $-$20$\pm$21 \\
 & & & & & & &  \\ 
A1835      & 31/07/99    & CO(1-0) & 0.8 &  1.08$\pm$0.13 & 4.5$\pm$0.3 & 227$\pm$38 & $-$105$\pm$12 \\
 & & & & & & &  \\ 
RXJ1532+30 & 12/08/99    & CO(1-0) & 0.7 &  0.34$\pm$0.09 & 1.5$\pm$0.6 & 217$\pm$60 & 97$\pm$30 \\
           & 16/08/99    & CO(1-0) & 0.6 &  0.73$\pm$0.11 & 1.4$\pm$0.3 & 472$\pm$85 & $-$169$\pm$37 \\ 
 & & & & & & &  \\ 
A2204      & 12/08/99    & CO(1-0) & 0.9 &  0.41$\pm$0.11 & 1.5$\pm$0.3 & 255$\pm$88 &  36$\pm$31 \\
           & 15/08/99    & CO(1-0) & 0.5 &  0.16$\pm$0.05 & 0.9$\pm$0.3 & 177$\pm$42 & $-$59$\pm$28 \\
 & & & & & & &  \\
Zw3916     & 13/08/99    & CO(1-0) & 0.8 &  $<$0.15 & $<$0.5 & 300Fixed & 0Fixed \\
 & & & & & & &  \\
Zw8193     & 13/08/99    & CO(1-0) & 0.6 &  0.51$\pm$0.08 & 1.5$\pm$0.3 & 313$\pm$50 & 57$\pm$23 \\
           &             &         &     &  0.52$\pm$0.07 & 1.8$\pm$0.3 & 268$\pm$35 & 460$\pm$19 \\
           & 14/08/99    & CO(1-0) & 1.2 &  0.72$\pm$0.15 & 2.1$\pm$0.7 & 313Fixed & 57Fixed \\
           &             &         &     &  $<$0.20 & $<0.6$ & 268Fixed & 460Fixed \\
 & & & & & & &  \\
A2390      & 13/08/99    & CO(1-0) & 0.8 &  0.30$\pm$0.08 & 1.4$\pm$0.5 & 203$\pm$61  & $-$88$\pm$30 \\
           &             &         &     &  0.13$\pm$0.08 & 1.9$\pm$0.8 & 60$\pm$38 & $-$410$\pm$11  \\
           & 15/08/99    & CO(1-0) & 0.7 &  0.37$\pm$0.08 & 1.5$\pm$0.5 & 230$\pm$48 & $-$3$\pm$24 \\
           &             &         &     &  0.33$\pm$0.07 & 1.7$\pm$0.6 & 179$\pm$50 & $-$454$\pm$22  \\
 & & & & & & &  \\
RXJ0821+07 & 14/08/99    & CO(1-0) & 1.1 &  1.30$\pm$0.14 & 9.7$\pm$1.0 & 126$\pm$16 & 256$\pm$6 \\
 & & & & & & &  \\
A2146      & 14/08/99    & CO(1-0) & 0.8 &  0.15$\pm$0.07 & 0.9$\pm$0.5 & 168$\pm$83 & $-$416$\pm$43 \\
           &             &         &     &  0.14$\pm$0.05 & 2.0$\pm$0.5 &  67$\pm$27 & 30$\pm$11 \\
 & & & & & & &  \\
Zw7160     & 14/08/99    & CO(1-0) & 0.6 &  0.41$\pm$0.09 & 1.0$\pm$0.3 & 380$\pm$77 & $-$233$\pm$43 \\
Zw7160     & 16/08/99    & CO(1-0) & 0.8 &  0.36$\pm$0.14 & 0.6$\pm$0.2 & 590$\pm$189 & $-$252$\pm$58 \\ 
 & & & & & & &  \\
Zw8197     & 14/08/99    & CO(1-0) & 0.7 &  0.32$\pm$0.07 & 1.3$\pm$0.4 & 240$\pm$55  & $-$75$\pm$276 \\
Zw8197     & 16/08/99    & CO(1-0) & 0.6 &  0.35$\pm$0.08 & 1.0$\pm$0.3 & 310$\pm$68 & $-$75$\pm$15 \\
 & & & & & & & \\ 
RXJ0352+19 & 15/08/99    & CO(1-0) & 2.4 &  $<$0.96       & $<$1.5 & 300Fixed & 0Fixed \\
 & & & & & & &  \\ 
A646       & 15/08/99    & CO(1-0) & 0.8 &  0.21$\pm$0.11 & 0.9$\pm$0.4 & 221$\pm$97 & 125$\pm$27 \\
\noalign{\smallskip \hrule}
\end{tabular}  
\end{table*}

\begin{table*}
\caption{Summary of {IRAM} results for 2000 data. The noise values are the rms in 8MHz bins}
\begin{tabular}{lcclclcc}
\noalign{\medskip \hrule \medskip}
Cluster & Date & Line & Noise & Area & Peak & Width & velocity shift \\
        &      &      & (mK)  & (K km~s$^{-1}$) & (mK) & (km~s$^{-1}$) & (km~s$^{-1}$) \\
 & & & & & & & \\
RXJ0821+07 & 19/04/00    & CO(1-0) & 1.1 &  1.64$\pm$0.12 & 8.1$\pm$2.1 & 191$\pm$17 & 260$\pm$7 \\
           & 19/04/00    & CO(2-1) & 3.1 &  2.90$\pm$0.19 & 20.0$\pm$3.0 & 137$\pm$11 & 260$\pm$5 \\
 & & & & & & & \\
A1068      & 19/04/00    & CO(1-0) & 1.2 &  1.75$\pm$0.17 & 7.3$\pm$2.0 & 226$\pm$25 & $-6\pm$11 \\
           & 19/04/00    & CO(2-1) & 4.0 &  5.74$\pm$0.31 & 22.2$\pm$3.0 & 243$\pm$13 & 2$\pm$6 \\
 & & & & & & & \\
A1664      & 19/04/00    & CO(1-0)  & 1.3 &  1.86$\pm$0.26 & 2.1$\pm$1.0 & 823$\pm$121 & $-52\pm$56 \\
A1664      & 16/06/00    &  CO(1-0)  & 1.1 &  1.10$\pm$0.18 & 1.7$\pm$0.4 & 592$\pm$99 & $-56\pm$76 \\ 
 & & & & & & & \\
A2390      & 20/04/00    & CO(1-0)  & 0.7 &  $<0.2$ & $<0.6$ & 300Fixed & -300--300 \\
           & 23/04/00    & CO(1-0)  & 1.0 &  $<0.2$ & $<1.0$ & 300Fixed & -300--300 \\
           & 25/04/00    & CO(1-0)  & 1.0 &  $<0.2$ & $<1.0$ & 300Fixed & -300--300 \\
           & 25/07/00    &  CO(1-0)  & 0.9 &  $<0.2$ & $<0.9$ & 300Fixed & -300--300 \\
 & & & & & & & \\
A2597      & 20/04/00    & CO(1-0)  & 0.9 &  0.33$\pm$0.11 & 1.1$\pm$0.4 & 278$\pm$88 & 108$\pm$48 \\ 
           & 21/04/00    & CO(2-1)  & 2.6 &  0.47$\pm$0.19 & 1.5$\pm$0.6 & 300Fixed & 130$\pm$67 \\ 
 & & & & & & & \\
RXJ0352+19  & 20/04/00    & CO(1-0)  & 0.8 &  0.41$\pm$0.11 & 1.1$\pm$0.3 & 350$\pm$76 & 21$\pm$42 \\
            & 23/04/00    & CO(1-0)  & 1.1 &  0.53$\pm$0.10 & 2.3$\pm$1.0 & 213$\pm$55 & 55$\pm$19 \\ 
            & 23/04/00    & CO(2-1)  & 3.1 &  1.20$\pm$0.24 & 3.7$\pm$1.4 & 304$\pm$65 & $-11\pm$31 \\ 
 & & & & & & & \\
RXJ0439+05  & 20/04/00    & CO(1-0)  & 1.0 &  $<0.3$ & $<1.0$ & 300Fixed & -300--300 \\
 & & & & & & & \\
RXJ0338+09  & 21/04/00    & CO(1-0)  & 1.1 &  1.29$\pm$0.14 & 3.2$\pm$1.7 & 376$\pm$40 & 176$\pm$21 \\
RXJ0338+09B & 23/04/00    & CO(1-0)  & 0.8 &  1.45$\pm$0.14 & 3.4$\pm$1.0 & 402$\pm$44 & 175$\pm$21 \\ 
            & 24/04/00    & CO(1-0)  & 1.8 &  1.05$\pm$0.25 & 3.8$\pm$1.2 & 256$\pm$62 & 170$\pm$31 \\ 
            & 24/04/00    & CO(2-1)  & 3.6 &  2.83$\pm$0.24 & 6.8$\pm$1.1 & 391$\pm$37 & 174$\pm$86 \\
 & & & & & & & \\
A262       & 23/04/00    & CO(1-0)  & 1.2 &  1.50$\pm$0.22 & 3.1$\pm$1.1 & 456$\pm$75 & 13$\pm$35 \\
           & 24/04/00    & CO(1-0)  & 1.2 &  1.68$\pm$0.20 & 3.3$\pm$1.0 & 473$\pm$60 & 18$\pm$27 \\
           & 24/04/00    & CO(2-1)  & 2.8 &  1.61$\pm$0.26 & 4.1$\pm$1.2 & 371$\pm$67 & 29$\pm$30 \\
 & & & & & & & \\
RXJ0747-19  & 23/04/00    & CO(1-0)  & 0.8 &  $<$0.30 & $<1.0$ & 300Fixed & -300--300 \\ 
            & 23/04/00    & CO(2-1)  & 5.9 &  $<$1.0 & $<3.0$ & 300Fixed & -300--300 \\
            & 24/04/00    & CO(1-0)  & 1.3 &  $<0.30$ & $<1.1$ & 300Fixed & -300--300 \\
 & & & & & & & \\
4C+55.16   & 23/04/00    & CO(1-0)  & 1.1 &  $<$0.30 & $<1.0$ & 300Fixed & -300--300 \\
 & & & & & & & \\
A478       & 24/04/00    & CO(1-0)  & 0.8 &  0.24$\pm$0.14 & 1.0$\pm$0.4 & 231$\pm$62 & $-154\pm$53 \\
           & 24/04/00    & CO(2-1)  & 2.9 &  0.54$\pm$0.17 & 4.4$\pm$1.0 & 120$\pm$16 & $-58\pm$18 \\
 & & & & & & & \\
Hydra-A    & 24/04/00    & CO(1-0)  & 1.3 &  $<0.3$ & $<1.1$ & 300Fixed & -300--300 \\ 
 & & & & & & & \\
A646       & 24/04/00    & CO(1-0)  & 0.6 &  $<0.2$ & $<0.5$ & 300Fixed & -300--300 \\
 & & & & & & & \\
Zw2089     & 16-17/06/00 &  CO(1-0)  & 0.5 &  0.11$\pm$0.04 & 1.5$\pm$1.0 & 67$\pm$33 & $-197\pm$13 \\ 
 & & & & & & & \\ 
A1795      & 17/06/00    &  CO(1-0)  & 1.0 &  0.23$\pm$0.06 & 2.1$\pm$1.1 & 100$\pm$33 & $-235\pm$15 \\
 & & & & & & & \\
Zw8276     & 25/07/00    &  CO(1-0)  & 0.6 &  0.60$\pm$0.09 & 1.3$\pm$0.3 & 440$\pm$62 & $-106\pm$34 \\ 
 & & & & & & & \\ 
A291       & 25/07/00    &  CO(1-0)  & 0.8 &  0.16$\pm$0.08 & 0.7$\pm$0.4 & 218$\pm$93 & 254$\pm$60 \\ 
 & & & & & & & \\
A11        & 02/08/00    &  CO(1-0)  & 1.1 &  0.46$\pm$0.12 & 2.2$\pm$1.1 & 191$\pm$56 & 103$\pm$24 \\ 
           &             &           &     &  0.45$\pm$0.13 & 2.2$\pm$1.0 & 188$\pm$72 & 535$\pm$31 \\  
           & 05/08/00    &  CO(1-0)  & 0.9 &  0.45$\pm$0.14 & 3.7$\pm$1.5 & 114$\pm$58 & 96$\pm$13 \\  
           &             &           &     &  0.44$\pm$0.14 & 2.0$\pm$0.9 & 209$\pm$88 & 676$\pm$24 \\
\noalign{\smallskip \hrule}
\end{tabular}  
\end{table*}

\begin{table*}
\caption{Summary of {IRAM} results for non-zero continuum levels. }
\begin{tabular}{llcc}
\noalign{\medskip \hrule \medskip}
Cluster   & Frequency & Continuum & Continuum\\
          &  (GHz)    &  (mK)     & (mJy)  \\
          &           & \\
RXJ0439+05 & \ \ 95.423 & 13$\pm$4 &  70$\pm$22 \\
RXJ0747-19 & 104.054 &  3$\pm$1 &  16$\pm$5 \\
A646       & 102.300 &  3$\pm$1 &  16$\pm$5 \\
4C+55.16   & \ \ 92.811 & 23$\pm$4 & 124$\pm$22 \\
Hydra-A    & 109.382 & 47$\pm$4 & 254$\pm$22 \\
Zw8193     & \ \ 97.210 &  3$\pm$1 &  16$\pm$5 \\
Zw8276     & 107.179 &  3$\pm$1 &  16$\pm$5 \\
A2390      & \ \ 93.853 &  3$\pm$1 &  16$\pm$5 \\
\noalign{\smallskip \hrule}
\end{tabular}  
\end{table*}

\section{Results}

In this section we discuss the individual detections and limits for each source
in order of first observation and then present results for global 
properties of the molecular gas in these systems. 

\noindent {\bf Zw3146} This source has our most comprehensive dataset and
is confirmed at several different observed frequencies and for CO(1-0), CO(3-2)
and CO(4-3). The line is 
relatively broad and slightly blue-shifted compared to the 
optically determined velocity (Allen et al. 1992). This apparent
blue-shift may result (at least in part) from the intrinsic uncertainties
in the optical spectra due to their low resolution. However,
there is a possibility that this velocity offset, and those
observed in other galaxies, is related to the discrete nature
of the molecular gas systems which are drawn from a broader velocity
range (i.e. the cluster core dispersion of 400--800 km~s$^{-1}$).
Combining the {JCMT} and {IRAM} results we derive a main beam brightness temperature
ratio for CO(3-2)/CO(1-0) of 0.83$\pm$0.20 (correcting for the beamsize
differences and beam efficiency) and CO(4-3)/CO(3-2) of 0.89$\pm$0.25. 
This is consistent with a temperature well in excess of 25~K.

\noindent {\bf RXJ1347$-$11} This is the most distant cluster observed
in this study but also the most massive known cooling flow (Allen 1998).
The high redshift precludes the observation of CO(1-0) with {IRAM}
as it falls below the lowest tunable frequency of the A/B100 receivers.
Our {JCMT} limit is particularly good thanks to the exceptionally 
favourable
conditions the observations were made under. There are several
factors that may weaken this limit. The optical published
spectroscopy for this galaxy is of low resolution (Schindler et al. 1995) so the 
true CO velocity may lie outside the narrow bandwidth used.
A velocity offset comparable to the largest we observe in CO(1-0)
between optical lines and CO could place the line outside the observed band. 
The only other published optical spectrum for the central galaxy is from
Sahu et al. (1998) and shows strong H$\alpha$ and [OI] 6300\AA.
From their published spectrum we make a crude estimate of the H$\alpha$ 
luminosity of 3$\pm2\times 10^{42}$ erg~s$^{-1}$.
Despite the non-detection of CO, the exceptional X-ray luminosity
and mass flow rate of this cluster make it an important testbed for cooling flow
predictions with future instrumentation (SOFIA, ALMA, {\it SIRTF}). 

\noindent {\bf RXJ1532+30} The operational difficulties with A3i 
during the JCMT observation of this galaxy prevent
any firm conclusions being drawn from the CO(3-2) data alone, but
when viewed with our {IRAM} CO(1-0) and {JCMT} CO(4-3) data,
a tentative CO(3-2) detection
can be claimed. Our limits on  CO(3-2)/CO(1-0) of $0.54\pm0.18$ 
and CO(4-3)/CO(3-2) of 0.77$\pm$0.33 are consistent with other
joint detections. This galaxy is our most massive molecular gas detection 
and is the second most optically line-luminous central cluster
galaxy in the Crawford et al. (1999) sample. This cluster has
erroneously been claimed to be an active galaxy (Fischer et al. 1998)
but {\it ROSAT HRI} imaging shows extended but strongly peaked emission 
consistent with a massive cooling flow (Crawford et al., in preparation).

\noindent {\bf A1835} The tuning range of A3i is just wide enough to
obtain CO(3-2) for this cluster in the upper side-band. Both the
{JCMT} and {IRAM} data show a consistent detection for this system.
These data give a ratio of  CO(3-2)/CO(1-0) of 0.47$\pm$0.12
and is consistent with an excitation temperature of $<$30~K. This
is less than the  temperature estimate
from dust emission from Edge et al. (1999) of 40$\pm$5~K.
Given the observational scatter in all the observations
and uncertainties of where the molecular emission lies
in the {JCMT} beam,
it would be premature to draw the conclusion that
the gas in A1835 is significantly colder than the dust
even if such differences are expected (Papadolpoulos et al. 2000).
However, the possibility that such low temperature 
components are present is an exciting one.
We also obtained a $^{13}$CO(3-2) spectrum for A1835 
(Figure 3) which shows no significant emission. The
data give a  $^{12}$CO/$^{13}$CO ratio of $>$5 which
rules out extremely optically thick clouds.

\noindent {\bf Zw7160} The {IRAM} data for this cluster are 
quite ambiguous. Our first CO(1-0) observation shows
a line at the $>3\sigma$ level which is also
present in a subsequent, velocity shifted observation
in better conditions but with quite different 
line properties.
Our {JCMT} data are not sensitive enough to detect the
CO(3-2) line as most of the data were taken in
comparatively poor conditions ($\tau_{CSO}>$0.12).
The ratio of CO(3-2)/CO(1-0) is $<0.55$ is
not restrictive. Our limit for CO(3-2) is consistent
with that of Chapman et al. (2000) for CO(4-3) using B3 on {JCMT}. 
The CO(4-3) data have a restricted bandwidth and the line is placed
close to the edge of the band making a confirmation of the CO
line impossible. Further {JCMT} B3 observations with a
wider bandwidth would provide a significantly better limit.
We claim a detection
for this source but wish to illustrate to the reader
the difficulty in determining reliable line properties
from such weak sources as we are working at the 
limits of single dish capabilities.

\noindent {\bf A1068} This is our strongest detection and
was significantly detected in the first 4 minutes of
data. We observed $^{13}$CO(1-0) and CO(2-1) for
this system with {IRAM}. The former observation gives
a firm limit on $^{12}$CO/$^{13}$CO of $>10$
indicating that the gas is not exceptionally
optically thick and the isotopic ratio
is comparable the value of 12 found in other galaxies 
(Young \& Sanders 1986, Aalto et al. 1995, Papadolpoulos \& Seaquist 1998). 
If this limit applies to all other $^{12}$CO detections then we predict
no $^{13}$CO detections are possible with currently available
instrumentation. The CO(2-1) data from August 1999 are likely
to give an underestimate of the total line 
intensity due to the narrow bandwidth used
(500~km~s$^{-1}$) but the April 2000 data was obtained
with the 1~GHz backend and is much more reliable.
The ratio of main beam brightness
temperatures for CO(2-1)/CO(1-0)
correcting for beam size and efficiency
is 0.71$\pm$0.08
so consistent {\it at face value} with a temperature of $<30$~K.

\noindent {\bf A2204} This is a relatively weak {IRAM} detection which
is clearer in the initial, shorter observation made in very
good conditions but is present in the follow-up observation
which was made in much poorer conditions.
Unfortunately the second observation was made without a 
frequency shift so our detection for this source not
as secure as some of the other weak detections. 
That caveat aside, this system is one of the strongest
optical line emitters within a redshift of 0.2 and
lies in the second most massive cooling flow in the
brightest 50 X-ray clusters (Peres et al. 1998),
so is of particular importance for future studies.

\noindent {\bf Zw3916} The {IRAM} data for this source
give a good upper limit. There is weak
excess at the low velocity end of the 
spectrum which could be a line offset from
the optical velocity by $>$600~km~s$^{-1}$
which is strongly affected by the baseline subtraction
but there was not sufficient time to 
make a velocity-offset observation
so this cannot be confirmed.

\noindent {\bf Zw8193} This peculiar source is the
most confused of the sample. The initial
optical spectrum of the central galaxy
gives significantly different redshifts for
the stellar features and the emission lines
(Allen et al. 1992). Recently K-band imaging and
spectroscopy indicates that the majority of
the line emission comes from the region of
a strong radio source offset by 2--3$''$
from the majority of the stellar continuum
and that this line emission shows a strong
velocity shear. There is also a strong flat
spectrum radio source present that 
shows up in the {IRAM} data as a baseline 
of 3mK (or 16mJy) implying that the flat ($\alpha=-0.5$)
spectrum continues to 100~GHz.
This intrinsic complexity
makes interpretation of the CO data 
non-trivial. The most that can be said from
our data is that there is no strong, narrow
($<300$~km~s$^{-1}$) component to the CO.
Our first spectrum (the longer of the two)
shows a very broad, flat-topped line. 
This line is consistent with two 300~km~s$^{-1}$
lines separated by 400~km~s$^{-1}$. Our second
spectrum shifts one of these lines to the
edge of our spectrum so the limits are
poor as the line is weaker 
but the data are consistent within the errors. 
However, the complexity of the line prevents
us claiming a CO detection in this system.
Given the high optical line luminosity of 
this galaxy, the upper limit derived is
of considerable importance (see Figure 9)
so this cooling flow demands further
attention at all wavelengths.

\noindent {\bf A2390} Given the wealth of data on this
cluster in the literature we were keen to 
obtain CO data for it. We were somewhat
surprised to find no detection of CO 
despite repeated, frequency-shifted observations
although the continuum level at 95~GHz is
close to that expected from the radio/sub-mm
spectrum of this galaxy in Edge et al.\ (1999).
One possibility is that the CO emission has
a very complex velocity structure.
The presence of multiple velocity 
components in CO is not unexpected if
the observed CO is related to discrete
clouds and is seen (or at least hinted at) in other distant 
galaxies (Papadopoulos et al. 2000). The
{\sl HST} image of the central galaxy in A2390
shows several `blue-knots' (Edge et al. 1999)
so the two lines could be related to those.
We note with interest the {\it STIS} spectrum of
this galaxy presented by Hutchings \& Balogh (2000)
which shows a strong velocity gradient over the
major axis of the central galaxy. The velocity
shifts between the `blue-knots' are around 400-600~km~s$^{-1}$
so we could be detecting CO associated with a number
of star-forming regions and no single narrow line is
found. Given the spatial separation
of the components in the {\it STIS} image (2$''$), it should
be possible to resolve these two components spatially
with current interferometers ({OVRO}, Plateau de Bure).

\noindent {\bf RXJ0821+07} Despite being made during some of the worst
conditions during the August 1999 run, the detection of this galaxy is
clear. The line is the narrowest and the most offset
in velocity from the optical value. The velocity offset
is at least in part due to the rounding introduced by
quoting redshifts to 3 decimal figures by Crawford et al (1995)
but is largely due to the strong velocity sheer
found in the optical lines themselves (Wilman \& Crawford, priv. comm.).
Significantly, this source is the third brightest {IRAS} source
in this study so there is a broad trend for the more luminous
far-infrared sources to be CO detections. Unfortunately the
{IRAS} data is not deep or uniform enough to provide a definitive
indicator.

\noindent {\bf A2146} The non-detection of this galaxy could be
due to the presence of a strong AGN component to the
ionization in this system (see Allen 1995). 
The X-ray emission around the galaxy is
extended so it is more reminiscent of Cygnus-A than
single, isolated AGN. The
X-ray imaging and optical spectroscopy for this cluster
indicate it has significantly different properties
from the majority of our other CO detections
in that the galaxy is not at the centre of the
cluster. The weak {IRAS} detection of the galaxy is
consistent with either interpretation.

\noindent {\bf Zw8197} This is one of our weakest CO detections
but is reproduced in a frequency-shifted observation
so we are confident of the detection. However, the
absolute line intensity is sensitive to the baseline
subtraction used so the error on the intensity is
large.

\noindent {\bf RXJ0352+19} This source is detected
in both CO(1-0) and CO(2-1) so we are confident of this
detection despite the small integrated intensity.

\noindent {\bf A646} There is  a very marginal detection of this
galaxy in one of our two observations so we do not
claim this as a detection.  
The continuum detection in this source
is consistent with a flat ($\alpha=-0.3$) source
from 1--100~GHz.

\noindent {\bf A1664}
This source is our most southerly target and was a challenging
observation. We obtained a frequency-shifted confirmation
of our April 2000 detection but could not obtain
a CO(2-1) confirmation to put this detection beyond doubt.

\noindent {\bf A2597}
This well-studied cooling flow has not been searched
for CO until now. We made both CO(1-0) and CO(2-1)
observations of A2597 and find a weak detection in
our first observation that has a marginal CO(2-1)
counterpart. The CO(1-0) detection is not confirmed
in our frequency shifted observation but it was
made in poorer conditions with the line close to the
edge of the bandwidth. This is one of our least
significant lines and needs much longer integrations
of CO(2-1).

\noindent {\bf RXJ0439+05}
This central galaxy contains one of the strongest radio sources
in the sample although the 1.4~GHz flux density is relatively
low due to the Giga-Hertz-peaked nature of the nucleus. We
detect the radio continuum at CO(1-0) and find no evidence for
a broad emission or narrow absorption lines.

\noindent {\bf RXJ0338+09}
The cluster was included in the study of Braine \& Dupraz (1994)
and a marginal line is visible in their figures. We observed this
galaxy in two positions; on the cD and a point 8$''$ NW of the
cD where the H$\alpha$ emission is strongest (Romanishin \& Hintzen 1988).
Both positions show evidence for a line but it is strongest
offset from the cD. The detection is confirmed at CO(2-1) and
has very recently been observed by {OVRO} so a clearer
picture of the extent of the CO emission will follow soon.

\noindent {\bf A262} 
Like RXJ0338+09, A262 is a cluster were we contradict the
non-detection of Braine \& Dupraz (1994). The CO(1-0) and
CO(2-1) detections give a relatively low ratio of C(2-1)/CO(1-0)
of 0.25 but this may be due to the CO emission overfilling the
CO(2-1) beam. 

\noindent {\bf RXJ0747-19 aka PKS0745-191}
This galaxy was observed on several occasions by us and has
been studied in CO by O'Dea et al. (1994) with {SEST}. The
smaller beamsize of {IRAM} 30m allows a significantly
better limit to be set in this paper. 
We find no evidence for a narrow line ($<500$~km~s$^{-1}$)
but a wide line is possible. The weak continuum is consistent
with the continuation of the steep radio spectrum.

\noindent {\bf A478}
This classic cooling flow has not been studied in any great
detail for CO  at least in part due to the poor
redshift of the central galaxy in the literature for
many years. Our CO(1-0) observation shows a marginal
detection but the CO(2-1) observation is more significant
and makes us confident of this detection.

\noindent {\bf 4C+55.16}
This cluster has only recently been recognised as a strong
cooling flow (Iwasawa et al.\ 1999) and contains
a remarkably bright, flat-spectrum radio source.
We clearly detect the continuum at CO(1-0) 
consistent with the 5--30~GHz spectral index of -1.3
but find no evidence for a line in emission or absorption.
This source may provide one of the strongest constraints
on CO in absorption given its strong, compact nucleus.

\noindent {\bf Hydra-A}
This galaxy was observed once by us and has
been studied in CO(2-1) by McNamara \& Jaffe (1994).
We find no evidence for a narrow line ($<500$~km~s$^{-1}$)
but a wide line is possible. The continuum level is consistent
with the lower frequency radio data but the published
continuum level of 17~mK, or 267~mJy, at 218~GHz by  McNamara \& Jaffe (1994)
implies that the radio/sub-mm spectrum may flatten around 150~GHz.
The continuum data implies that {SCUBA} would detect around 100~mJy
at 850~$\mu$m.

\noindent {\bf Zw2089}
This BCS-selected cooling flow is not part of the Crawford
et al.\ (1999) spectroscopic sample but is a strong optical
line-emitter. Our observations show no strong line
but do indicate a repeated narrow line. We do not
claim any detection but further CO observations
may be justified in the future.

\noindent {\bf A1795} 
This galaxy was observed once by us and has
been studied in CO by two of groups (Ulmer \& Grabelsky 1990;
Braine \& Dupraz 1994).
We find a marginal line in a single service observation
but this claim requires a frequency-shifted reconfirmation
and/or CO(2-1) observations before it can be regarded as
secure. This galaxy has an unusual optical line emission
filament (see Fabian et al.\ 2001a) so further CO observations
are necessary to search for similarly extended molecular gas.

\noindent {\bf Zw8276}
Again, this galaxy has a detection in a single service observation
but is strong enough to lead us to claim without confirmation. 
The continuum level detected implies a flat ($\alpha=-0.4$)
radio source from 1--100~GHz.

\noindent {\bf A291} No line was found for this relatively anonymous system which is
amongst the lowest optical line luminous systems observed. Our upper limit lies 
above the line that marks the highest molecular mass detections.

\noindent {\bf A11} This cluster was included in this study due to its remarkably
strong optical emission line spectrum (Perlman et al. 1998). The {IRAM} data
were taken in service during relatively poor conditions but show a line at the same frequency
at two different tunings.

Overall we have seven high significance CO(1-0) detections,  which
have additional CO(2-1), CO(3-2) or CO(4-3) detections, and
another nine CO(1-0) detections which close to the limit of detectability
with {IRAM} but confirmed with repeat observations. 
Even if several of our tentative detections prove
to be spurious, these observations represent more than an order of
magnitude increase in the number of CO detections for cooling flows.
The prospects for increasing this number of detections further may
be limited as the number of systems with strong optical emission
lines is relatively small and we have observed the vast majority of 
these in this study.

We have calculated molecular gas masses from the
relationship used in Sanders, Scoville \&  Soifer (1991):
which includes 1.36 factor to account for the contribution of
Helium to the total gas mass:

\begin{equation}
\ \ \ \ \ \ \rm M(H_2)= 1.18 \times 10^{4} S(CO) d_{Mpc}^2  M_\odot
\end{equation}

where S(CO) is the integrated flux density of the line
(Jy km~s~$^{-1}$) determined from the measured antenna 
temperature (T$_{\rm A}^*$)
using 6.8 $(1+z)^{-\frac{1}{2}}$ Jy~K$^{-1}$ for {IRAM}
and $d_{Mpc}$ is the luminosity distance to the cluster.
This conversion excludes the factor of 1.36 to account for 
the contribution of Helium to the total gas mass. We
use molecular gas mass throughout this paper.
With one exception,
this formula agrees with all other quoted conversion factors in 
previous cooling flow CO search papers 
(Ulmer \& Grabelsky 1990; McNamara \& Jaffe 1994; O'Dea et al.\ 1994;
Braine \& Dupraz 1994; Fujita et al.\ 2000) to within 20\%
taking into account differences in H$_0$, 
use of simple approximation about beam area that doesn't
correctly account for (1$+z$) effects,
number of $\sigma$ the results are quoted to,
the assumed width of the undetected line,
beam size, the Galactic CO to H$_2$ conversion factor
(Sanders et al.\  use $3 \times 10^{20}$ cm$^{-2}$ (K km~s$^{-1}$)$^{-1}$)
and use of beam efficiencies (0.842 used for all {IRAM} 30m 
CO(1-0) observations).
The exception is O'Dea et al.\ (1994) where the gas mass limits
are a factor of four too low given the formula they quote in
Equation 4 compared to the comparable one in Braine \& Dupraz (1994).
This discrepancy arises due to the incorrect use of a radius
rather than a diameter in these calculations (O'Dea, priv. comm.).
Given the cumulative
uncertainties in all these previous papers, we present a direct
calculation {\it using the same assumptions} for all previous papers
(including  O'Dea et al.\ 1994) in Table~9 for comparison to our results.

Clearly these crude conversions may not be appropriate in the cooling flow
systems so they should be viewed very much as a guideline.
For warmer gas than in the Galaxy we could
overestimate (e.g. in ULIRGs Solomon et al.\ (1997)
claim a factor of four overestimate). If the gas
is colder ($<10$~K), then this relationship could be
substantially underestimated.
Whichever is the case, the overall proportions will scale between
the galaxies in the sample.
The values we calculate
are presented in Table 8 with other relevant 
parameters and other data from the literature.
Overall these observations give molecular gas mass estimates
in the range 8$\times 10^8$ to 2.5$\times 10^{11}$~M$_\odot$. 

\scriptsize

\begin{table*}
\caption{Summary of derived parameters. All upper limits are 3$\sigma$.
Mass flow rates are from Allen et al. (1995) and Allen (2000) and are corrected for
excess absorption. The gas mass for RXJ1347$-$11 is derived from our CO(3-2) limit assuming CO(3-2)/CO(1-0) of 0.4.
The optical line luminosities are from Crawford et al.\ (1999) apart from A11 (Perlman, priv. comm.) and estimates
for 4C+55.19, Zw2089, I0910+41, 3C48 and A1367/3C264 from inspection of published spectra. The dust masses are calculated
for a dust temperature of 40~K. Note gas-to-dust ratios are for molecular gas only and an additional factor of 1.36
is required for total gas-to-dust ratios.}
\begin{tabular}{llccccccc}
\noalign{\medskip \hrule \medskip}

\scriptsize

cluster   &  redshift &  optical line &  Mass flow rate & {SCUBA} flux         & {IRAS} flux  & dust mass  & Molecular gas   & gas-to-dust \\
          &           &  luminosity   & (M$_\odot$ yr$^{-1}$) & at 850$\mu$m & at 60$\mu$m  & (M$_\odot$)& mass estimate & ratio       \\
          &           & (erg s$^{-1}$) &             &    (mJy)               & (mJy)      &           & (M$_\odot$)     & \\
 & & & & & & &  \\ 
A11        & 0.1503   & 1.0$\times 10^{42}$ &   -- & --   & 100$\pm$40 & 1.4$\times 10^7$  & 2.6$\pm0.7\times 10^{10}$  & 1820    \\
A262       & 0.0171   & 6.0$\times 10^{39}$ &   27 & --   & 290$\pm$24 & 4.4$\times 10^5$  & 9.0$\pm1.3\times 10^{8}$   & 2040    \\
A291       & 0.196    & 4.6$\times 10^{41}$ &  --  & --   & 110$\pm$35 & 2.9$\times 10^7$  & $<2.3\times 10^{10}$       & $<$795  \\
RXJ0338+09 & 0.0338   & 1.0$\times 10^{41}$ &  325 & --   & 120$\pm$33 & 7.3$\times 10^5$  & 3.9$\pm0.4\times 10^{9}$   & 5350    \\
RXJ0352+19 & 0.109    & 5.8$\times 10^{41}$ &  --  & --   & $<$99      & $<6.9\times 10^6$ & 1.2$\pm0.3\times 10^{10}$  & $>$1700 \\
A478       & 0.0882   & 1.1$\times 10^{41}$ &  616 & --   & 230$\pm$56 & $1.0\times 10^7$  & 4.5$\pm2.6\times 10^{9}$   & 450     \\
RXJ0439+05 & 0.208    & 1.1$\times 10^{42}$ &  --  & --   & $<$102     & $<3.1\times 10^7$ & $<3.3 \times 10^{10}$      & --      \\
RXJ0747$-$19 & 0.1028   & 1.4$\times 10^{42}$ & 1038 & --   & $<$288     & $<1.8\times 10^7$ & $<7.6 \times 10^{9}$       & --      \\
A646       & 0.1268   & 1.6$\times 10^{41}$ &  --  & --   & $<$132     & $<1.3\times 10^7$ & $<1.3 \times 10^{10}$      & --      \\
RXJ0821+07 & 0.110    & 3.0$\times 10^{41}$ &  --  & --   & 300$\pm$30 & $2.2\times 10^7$  & 3.9$\pm0.4 \times 10^{10}$ & 1750    \\
4C+55.19   & 0.242    & $\approx 10^{42}$   &  --  & --   & $<$105     & $<4.6\times 10^7$ & $<4.5 \times 10^{10}$      & --      \\
Zw2089     & 0.235    & $\approx 10^{42}$   &  --  & --   & $<$105     & $<4.2\times 10^7$ & $<4.6 \times 10^{10}$      & --      \\
Hydra-A    & 0.052    & 1.6$\times 10^{41}$ &  264 & --   & 90$\pm$33  & $1.3\times 10^6$  & $<2.0 \times 10^{9}$       & $<$1540 \\
Zw3146     & 0.2906   & 7.0$\times 10^{42}$ & 1358 & 6.6  & 80$\pm$30  & $2.2\times 10^8$  & 1.6$\pm0.3 \times 10^{11}$ & 740     \\
A1068      & 0.1386   & 1.7$\times 10^{42}$ &  937 & --   & 650$\pm$39 & $7.7\times 10^7$  & 8.5$\pm0.6 \times 10^{10}$ & 1110    \\
Zw3916     & 0.204    & 3.0$\times 10^{41}$ &  --  & --   & $<$123     & $<3.5\times 10^8$ & $<1.6 \times 10^{10}$      & --      \\
A1664      & 0.1276   & 1.1$\times 10^{42}$ &  260 & --   & $<$159     & $<1.6\times 10^7$ & 4.4$\pm0.7 \times 10^{10}$ & $>$2720 \\
RXJ1347$-$11 & 0.4503   & 3.0$\times 10^{42}$ & 1790 & 3.5  & $<$132     & $1.8\times 10^8$  & $<6.8 \times 10^{10}$      & $<$375  \\
A1795      & 0.0620   & 1.1$\times 10^{41}$ &  381 & --   & $<$147     & $<3.1\times 10^6$ & $<2.7 \times 10^{9}$       & ---     \\
A1835      & 0.2523   & 1.4$\times 10^{42}$ & 2111 & 4.4  & 330$\pm$69 & $1.0\times 10^8$  & $1.8\pm0.2 \times 10^{11}$ & 1760    \\
Zw7160     & 0.2578   & 5.0$\times 10^{41}$ & 1227 & 5.3  & $<$87      & $1.5\times 10^8$  & $6.1\pm2.4 \times 10^{10}$ & 410     \\
RXJ1532+30 & 0.3615   & 4.2$\times 10^{42}$ &  --  & --   & $<$99      & $<1.2\times 10^8$ & $2.5\pm0.4 \times 10^{11}$ & $>$2110 \\
A2146      & 0.2343   & 1.4$\times 10^{42}$ &  --  & --   & 140$\pm$26 & $5.6\times 10^7$  & $<3.5 \times 10^{10}$      & $<$620  \\
A2204      & 0.1514   & 1.6$\times 10^{42}$ & 1660 & --   & $<297$     & $<4.3\times 10^7$ & $2.3\pm0.6 \times 10^{10}$ & $>$540  \\
Zw8193     & 0.1825   & 1.5$\times 10^{42}$ &  --  & --   & $<$99      & $<2.2\times 10^7$ & $<4.3 \times 10^{10}$      & --      \\
Zw8197     & 0.1140   & 1.6$\times 10^{41}$ &  --  & --   & $<$87      & $<6.7\times 10^6$ & $1.1\pm0.3 \times 10^{10}$ & $>$1640 \\
Zw8276     & 0.0757   & 1.3$\times 10^{41}$ &  --  & --   & 80$\pm$22  & $2.6\times 10^6$  & $8.2\pm1.2 \times 10^{9}$  & 3140    \\
A2390      & 0.2328   & 6.2$\times 10^{41}$ &  600 & 4.8  & $<$162     & $9.0\times 10^7$  & $<4.9 \times 10^{10}$      & $<$540  \\
A2597      & 0.0852   & 5.2$\times 10^{41}$ &  271 & --   & $<$100     & $<4.1\times 10^6$ & $8.1\pm3.3 \times 10^{9}$  & $>$1990 \\
           &          &                     &      &      &            &                   &                            &         \\
NGC1275    & 0.0184   & 4.7$\times 10^{42}$ &  556 & --   & 35000      & $5.3\times 10^7$  & $1.7\pm0.2 \times 10^{10}$ & 323     \\
I09104+41  & 0.4420   &  $>1\times 10^{42}$ & 1060 & $<6.4$ &  400       & $1.6\times 10^8$  & $<5.1 \times 10^{10}$      & $<319$  \\
3C48       & 0.3695   & $>1\times 10^{42}$  &  300 & --   & 761        & $2.0\times 10^8$  & $1.6\pm0.6 \times 10^{10}$ & 80      \\
R0107+32   & 0.0175   & 6.0$\times 10^{39}$ &  --  & --   & 360$\pm$63 & $5.8 \times 10^5$ & $2.2\pm0.3 \times 10^{9}$  & 3793    \\
A1367      & 0.0218   & 5.0$\times 10^{39}$ &    0 & --   & 160$\pm$58 & $4.1 \times 10^5$ & $5.2\pm1.0 \times 10^{8}$  & 1268    \\

\noalign{\smallskip \hrule}
\end{tabular}
\end{table*}

\begin{table*}
\caption{Summary of derived parameters from previous papers. All intensities are corrected for beam efficiency and assume
a line width of 300~km~s$^{-1}$. The limits derived from CO(2-1) data assume CO(2-1)/CO(1-0) of 0.6 as used by
McNamara \& Jaffe (1994). The mass flow rates are taken from Peres et al.\ (1998) or White, Jones \& Forman (1997).
The optical line luminosities are from Crawford et al.\ (1999), the compilation  of Heckman et al. (1989)
or Owen et al.\ (1995).
The references are J87 - Jaffe (1987); BH88 - Bregman \& Hogg (1988) - BH88; Ulmer \& Grabelsky (1990) - UG90; 
McNamara \& Jaffe (1994) - MJ94; Braine \& Dupraz (1994) - BD94; O'Dea et al.\ (1994) - O94;
Fujita et al.\ (2000) - F00; Lim et al.\ (2001) - L01.}
\begin{tabular}{lllcccccc}
\noalign{\medskip \hrule \medskip}

cluster & reference & Telescope & line    & beam          & 3$\sigma$  intensity & molecular  & mass flow & optical line \\
        &           &           &         & diameter      & limit       & gas mass      & rate &  luminosity\\
        &           &           &         & (arcsec)      & (K~kms$^{-1}$)                      & & (M$_\odot$~yr$^{-1}$) & (erg~s$^{-1}$)\\
        &           &           &         &               &                      &               & & \\
M87     & J87       & NRAO-12m  & CO(1-0) &  56           & $<$1.35                & $<1.3\times 10^{8}$              & -- & 1.1$\times 10^{40}$\\
        &           &           &         &               &                      & & & \\
MKW1    & BH88      & NRAO-12m  & CO(1-0) &  56           & $<$0.35                & $<1.6\times 10^{9}$              & -- & -- \\
R0338+096 & BH88    & NRAO-12m  & CO(1-0) &  56           & $<$0.45                & $<5.5\times 10^{9}$              & 325 & 1.0$\times 10^{41}$\\
A1126   & BH88      & NRAO-12m  & CO(1-0) &  59           & $<$0.12                & $<3.6\times 10^{9}$             & -- & 5.4$\times 10^{41}$\\
A2199   & BH88      & NRAO-12m  & CO(1-0) &  56           & $<$0.30                & $<2.9\times 10^{9}$             & 154 &  3.5$\times 10^{40}$\\
        &           &           &         &               &                        & & \\
A262    & UG90      & NRAO-12m  & CO(1-0) &  56           & $<$1.24                & $<3.3\times 10^{9}$              & 27 & 6.0$\times 10^{39}$\\
A496    & UG90      & NRAO-12m  & CO(1-0) &  56           & $<$0.81                & $<9.3\times 10^{9}$              & 95 & 3.4$\times 10^{40}$ \\
A978    & UG90      & NRAO-12m  & CO(1-0) &  59           & $<$1.27                & $<4.8\times 10^{10}$              & -- & $<2\times 10^{40}$\\
A1126   & UG90      & NRAO-12m  & CO(1-0) &  59           & $<$0.65                & $<3.8\times 10^{10}$              & -- & 5.4$\times 10^{41}$ \\
A1185   & UG90      & NRAO-12m  & CO(1-0) &  57           & $<$0.83                & $<1.1\times 10^{10}$              & 0 & $<5\times 10^{39}$\\
A1795   & UG90      & NRAO-12m  & CO(1-0) &  58           & $<$0.57                & $<2.4\times 10^{10}$              & 381 & 1.1$\times 10^{41}$ \\
A1983   & UG90      & NRAO-12m  & CO(1-0) &  57           & $<$1.00                & $<2.2\times 10^{10}$              & 6 & $<7\times 10^{39}$\\
A2052   & UG90      & NRAO-12m  & CO(1-0) &  57           & $<$1.78                & $<2.3\times 10^{10}$              & 125 & 4.8$\times 10^{40}$\\
A2199   & UG90      & NRAO-12m  & CO(1-0) &  56           & $<$1.49                & $<1.4\times 10^{10}$              & 154 & 3.5$\times 10^{40}$\\
A2319   & UG90      & NRAO-12m  & CO(1-0) &  58           & $<$0.56                & $<1.7\times 10^{10}$              & 20 & $<1\times 10^{41}$\\
R0338+096 & UG90    & NRAO-12m  & CO(1-0) &  56           & $<$0.59                & $<7.1\times 10^{9}$              & 325 & 1.0$\times 10^{41}$\\
        &           &           &         &               &                      & & & \\
Hydra-A & MJ94   & JCMT-15m  & CO(2-1) & 22 & $<$0.55                 &  $<4.5\times 10^{9}$            & 264 & 1.6$\times 10^{41}$\\
A1060   & MJ94   & JCMT-15m  & CO(2-1) & 21 & $<$0.96                 &  $<4.1\times 10^{8}$            & 15  & \\
MKW3s   & MJ94   & JCMT-15m  & CO(2-1) & 22 & $<$0.42                 &  $<2.3\times 10^{9}$            & 175 & $<5\times 10^{39}$\\
A2151   & MJ94   & JCMT-15m  & CO(2-1) & 22 & $<$0.61                 &  $<2.1\times 10^{9}$            & 166 & $<5\times 10^{39}$\\
A2256   & MJ94   & JCMT-15m  & CO(2-1) & 22 & $<$0.94                 &  $<9.5\times 10^{9}$            & 0   & $<5\times 10^{40}$ \\
Cygnus-A & MJ94  & JCMT-15m  & CO(2-1) & 22 & $<$0.99                 &  $<9.0\times 10^{9}$            & 244 & 6.5$\times 10^{42}$\\
        &           &           &         &               &                      & & & \\
A262    & BD94    & IRAM-30m  & CO(1-0) & 21 & $<$0.60                & $<3.1\times 10^{8}$              & 27 & 6.0$\times 10^{39}$\\
A262    & BD94    & IRAM-30m  & CO(2-1) & 11 & $<$1.05                & $<2.3\times 10^{8}$              & 27 & 6.0$\times 10^{39}$\\
R0338+096 & BD94  & IRAM-30m  & CO(1-0) & 22 & $<$0.66                & $<1.5\times 10^{9}$              & 325 & 1.0$\times 10^{41}$\\
R0338+096 & BD94  & IRAM-30m  & CO(2-1) & 11 & $<$0.93                & $<9.3\times 10^{8}$              & 325 & 1.0$\times 10^{41}$\\
A478    & BD94    & IRAM-30m  & CO(1-0) & 23 & $<$0.45                & $<7.3\times 10^{9}$              & 616 & 1.1$\times 10^{41}$\\
A478    & BD94    & IRAM-30m  & CO(2-1) & 11 & $<$0.90                & $<6.1\times 10^{9}$              & 616 & 1.1$\times 10^{41}$\\
Hydra-A & BD94    & IRAM-30m  & CO(1-0) & 22 & $<$0.66                & $<3.9\times 10^{9}$              & 264 & 1.6$\times 10^{41}$\\
Hydra-A & BD94    & IRAM-30m  & CO(2-1) & 11 & $<$0.87                & $<2.1\times 10^{9}$              & 264 & 1.6$\times 10^{41}$\\
A1795   & BD94    & IRAM-30m  & CO(1-0) & 22 & $<$0.54                & $<4.4\times 10^{9}$              & 381 & 1.1$\times 10^{41}$\\
A1795   & BD94    & IRAM-30m  & CO(2-1) & 11 & $<$0.87                & $<3.0\times 10^{9}$              & 381 & 1.1$\times 10^{41}$\\
A2029   & BD94    & IRAM-30m  & CO(1-0) & 23 & $<$0.45                & $<5.7\times 10^{9}$              & 556 & $<8\times 10^{39}$\\
A2029   & BD94    & IRAM-30m  & CO(2-1) & 11 & $<$1.11                & $<5.8\times 10^{9}$              & 556 & $<8\times 10^{39}$\\
A2052   & BD94    & IRAM-30m  & CO(1-0) & 22 & $<$0.39                & $<9.7\times 10^{8}$              & 125 & 4.8$\times 10^{40}$\\
A2052   & BD94    & IRAM-30m  & CO(2-1) & 11 & $<$0.24                & $<2.5\times 10^{8}$              & 125 & 4.8$\times 10^{40}$\\
A2199   & BD94    & IRAM-30m  & CO(1-0) & 22 & $<$0.54                & $<1.0\times 10^{9}$              & 154 & 3.5$\times 10^{40}$\\
A2199   & BD94    & IRAM-30m  & CO(2-1) & 11 & $<$0.69                & $<5.3\times 10^{8}$              & 154 & 3.5$\times 10^{40}$\\
        &           &           &         &               &                      &  & & \\
P0745-19 & O94    & SEST-15m  & CO(1-0) & 47 & $<$0.57                & $<4.2\times 10^{10}$             & 1038 & 2.9$\times 10^{42}$\\
Hydra-A & O94     & SEST-15m  & CO(1-0) & 45 & $<$0.66                & $<1.3\times 10^{10}$             & 264  & 1.6$\times 10^{41}$\\
A3526   & O94     & SEST-15m  & CO(1-0) & 43 & $<$0.66                & $<5.0\times 10^{8}$              & 30   & 5.0$\times 10^{39}$\\
A3526   & O94     & SEST-15m  & CO(2-1) & 22 & $<$1.80                & $<5.6\times 10^{8}$              & 30   & 5.0$\times 10^{39}$\\
        &           &           &         &               &                      &  & & \\
AWM7    & F00     & NRO-45m   & CO(1-0) & 15 & $<$1.39                & $<3.5\times 10^{8}$              & 41   & $<$5.0$\times 10^{39}$\\
        &           &           &         &               &                      &  & & \\
RXJ0107+32   & L01     & IRAM-30m  & CO(1-0) & 22 & 3.74$\pm$0.49 & 2.2$\pm0.3 \times 10^{9}$ &  -- & 5.0$\times 10^{39}$\\
A1367        & L01     & IRAM-30m  & CO(1-0) & 22 & 0.55$\pm$0.10 & 5.2$\pm1.0 \times 10^{8}$ &  -- & 4.0$\times 10^{39}$\\
\noalign{\smallskip \hrule}
\end{tabular}
\end{table*}

\normalsize


\section{Discussion}

The detection of CO in sixteen massive cooling flows and 
sensitive limits on a further thirteen has significant 
implications for our understanding of the deposition of matter in the cores
of these systems. 

\subsection{Previous observations}

The first question posed by our results is why haven't more
cooling flows been found to contain molecular gas? The
principle reason for this lies in the small number of extreme
cooling flows that lie within a redshift of 0.1. Perseus/NGC1275
is the only cooling flow with $z<0.1$ with a central galaxy with
an optical line luminosity in excess of $10^{42}$ erg s$^{-1}$
to have been observed and detected at CO(1-0) (see Bridges \& Irwin 1998). 
The only other such system
to have been studied is PKS~0745-191 (O'Dea et al. 1994).
 If a common mechanism lies behind the excitation of optical lines
and warming of molecular gas, either by the UV continuum of massive
stars (Allen 1995; Crawford et al. 1999) or X-ray excitation
(Voit \& Donahue 1995; Wilman et al. 2001), then 
the luminosity of the two should be related linearly.
Figure~9 shows the mass of molecular gas plotted against
H$\alpha$ luminosity for our detections and upper limits plus NGC1275 and
3-$\sigma$ upper limits from Grabelsky \& Ulmer (1990),
McNamara \& Jaffe (1992), O'Dea et al. (1994)
and Braine \& Dupraz (1994). We also plot the recent detections
of CO in 3C31 (the central galaxy in an X-ray luminous group and
part of the BCS sample) and 3C264 (the dominant galaxy in A1367,
also in the BCS) of Lim et al.\ (2001) which are
very different in appearance (double-peaked velocity profiles)
but comparable in molecular Hydrogen mass. While
the large majority of the plot is unconstrained, it is
clear that the most line luminous sources are more likely
to be found in current CO searches and a linear relation
between molecular gas and emission line luminosity is
consistent with the data.  The relationship plotted in Figure~9
is of one luminosity against another so care must be taken in
interpreting it. That said, this paper presents data on a
complete sample of luminous optical line emission systems 
($L_{H\alpha}>10^{41.5}$ erg~s$^{-1}$) where 11 out of 21 are detected
and only 2 systems have with upper limits below scatter in the
detected objects (PKS~0745-191 and RXJ0439+05, which both
host powerful radio sources). Therefore the
observed correlation holds for the majority of luminous
optical line systems and is not an artifact.
If this relation holds
then previous surveys may have only {\it just} missed
detecting CO. The effect of the beam size 
compared to the possible size of the CO emission
($<40$~kpc for z$<$0.1 for most observations)
and use of smaller telescope diameter
also acts to weaken the limits set by previous
observations.  
The high detection rate for CO in this paper
is largely due to the selection of the newly discovered
line-luminous systems found in the {\it ROSAT} All-Sky Survey
(Crawford et al. 1999) which draws massive cooling flows
from a much larger volume and the smaller beamsize
provided by the {IRAM} 30m. 

An additional factor that can account for
some of the observed differences is the velocity
shifts seen in our data were not accounted for
in previous studies so some detections may have been
overlooked in the past.
Indeed, inspecting the spectra
of Braine \& Dupraz (1994) shows possible detections
of A262, A478 and 3A0335+09 with $\approx200$~km~s$^{-1}$
offset from zero velocity. All three of these galaxies
are detected in our {IRAM} 30m data. Although A478 is only
a marginal detection, both A262 and  3A0335+09 are
detected at 5--7 times the level quoted as a limit
by Braine \& Dupraz (1994). We can only account for
this discrepancy if the authors subtracted any velocity
offset emission in the baseline subtraction. 

{ As global mass deposition rates are known for the majority of the
sample presented here and all the published upper limits,
we can plot the molecular gas mass against mass deposition rate
(Figure 10). This plot illustrates that there are a 
number of very massive cooling flows without strong optical
emission lines (e.g. A478, A1795 and A2029)
where the upper limits/detections on the molecular gas mass are restrictive.
Figure 10 implies that the observed mass of molecular gas
is {\it not directly} related to the total mass deposited. 
There may be a correlation with the mass deposited on smaller scales
($<$50kpc) but the high resolution X-ray data has not
yet been obtained to test this. However, the
upper bound on the molecular gas mass is correlated with the
mass deposition rate. This is expected if the observed molecular gas
is a small percentage of the total deposited gas that is warmed
by young stars or an active nucleus.}

The relationship plotted in Figure~9, although a
luminosity-luminosity plot, has some predictive 
power if the optical line luminosity and
Balmer decrement are known. For instance all
sources with an optical line luminosity
less than 10$^{41}$ erg~s$^{-1}$ in 
Crawford et al. (1999) are likely to have
a CO(1-0) line intensity below $0.3$K~km~s$^{-1}$ 
so beyond the observable limits of {IRAM} 30m. 
More distant radio galaxies and other
starbursts with CO detections should have
a similar ratio of CO to H$\alpha$ observed
here.

There is one other factor that also plays a role; the overall
radio power. There are a number of 
central cluster galaxies with luminous optical line emission
that should be detectable using the
relationship in  Figure~9 (e.g. PKS0745-191, Hydra-A, Cygnus-A) but
are so far undetected. Of the eight most radio powerful
radio sources ($>2\times10^{25}$~W~Hz$^{-1}$ at 1.4~GHz), only 3C48
is detected even though several of them have CO
limits an order of magnitude below the level of
detections at the same H$\alpha$ luminosity.
This could be caused by an additional
optical line contribution powered by the radio source
as is clearly the case in 3C48 which has broad H$\alpha$
(Jackson \& Browne 1991). In the less extreme
cases the radio source and optical line morphologies
do correspond (e.g. PKS0745-191, Donahue et al.\ 2000) so the line
emission in these systems may be directly
powered (at least in part) by the radio source
and hence the relationship in Figure 9 does not
hold. Alternatively, these powerful radio systems could
have broader line widths ($>600$~km~s$^{-1}$) making detection in 
a 1000~km~s$^{-1}$ bandwidth virtually impossible. A
broad line with a substantial integrated intensity
would barely peak above the noise so the quoted limits
are for a narrow line {\it only} and a much larger
molecular gas mass could be present.
Importantly, these powerful radio galaxies
are still detected in ro-vibrational H$_2$ lines
in the infrared (Jaffe \& Bremer 1997, Falcke et al.\ 1998,
Wilman et al.\ 2000),
so molecular gas is present despite the non-detection
in CO. One final explanation is that
the radio source evaporates the cold clouds
as proposed by Soker et al.\ (2000). 
While this view is difficult to reconcile
with the detection of IR H$_2$ lines, it
cannot be dismissed with the current data.
Future broad bandwidth CO observations will help
differentiate between these possibilities.
The lack of CO in powerful radio 
systems accounts for many of the previous non-detections.

\subsection{Are we detecting the cold `sink'?}

The relationship between the observed warm molecular
gas and the `sink' of cold material that should
form out of the cooling flow is not immediately
obvious from our observations and the wealth
of previous limits on `cold' material in cluster
cores. Given recent {\it XMM-Newton} X-ray  observations
that appear to show a deficit of X-ray line emission from
the coolest gas components in several strong cooling flows 
(Tamura et al.\ 2001, Peterson et al.\ 2001)
and recent {\it Chandra} observations indicating that
cooling flows do not extend to large radii and
have characteristic timescales of $\sim10^9$~years
(Allen et al. 2000)
there is a renewed debate on this issue (see Fabian et al.\ 2001b).

Taking previous X-ray observations at face value 
and assuming the observed
molecular gas is the only molecular component in the cooling flow
then it is not possible to account for
more than 5--10\% of the deposited material in
any of our 16 CO-detected cooling flows. 
{ This apparent contradiction can be explained
if the observed molecular gas is only visible due to the 
warming influence of either star-formation in the central
galaxy or the strongly peaked X-ray emission. In these
two cases, the vast majority of the material
would remain unobservable in clouds of
very low temperature ($<10K$) in systems
without star-formation or a strong X-ray peak
and/or at larger radii.}
The theoretical support for this view
is divided (Ferland, Fabian \& Johnstone 1994;
Voit \& Donahue 1995) and 
the most direct evidence for the cold sink is
the observed intrinsic X-ray absorption
seen in clusters (White et al. 1991, Allen 2000).
This absorption is
consistent with the presence of a column
density of 10$^{20.5-22}$cm$^{-2}$. These 
levels are within an order of magnitude
of the derived column densities
of H$_2$ from our CO detections and our
stronger CO detections are in systems that
are known to have intrinsic X-ray absorption
(Allen 2000).  
Allen et al. (2001) and Allen (2000) discuss
the fate of the luminosity absorbed and
re-processed by this intrinsic X-ray absorbing column.

On the other hand, the new X-ray observations
point to lower mass deposition rates by 
factors of 2 to 5 and shorter timescales
by factors of 2 to 4 so the estimates
of deposited mass could come down by factor as 
much as 20. This would then give, for the
first time, a match in the predicted and 
observed cold gas masses even with the
substantially uncertainties introduced
using a standard CO-to-H$_2$ conversion.
Without obtaining
the spatial extent of the CO emission it
is not yet possible to prove the cold
gas found here is as concentrated as
X-ray observations predict ($<50$~kpc).
Future X-ray observations and mm-interferometry will
address this question directly. There
is a possibility that the molecular gas detected
in this study is the sum total of the mass 
deposited in a cooling flow but a great
deal of work remains to be done before
this can be proven beyond reasonable doubt.

\subsection{NGC1275 - the prototype?}

The detection of molecular gas in systems other than
NGC~1275 throws the argument that the presence of
molecular gas is unrelated to the cooling flow into
question (see Bridges \& Irwin 1998 for discussion).
If the molecular gas found in cooling flows is 
deposited through mergers with gas-rich 
members, how can 10$^{11}$~M$_\odot$ of gas
be present in  the most massive flows
observed? When only one system is known,
particularly one as outwardly peculiar as NGC~1275,
then a merger is consistent with the available
facts. However, when a set of galaxies with the
same X-ray characteristics share the same 
properties (strong optical emission lines and
CO emission), then some common causal relationship 
other than infrequent mergers must be invoked. 
The connection between low central cooling
times measured from X-ray imaging and the 
presence of optical lines (Peres et al. 1998) 
points to a causal link between cooling gas
and the presence of irradiated cold clouds.
Any model invoking mergers
in all cases has to explain the high fraction of
massive cooling flows with CO emission in
clusters with relatively few spiral galaxy members.

Our results should also prompt us to reconsider the
properties of NGC~1275. The presence of dust,
strong optical emission lines and a powerful
radio source are features found in many
central clusters galaxies in the most massive
cooling flows. The presence
of young star clusters like the ones found in NGC1275
(Holtzmann et al. 1992, Zepf et al. 1995) is 
hinted at in the strong Balmer series seen in 
the integrated spectra of some galaxies (Crawford et al. 1999).
As yet imaging data hasn't resolved these young stars
into clusters but further HST imaging could do so.

\subsection{Line widths}

The velocity width of the detected CO lines is
another important factor. Our detections are
all of systems with widths less than 500~km~s$^{-1}$
implying the gas is localised to galaxy-scale
regions and is not virialised in the cluster core.
Taking all sixteen of our detections, the
systemic velocity shifts relative to the published optical
line redshift have a mean of
-7~~km~s$^{-1}$ and an r.m.s. of 130~km~s$^{-1}$.
Given the errors in the optical velocities, these
values are consistent
with the warm molecular gas coming from a small
number of regions of intense star-formation
drawn from the underlying velocity dispersion of
the central galaxy. 
The velocity discrepancies and narrow line widths
are similar to those seen in other radio galaxies
with CO detections (e.g. 3C~293 Evans et al. 1999).

The underlying dispersion could be underestimated  
if the restricted bandwidth available results
individual clouds on a high velocity `tail' of this
distribution being missed entirely thus lowering the
r.m.s. calculated above.  Future wide bandwidth
observations would detect any such high velocity clouds.

The line widths can also be used to determine crude
dynamical mass estimates on the somewhat sweeping 
assumption that the molecular gas is in a disk
(Papadopoulos et al. 2000, Carilli, Menten \& Yun 1999).
Without direct measurement of the extent of the
CO emission, the disk diameter is a free parameter.
For M$_{\rm dyn}\approx$M$_{\rm H_2}$ the 
disk diameters would range from 1 to 5~kpc. 
If future interferometry of these systems finds
larger sizes then this would be another
example of the ``excess mass'' problem 
(Carilli, Menten \& Yun 1999) where the
gas mass exceeds the dynamical mass in
nuclear starbursts. 

At the opposite extreme, lines with large
velocity widths ($>$800 km~s$^{-1}$) would be
difficult to detect even with our 1.8~GHz bandwidth {JCMT}
data due to the baseline subtraction method used.
There is no significant difference in the 
results for the narrow lines changing velocity ranges
used in the baseline subtraction but the
possibility of missing some molecular gas remains 
if a flat but broad emission line were present. 
Such broad lines, though weak in 
peak intensity, could easily reach  2--5 K km~s$^{-1}$ and
account for in excess of $5\times10^{11}$ M$_\odot$
of gas if warm. { Any broad lines are likely to
be from colder gas clouds further from any star-formation
or intense X-ray emission, 
so the CO(1-0) emission will be considerably weaker than 
that from the $>$30~K gas. }
Work by Antonucci \& Barvainis (1994) 
demonstrates that broad bandwidth observations
can be made by splicing together smaller
bandwidths but none of the five clusters
they observed contained a strong optical line
emission system so it is difficult to compare
their results with ours.
Future wide bandwidth receivers will
provide the data required to detect such broad features
although the low temperature of the gas may
limit the chances of detection. 
Current technology is well suited to detecting
narrower CO lines which are reasonably warm.

\subsection{Timescales}

If one assumes the systems we are observing are
in a steady-state then
the observed molecular gas mass should reflect
the balance of mass deposition and star-formation
rates. Following the argument of Braine \& Dupraz
(1994), one can determine the likely timescale
for gas to be held in clouds before forming stars.
While most of the clusters in this sample
do not have accurate and high resolution X-ray
imaging and the true extent of the 
molecular gas in unknown, it is difficult to 
derive the true ratio of total mass to mass
deposition rate. However, in the case of A1068 and
A1835 we can derive values of mass deposition rates of 150 and
500~M$_\odot$~yr$^{-1}$ at 50~kpc
from deprojection results of Allen (2000). 
In both of these systems, Crawford
et al. (1999) quote apparent star formation
rates (31 and 125~M$_\odot$~yr$^{-1}$) which imply that the efficiency 
of forming stars ($\eta_{SF}$) is of the order
of 0.2. This efficiency is close to that
used in Braine \& Dupraz (1994). Given the total
molecular mass present of 8$\times 10^{10}$ and
1.8$\times 10^{11}$~M$_\odot$, the implied
timescales for gas consumption are therefore
5$\times 10^{8}$ and 4$\times 10^8$ years.
So, in these very central regions, the
directly observed star formation rates imply
that {\it with no further deposition} the
detected molecular gas would be used up
on timescale of  less than $10^{9}$~years but in every
case the observed deposition exceeds the
star formation rate so the timescales could
be comparable to the Hubble time.
Therefore, the extreme star-formation seen in this
sub-sample of central cluster galaxies could
be a long-lived phenomenon in which a substantial
fraction of the stellar population is produced.
Only when CO mapping and high resolution X-ray 
imaging are available for more of this sample
will the issue of timescales be clarified.

\subsection{Dust}

When viewed with the recent detection of dust emission in
the sub-mm of A1835 and A2390 (Edge et al. 1999), the
masses  of the  molecular gas implied here 
for them are slightly higher than galactic gas-to-dust ratios 
(900$\pm$200 and 350$\pm$150).
Recent {SCUBA} detections of Zw3146 and Zw7160
imply dust masses of 2.2$\times 10^{8}$M$_\odot$
and detection of  1.5$\times 10^{8}$M$_\odot$ respectively
(Chapman et al., 2001), giving
 gas-to-dust ratios of 740$\pm$150 and 410$\pm$160. 
The claimed Sunyaev-Zel'dovich effect detection at 850~$\mu$m with 
{SCUBA} of RXJ1347$-$11
(Komatsu et al. 1999) is complicated by the detection
of a central source of 3.5~mJy or 1.8$\times 10^{8}$M$_\odot$
of dust. This implies a gas-to-dust ratio of $<500$
from our {JCMT} CO(3-2) limit (assuming CO(3-2)/CO(1-0)$=$0.4
and no radio source or S-Z contribution).
Finally, IRAS~09104+4109  was recently observed with SCUBA by
Deane \& Trentham (2001) who claim this source lacks
cold dust. Their $3\sigma$ upper limit of 6.4~mJy at 850$\mu m$
is consistent with $<3.2\times 10^8$M$_\odot$. This
limit would not be sufficient to detect any other known
cooling flow with dust so, until substantially deeper
SCUBA photometry is obtained, it cannot be claimed that
IRAS~09104+4109 is deficient in cold dust.

The gas-to-dust ratios derived for this sample
show a similar dispersion to that seen
in other classes of galaxies.
Assuming a dust temperature of 40~K, we can also
estimate dust masses from the few {IRAS} detections
and upper limits derived from  the {\it XSCANPI} utility
at {IPAC} (Table 8). The derived gas-to-dust ratios 
are strongly dependent on the dust temperature used
but agree with those from sub-mm detections and limits.
For the joint CO and {SCUBA/IRAS} detections including NGC~1275
but excluding 3C48,
the average gas-to-dust ratio is 1720 with a dispersion of 1410.
The few very high gas-to-dust ratios (e.g. RXJ0338+09 at 5350)
is probably an indication that the assumption of a dust temperature
of 40~K is incorrect and to obtain ratios of 500--1000 requires
dust temperatures of around 30~K.
Given the number of CO detections presented here, it will be 
important to obtain {SCUBA} limits on dust masses for
them.  The dust temperature
derived by Edge et al. (1999) of 30--50~K is 
comparable to the observed excitation temperatures 
of CO and those expected
in other extra-galactic systems undergoing star-formation.
The two possible exceptions to this are IRAS~09104+4109 
(see Deane \& Trentham 2001) and 3C48
where the dust temperature is probably substantially higher
than 40~K so the majority of the 60~$\mu$m flux is probably
from much  a smaller mass ($<10^6$M$_\odot$) of 
hotter ($\sim$200~K) dust heated by the QSO 
which is hidden in the case of IRAS~09104+410 (Hines \& Wills 1993)
or seen directly in 3C48. Therefore the gas-to-dust ratios
for these objects in Table~8 are severely underestimated
as a much lower dust temperature is assumed.

The question of the origin of the observed dust is still an
open one, with injection from star-formation, stellar mass loss
and liberation from cold clouds all plausible
explanations. The spluttering timescale due
to X-rays can be circumvented in these models
through rapid generation or shielding. Given the
extreme properties of cluster cores, it would
be surprising if galactic gas-to-dust ratios
applied in these environments.

One additional point to note with respect to the dust emission is
that there are several CO-detected central cluster galaxies
with relatively weak radio sources (e.g. A1068 and RXJ0821+07).
Assuming a gas-to-dust ratio of 1000 for both
systems we predict 850~$\mu$m (350~GHz) flux densities
of 6 and 4 mJy respectively compared to 1.4~GHz flux
densities of 8.7 and 2.4 mJy. If one takes the radio/sub-mm 
spectral index, $\alpha^{350}_{1.4}$,
used by Carilli \& Yun (1999) then these
galaxies are comparable to the ULRIGs found at low
redshifts ($\alpha^{350}_{1.4}=$-0.2 and -0.1 compared to -0.1 to 0.2).
This may imply that the majority of the radio 
emission in these radio-weak central cluster
galaxies is related to the star-formation and 
{\it not} a central active nucleus. On the other hand, most
of our detected targets have moderately bright
radio sources (20 to 100~mJy). For these, the probable
$\alpha^{350}_{1.4}$ values fall in the range -0.4 to -1.0,
well below the relation for starbursts presented
in Carilli \& Yun (1999), so these radio-weak cases
are interesting exceptions and not the norm.

\subsection{Comparison to other star-forming systems}

If the sub-mm properties of these line-emitting central cluster
galaxies are compared to nearby starbursts, ULIRGs and more
distant sub-mm selected galaxies then they appear to 
share a great many properties. Perhaps most importantly,
the evidence from optical spectroscopy (Allen 1995, Crawford et al. 1999)
in terms of the excess blue light and power of the low ionization emission
lines, points to substantial and on-going star-formation (10--100 M$_\odot$ yr$^{-1}$).
Therefore, the cores of cooling flows are forming stars at a rate comparable
to that deduced from mass flow rate derived from X-ray observations.
The possibility that more distant sub-mm and infrared selected galaxies
are in fact in cooling flows (c.f. IRAS~09104+4109 Kleinmann et al. 1988,
Fabian \& Crawford 1995, Evans et al. 1998) should be considered in the interpretation
of these galaxies. Taking  IRAS~09104+4109 as a case in point, it shows
strong H$\alpha$ emission (Evans et al. 1998) and a strong cooling
flow (Fabian \& Crawford 1995) but no CO detection (Evans et al. 1998).
This non-detection of CO is consistent with the correlation
given in Figure~9 as long as the H$\alpha$ luminosity does
not exceed 10$^{42}$ erg~s$^{-1}$. 


Looking to more distant, sub-mm selected galaxies it is possible that
galaxies such as 4C~41.17 (Dunlop et al 1994) and 
8C~1435+635 (Ivison et al 1998) have more in common with central cluster
galaxies than the more commonly assumed archetypes of Arp220 and M82.
The combination of active nucleus and starburst in NGC1275 illustrates
the difficulty in classifying very distant galaxies into ``monster''
or ``starburst''.  We are probably witnessing the on-going formation 
of a giant elliptical galaxy in these systems and the process
that is postulated to occur in the more distant galaxies. The
study of these rare, low redshift galaxies could provide important
clues to the nature of distant systems.

\section{Conclusions}

We have, through the combination of better receivers and selection of
more extreme cooling flows, succeeded in detecting molecular emission
in as many as sixteen  central cluster galaxies. These detections are consistent
with molecular gas warmed to 20--40~K by young stars. While the
mass of molecular gas is 5--10 per cent of that expected to 
have been deposited in these cooling flows in total, it may only fill a relatively
small volume of the core so may either be on the warm tip of a cold
iceberg or the sum total of the deposited mass if cooling flows
aren't as large as previously thought.

The commonly held view that central cluster elliptical galaxies are
the oldest and most quiescent of galaxies is difficult to square
with the variety of observations showing a dusty, gas-rich environment
fuelling substantial star-formation. The common feature amongst these
peculiar galaxies is that they all lie in the cores of massive
cooling flows and have relatively low radio powers.
The exact mechanism that links these observations
is not clear but the association is too strong to be dismissed.
This possible causal link between cooled gas and star-formation is one that
has substantial implications for the interpretation of distant radio
and sub-mm selected galaxies.

The results presented in this paper
indicate that observations with current and future
instrumentation will be very productive. With {SCUBA} capable of
detecting dust in the more extreme systems ($>5\times 10^7$ M$_\odot$)
and {JCMT} and {IRAM} available to detect several CO transitions and their isomers and other 
atomic and molecular lines such as CI, HCN, CS and H$_2$O, the immediate prospects are
excellent. We have also recently obtained {OVRO} observations of CO(1-0) for
A1068, A1835, Zw3146, RXJ0338+096  and RXJ0821+07 and these results will be presented separately
(Edge \& Frayer, in preparation). In the longer term,
the development of millimetre arrays will allow the
molecular gas to be spatially resolved and
missions such as {SOFIA,\it SIRTF} and {\it FIRST} offer the opportunity
to sample far-infrared lines from the gas sampled by CO and any that is colder.
While the ultimate fate of gas being deposited in a cooling flow is 
not resolved, it is encouraging to find at least some molecular gas in the cores of
cooling flows.

\subsection*{ACKNOWLEDGEMENTS}

ACE thanks the Royal Society
for generous support. Particular thanks go to
Harald Ebeling, Andy Fabian, Carolin Crawford and Steve Allen
without whom the selection of the BCS cooling flows
would not have been possible.
Rob Ivison, Ian Smail, Dave Frayer, Roderick Johnstone, Glenn Morris,
Richard Wilman and Terry Bridges are
thanked for help and encouragement.
Scott Chapman is thanked for communication of {SCUBA}
results before publication. The success of the observations
presented in this paper is due largely to the excellent
facilities and staff of the  {JCMT} and {IRAM}
telescopes. We are especially grateful
to Per Frieberg, Gilles Niccolini, Gabriel Paubert and Teresa Gallego
for performing the first {JCMT} and additional {IRAM} observations
in service mode. Per deserves particular thanks for
making the observations during the Festive Season.
Finally, thanks to the referee Chris O'Dea for being far
prompter and less verbose than the author.
{JCMT} is operated by the Joint Astronomy Centre on behalf of the
{\it PPARC}, the Netherlands Organisation for Scientific Research and the
National Research Council of Canada.  
This research has made use of the NASA/IPAC Extragalactic Database (NED) which is operated by the Jet Propulsion
Laboratory, California Institute of Technology, under contract with the National Aeronautics and Space Administration
and the Canadian Astronomy Data Centre, which is operated by the Herzberg Institute of Astrophysics, National Research Council of Canada.

\newpage

%
%

\begin{figure*}

\centerline{
\psfig{file=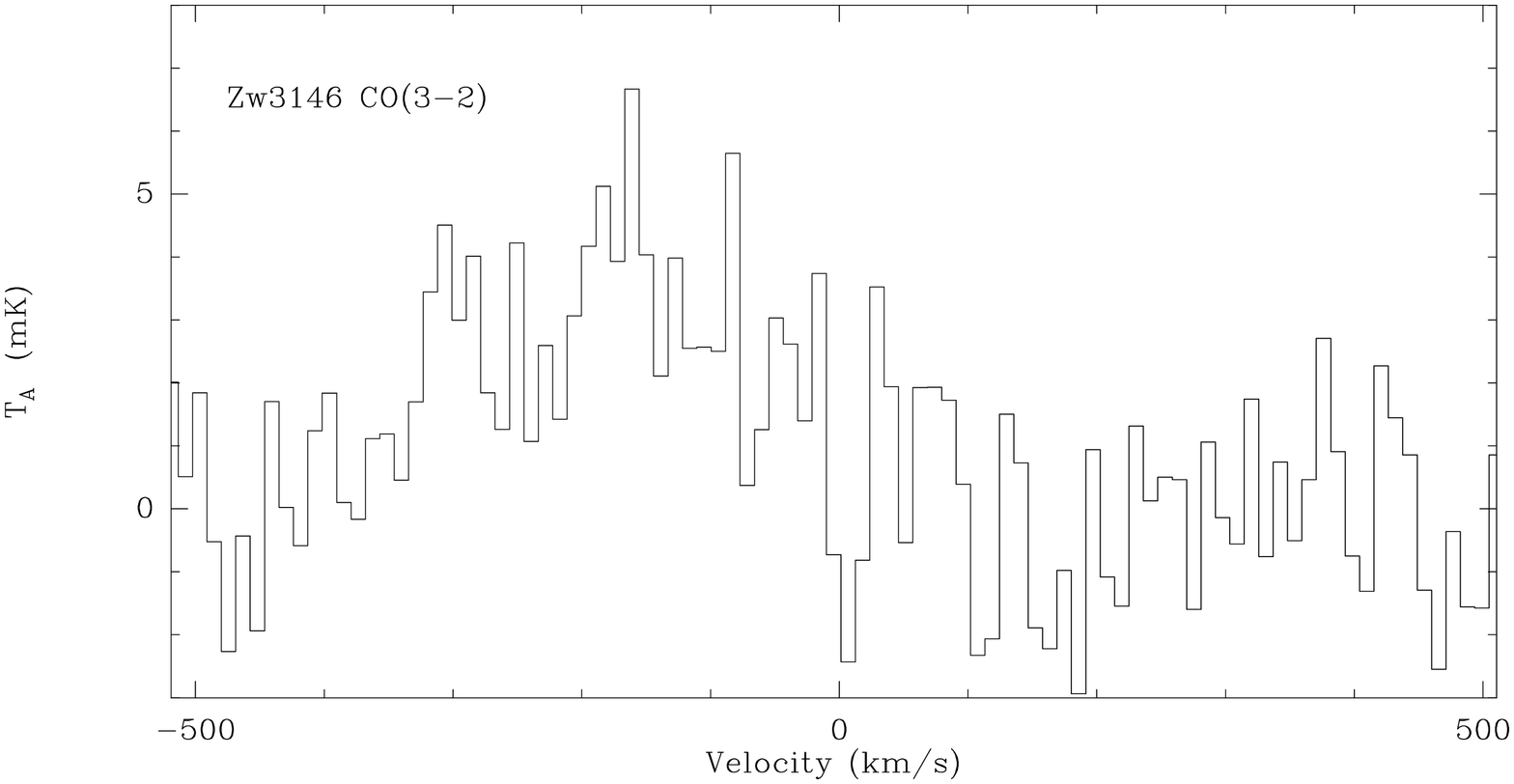,angle=270,width=8cm}
\psfig{file=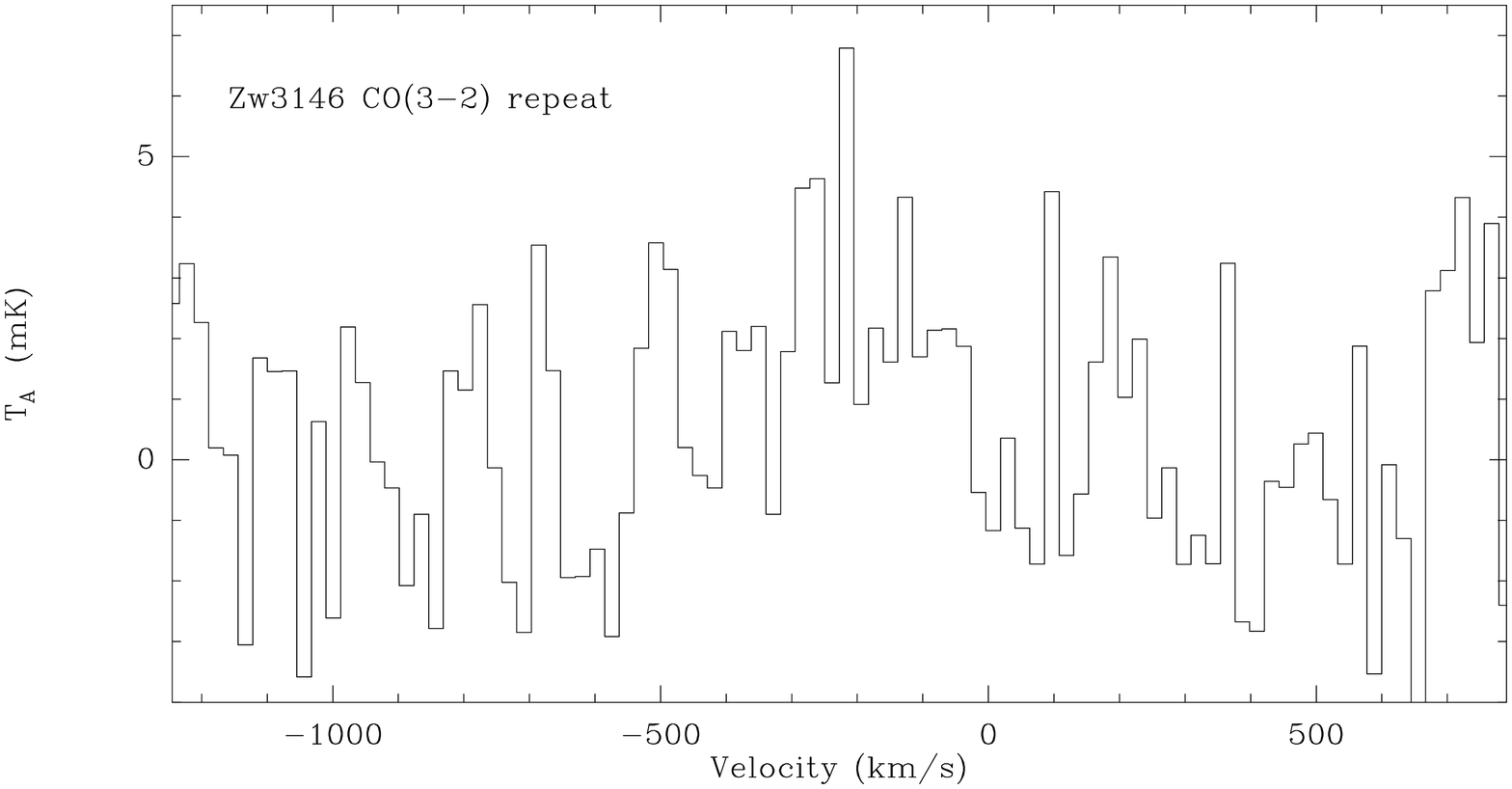,angle=270,width=8cm}}
\centerline{
\psfig{file=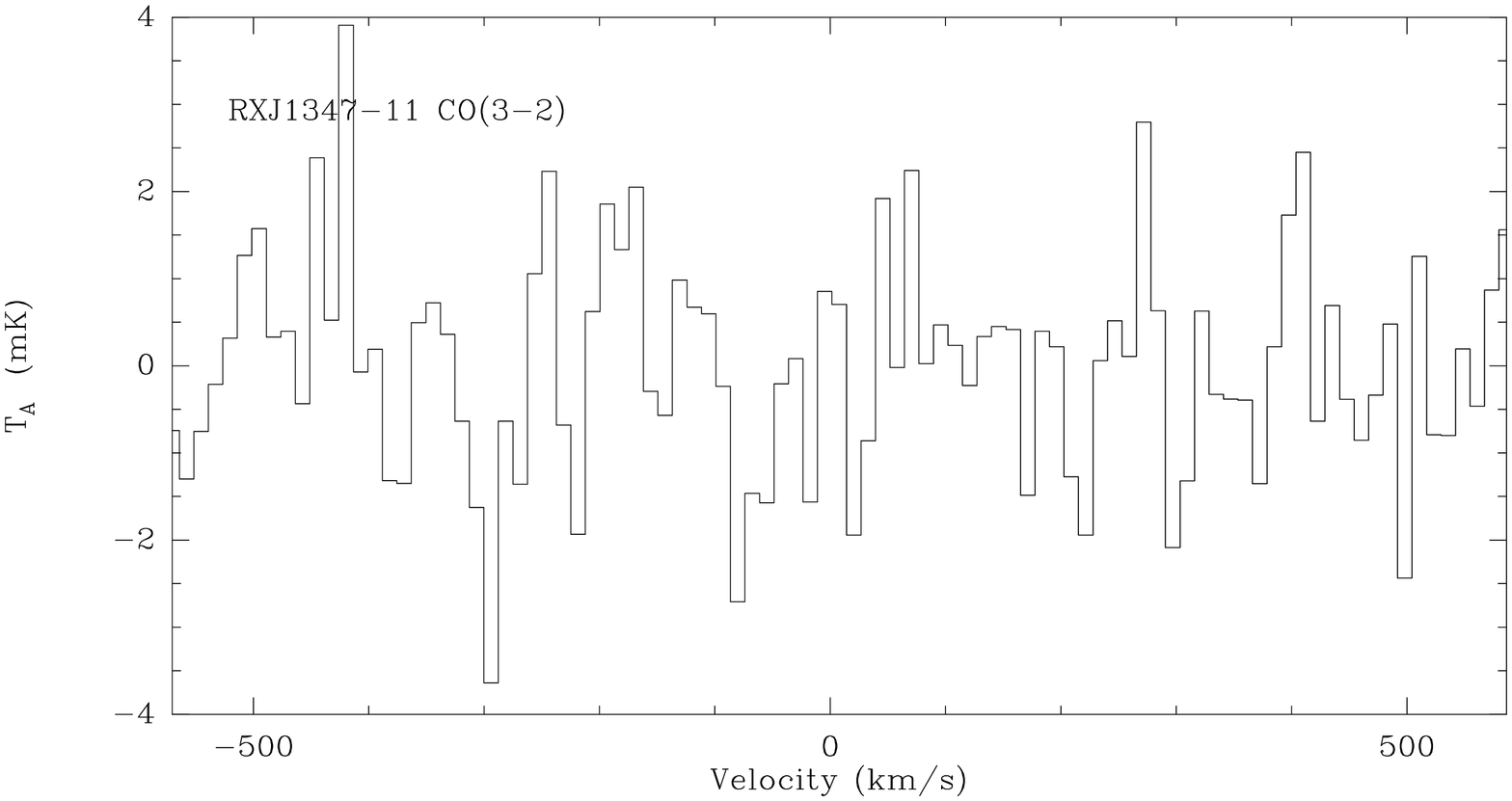,angle=270,width=8cm}
\psfig{file=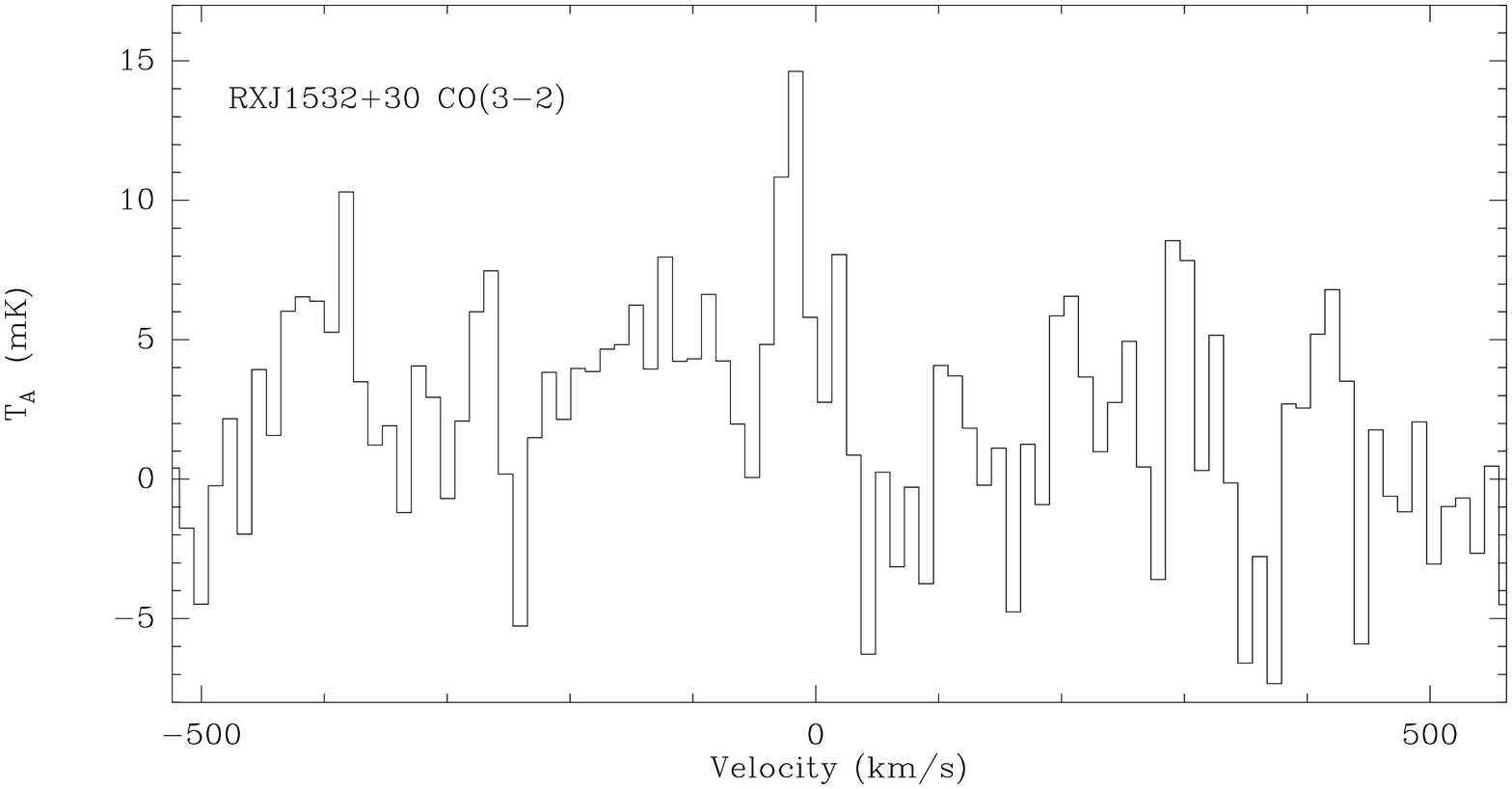,angle=270,width=8cm}}
\centerline{
\psfig{file=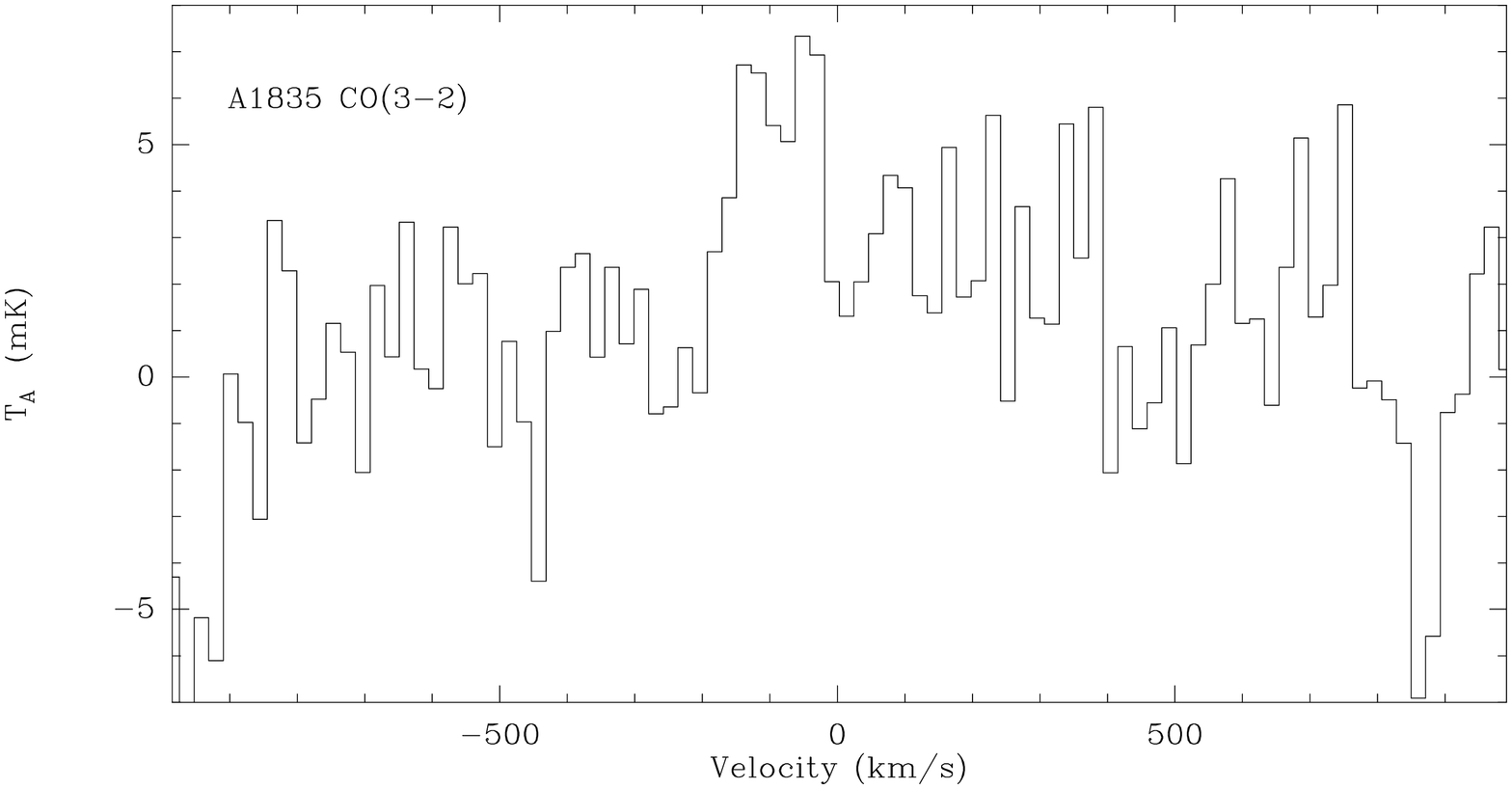,angle=270,width=8cm}
\psfig{file=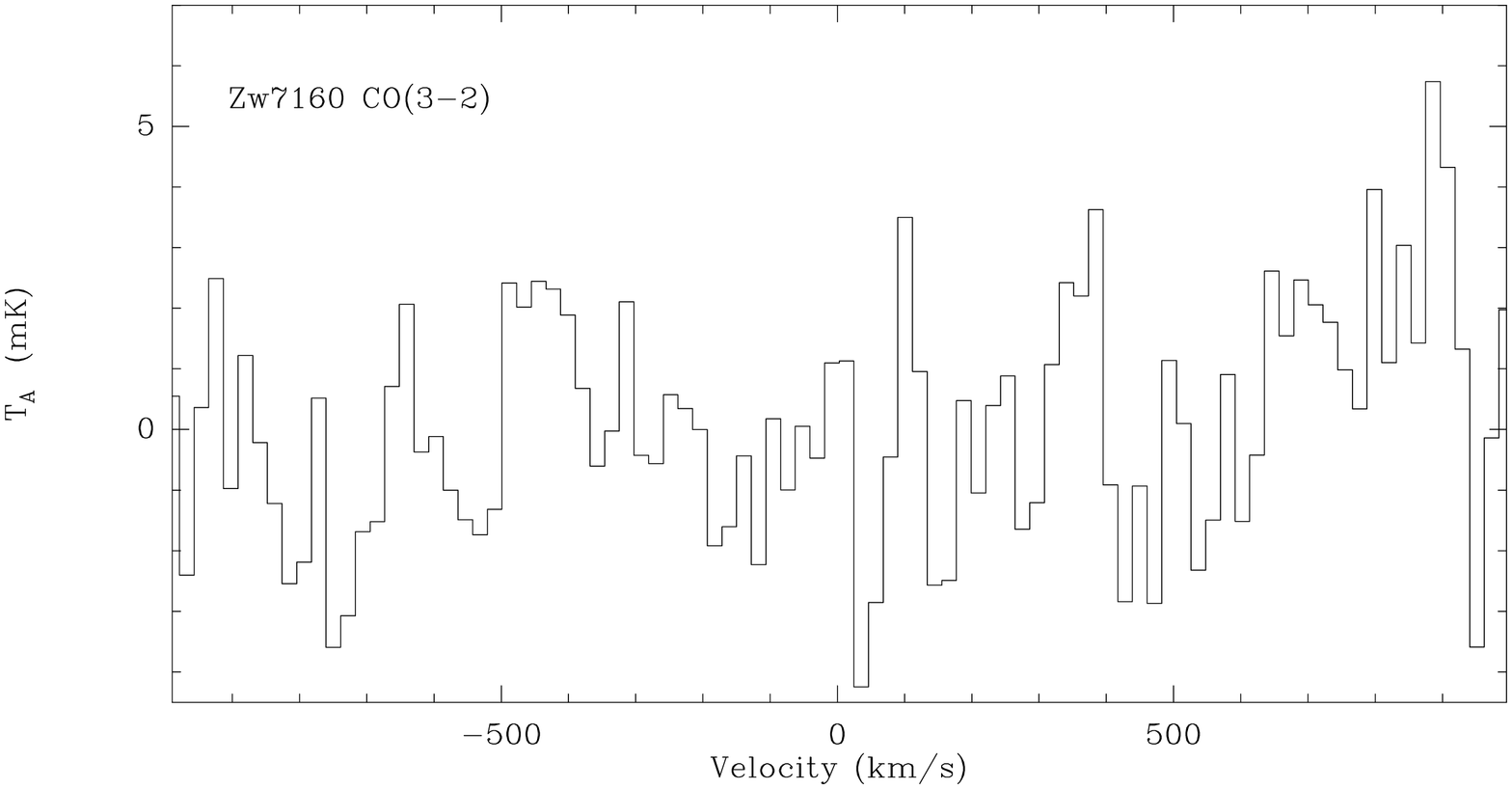,angle=270,width=8cm}}

\caption{{JCMT} CO(3-2) spectra for Zw3146, RXJ1347$-$11, RXJ1532+30, A1835 and Zw7160.}
\end{figure*}

\begin{figure*}
\centerline{
\psfig{file=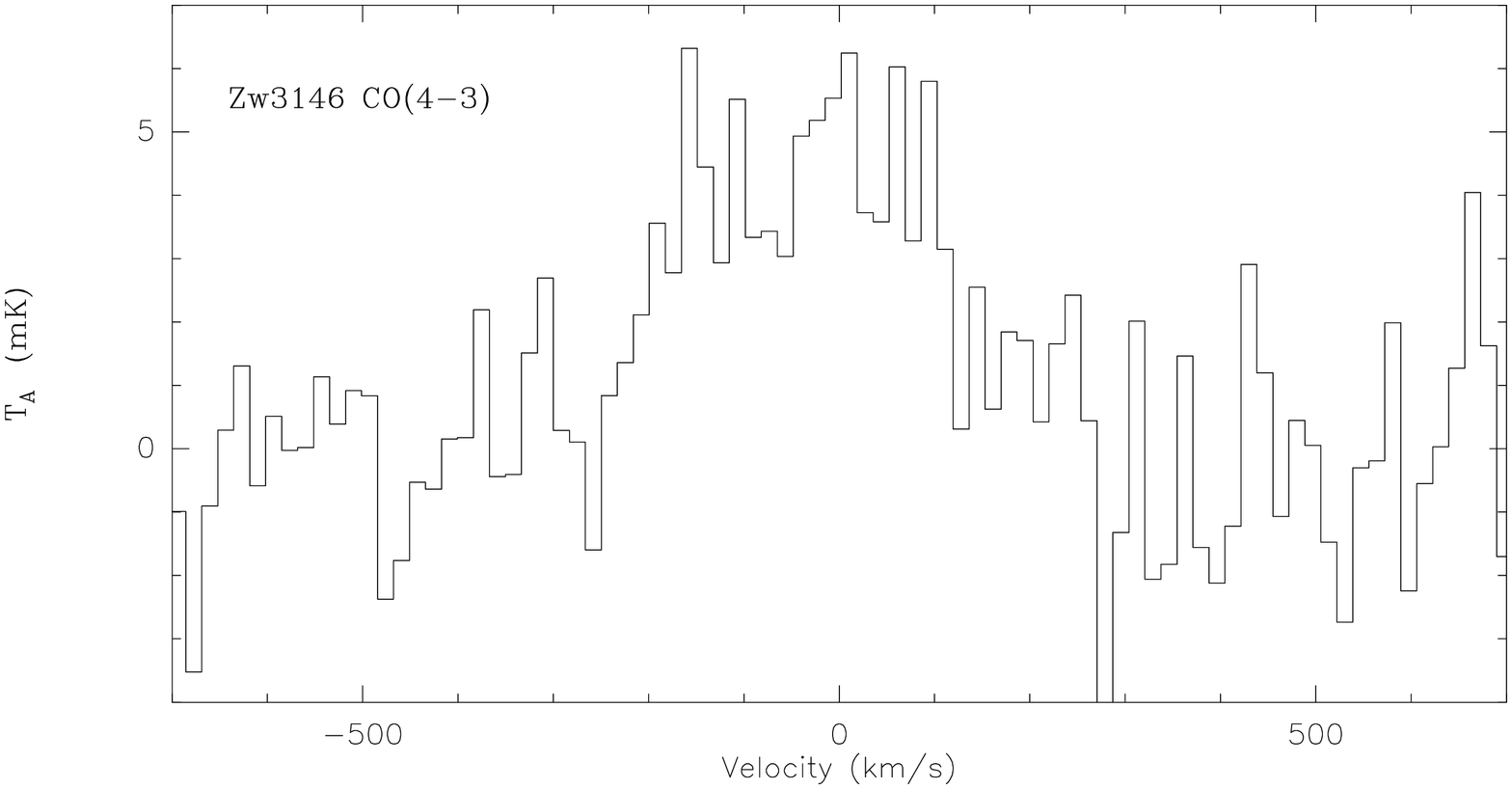,angle=270,width=8cm}
\psfig{file=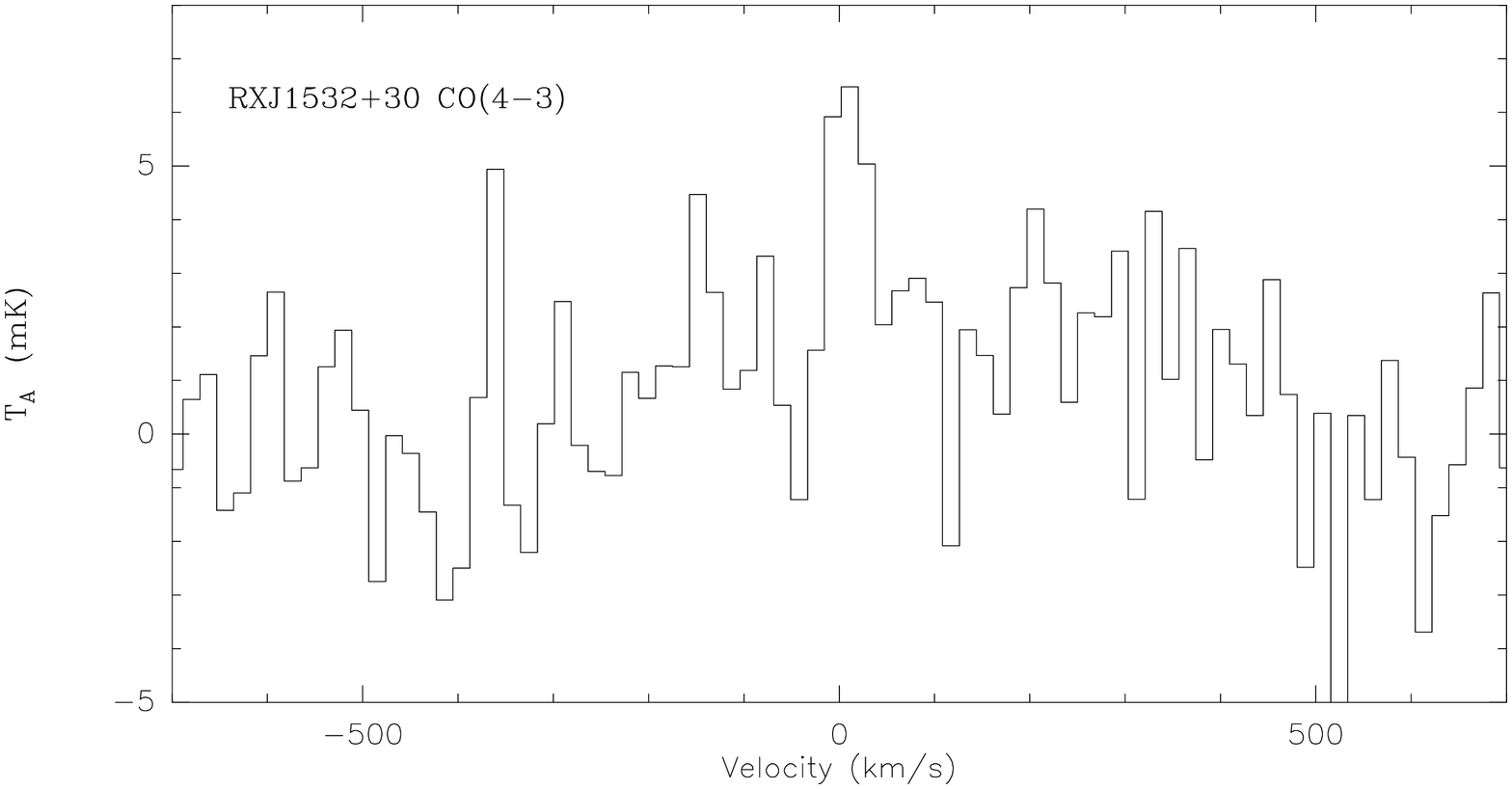,angle=270,width=8cm}}
\centerline{
\psfig{file=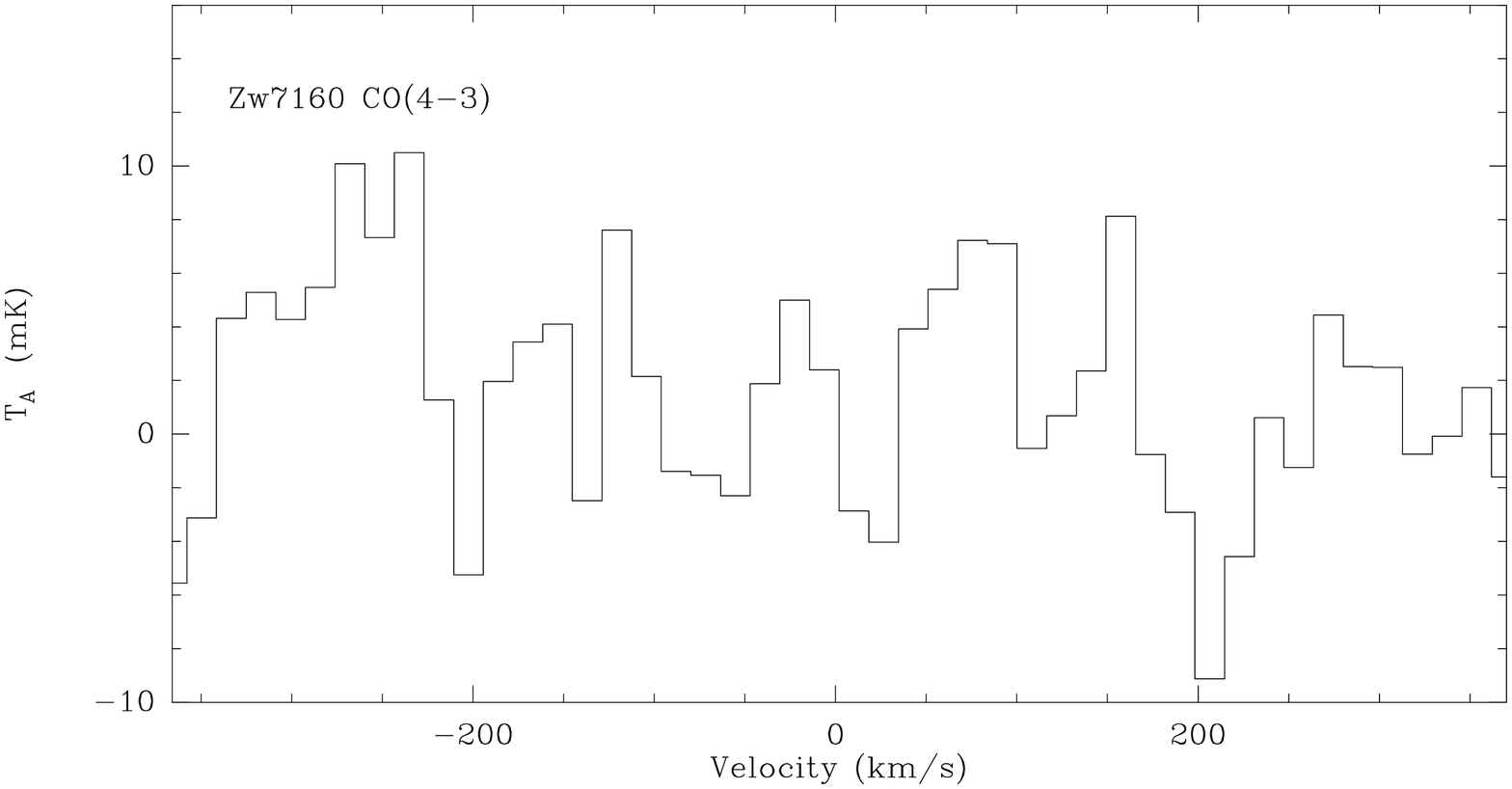,angle=270,width=8cm}}

\caption{{JCMT} CO(4-3) spectra for Zw3146, RXJ1532+30 and Zw7160. Note the
much smaller bandwidth used in the archival Zw7160 observation.}
\end{figure*}


\begin{figure*}
\centerline{\psfig{file=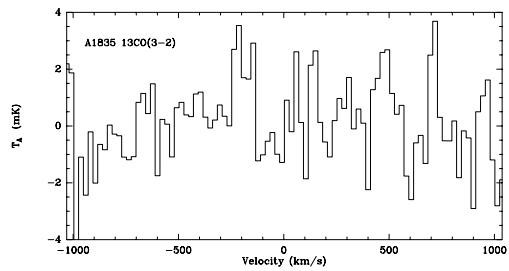,angle=270,width=7.2cm}}
\caption{{JCMT} spectrum for A1835 $^{13}$CO(3-2)}
\end{figure*}

\begin{figure*}
\centerline{
\psfig{file=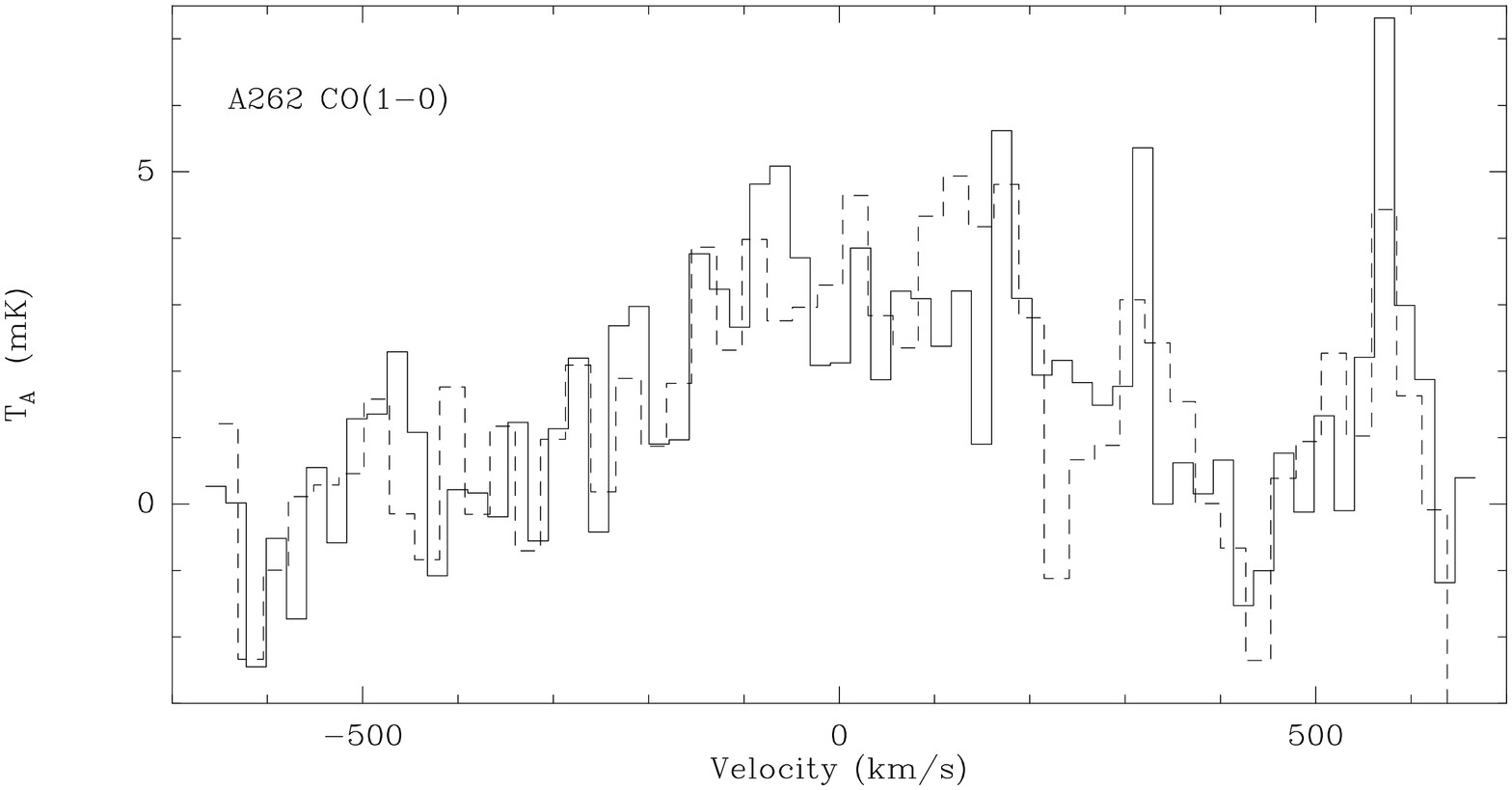,angle=270,width=8cm}
\psfig{file=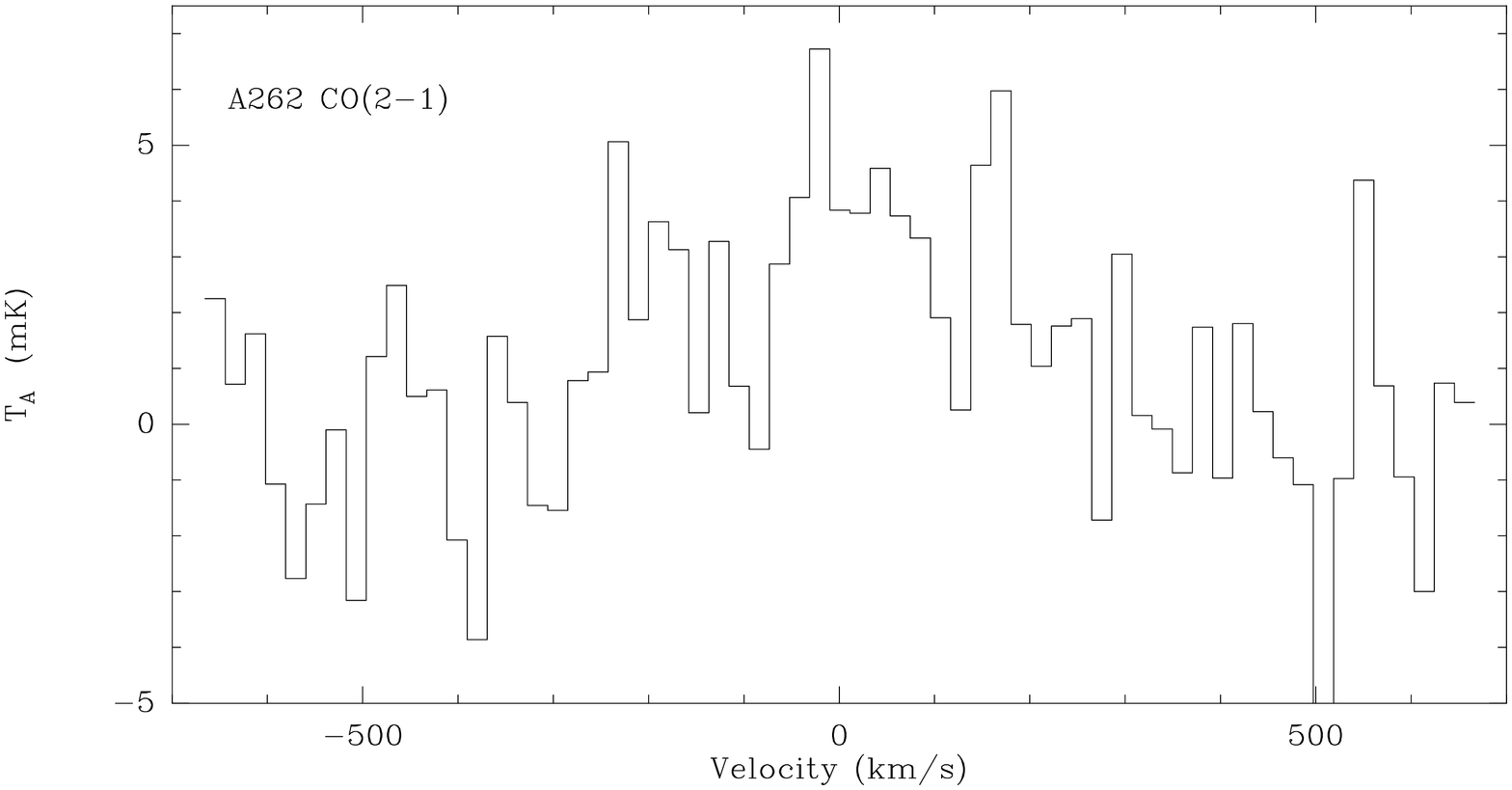,angle=270,width=8cm}}
\centerline{
\psfig{file=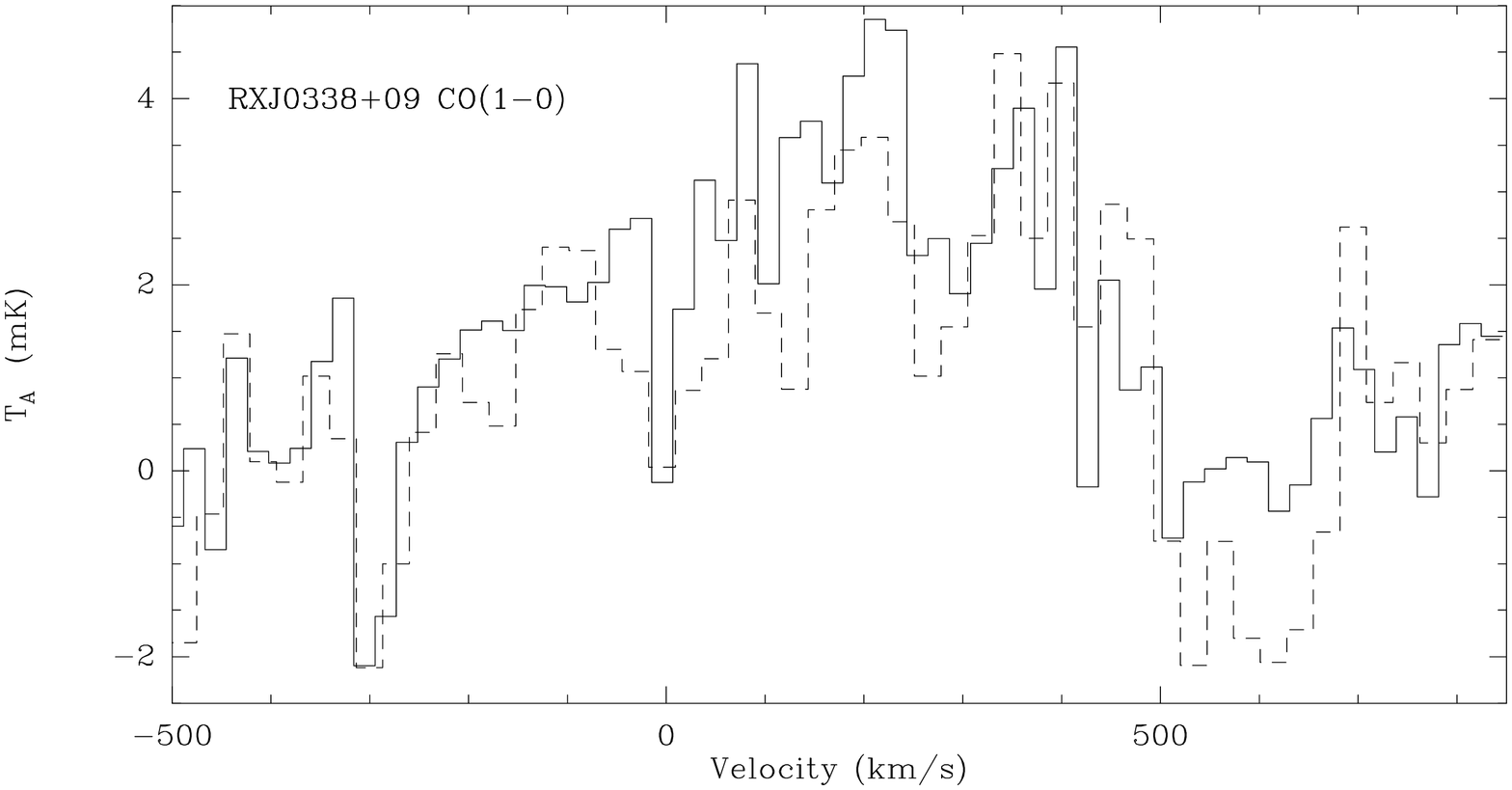,angle=270,width=8cm}
\psfig{file=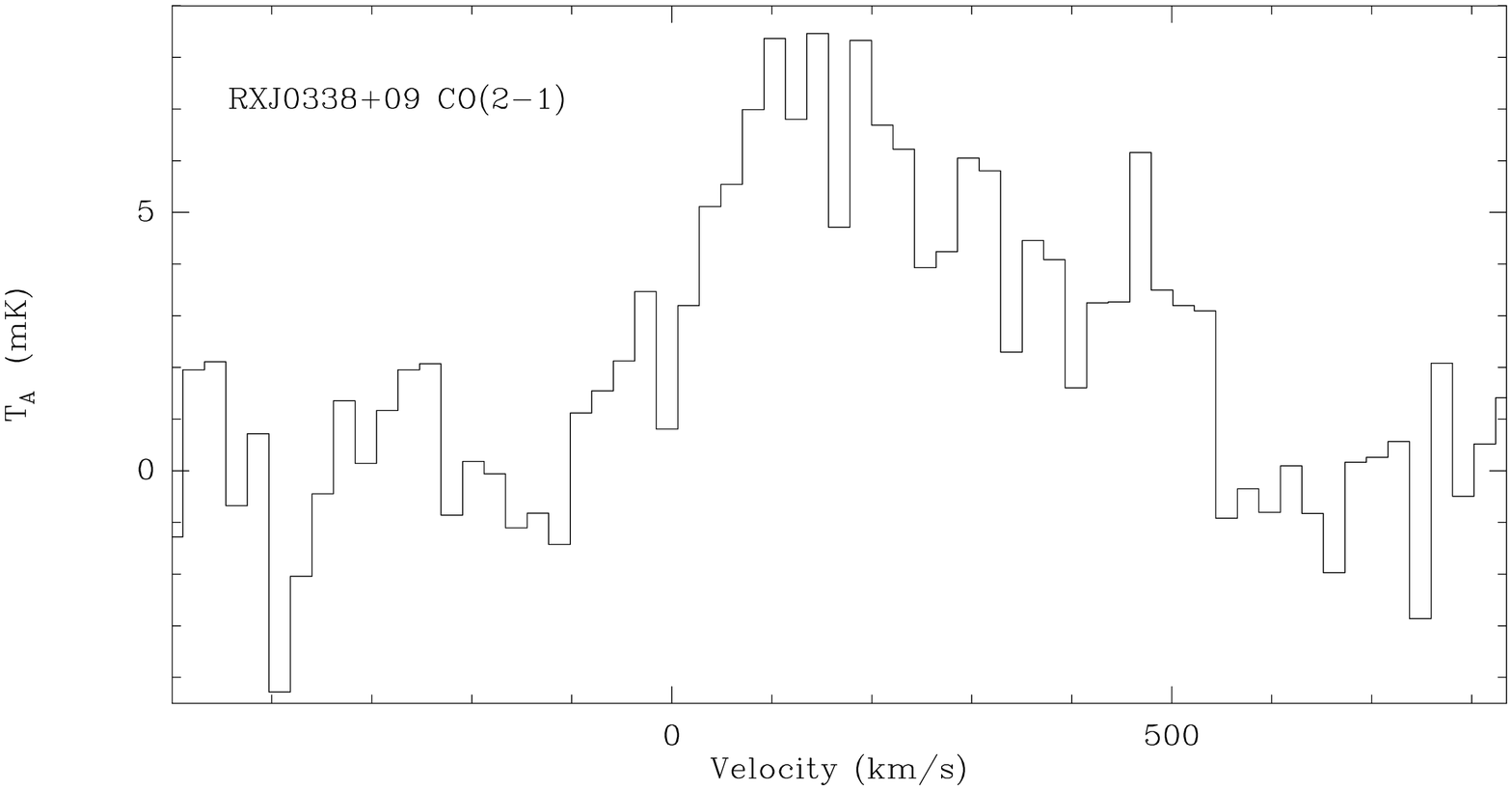,angle=270,width=8cm}}
\centerline{
\psfig{file=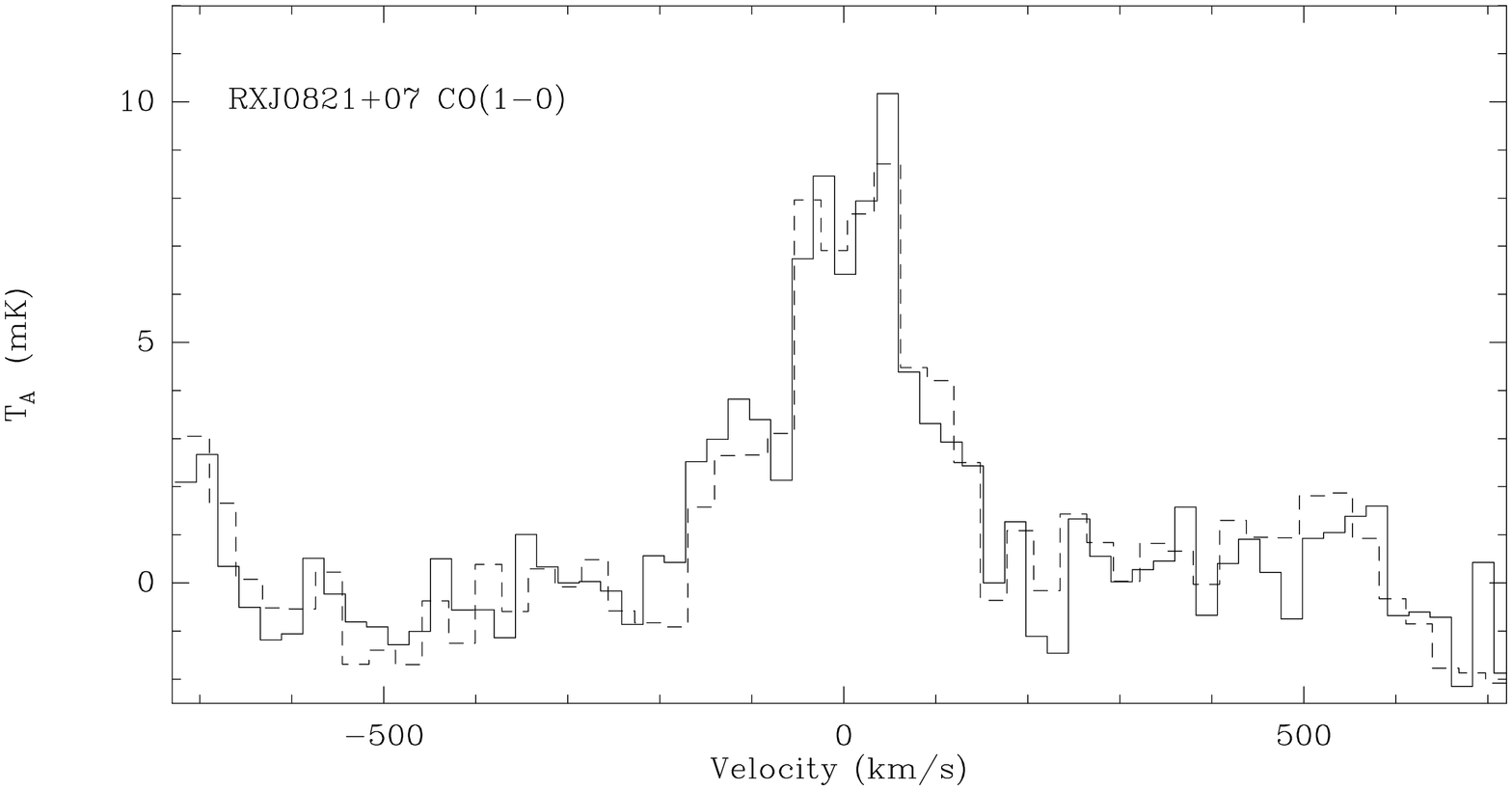,angle=270,width=8cm}
\psfig{file=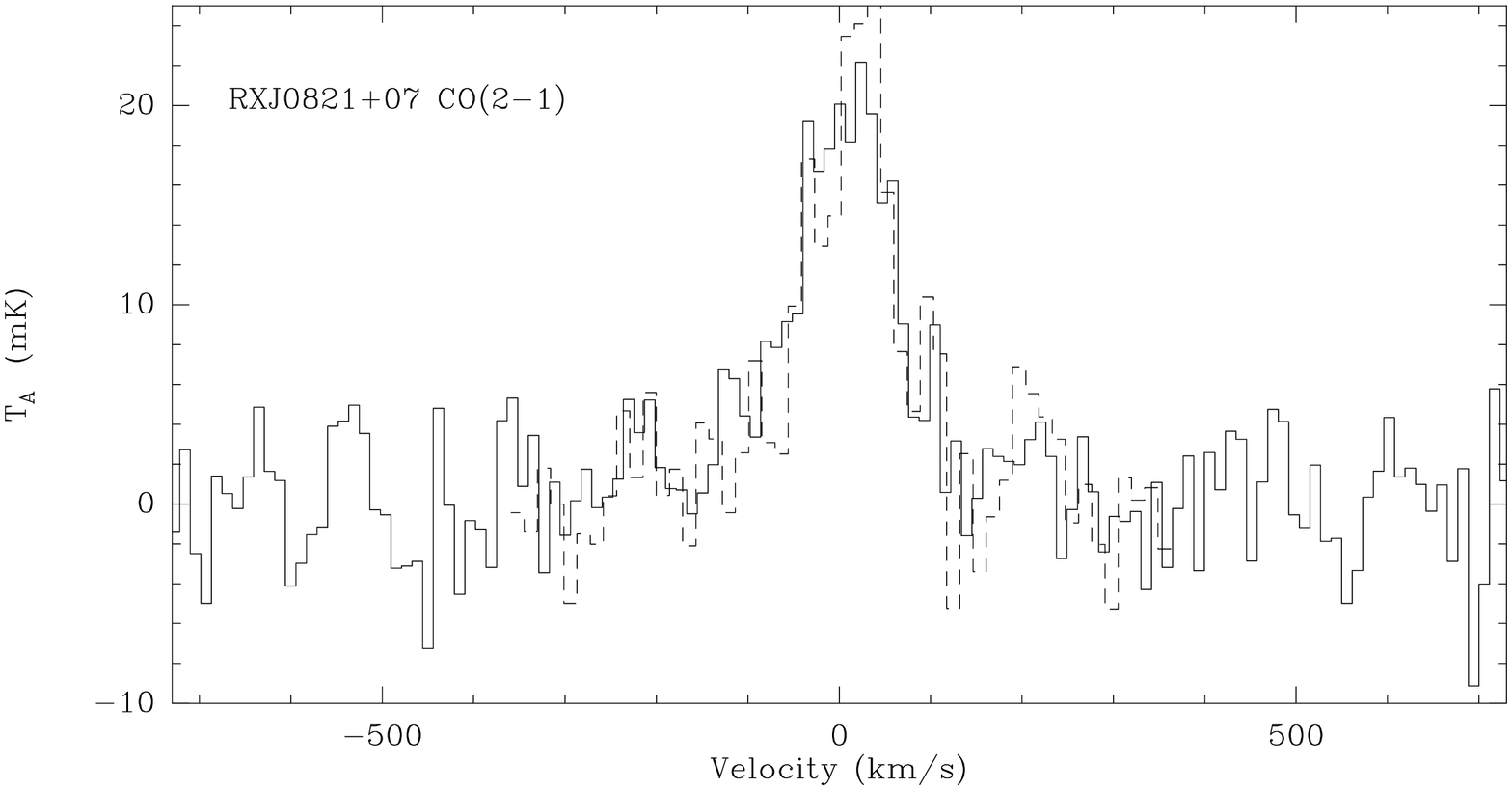,angle=270,width=8cm}}
\centerline{
\psfig{file=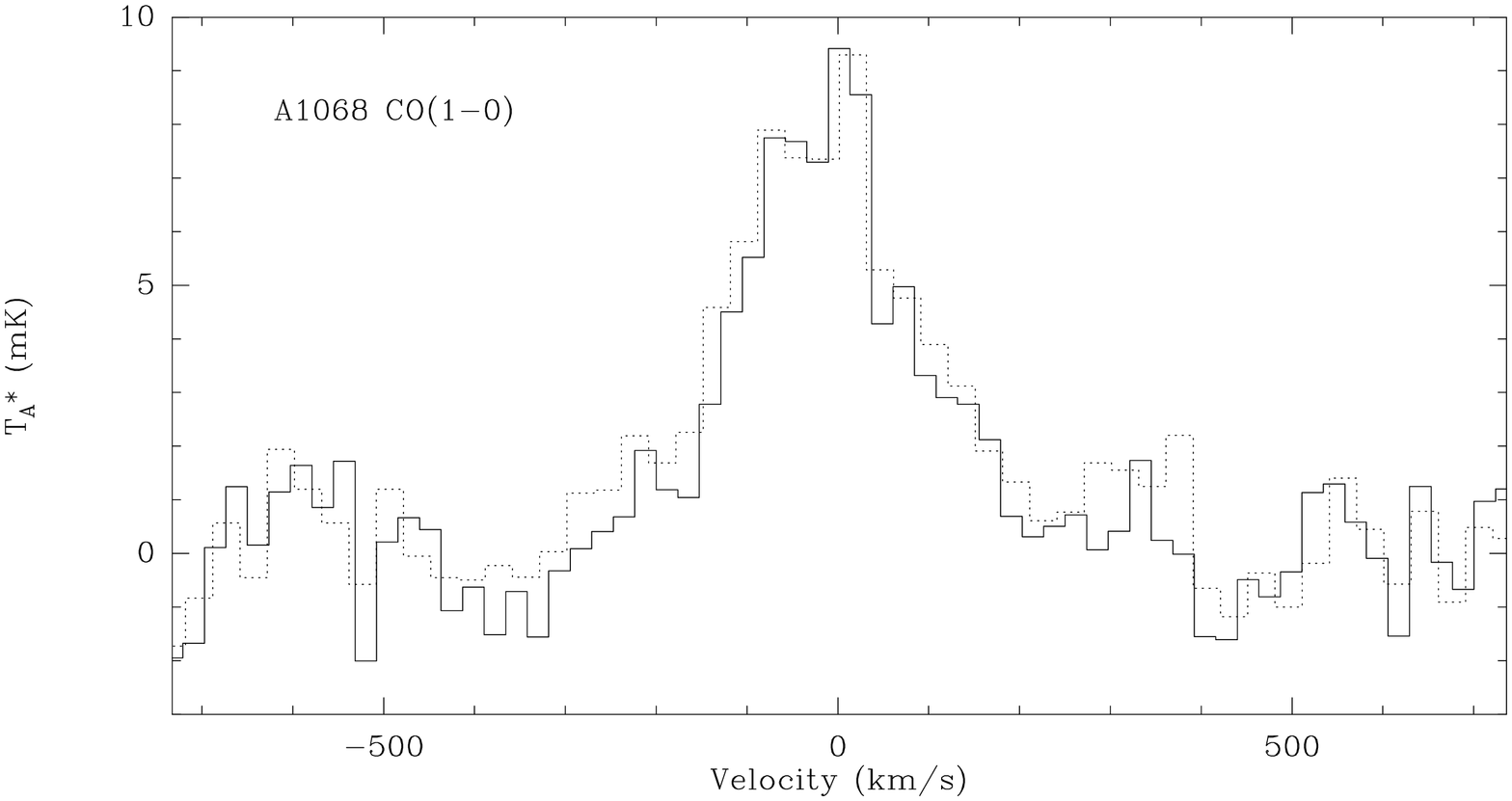,angle=270,width=8cm}
\psfig{file=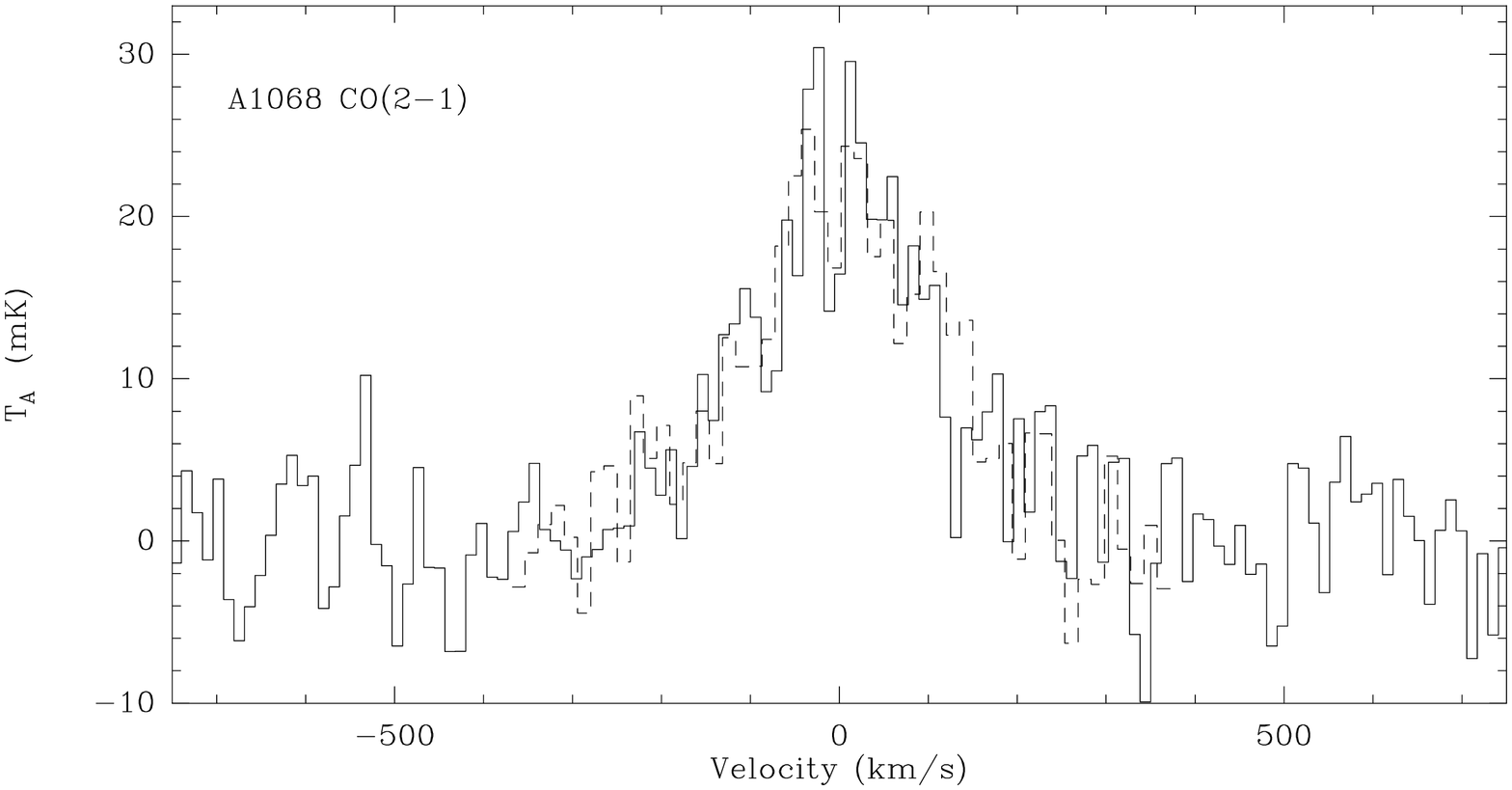,angle=270,width=8cm}}

\caption{{IRAM} 30m spectra for CO(1-0) and CO(2-1) for A262, RXJ0338+09, 
RXJ0821+07 and A1068. The solid line is the
coadded 500~MHz data from A100 and B100 and the dashed line is the coadded
Autocorrelator data from A100 and B100. }

\end{figure*}

\vfill
\newpage

\begin{figure*}

\centerline{
\psfig{file=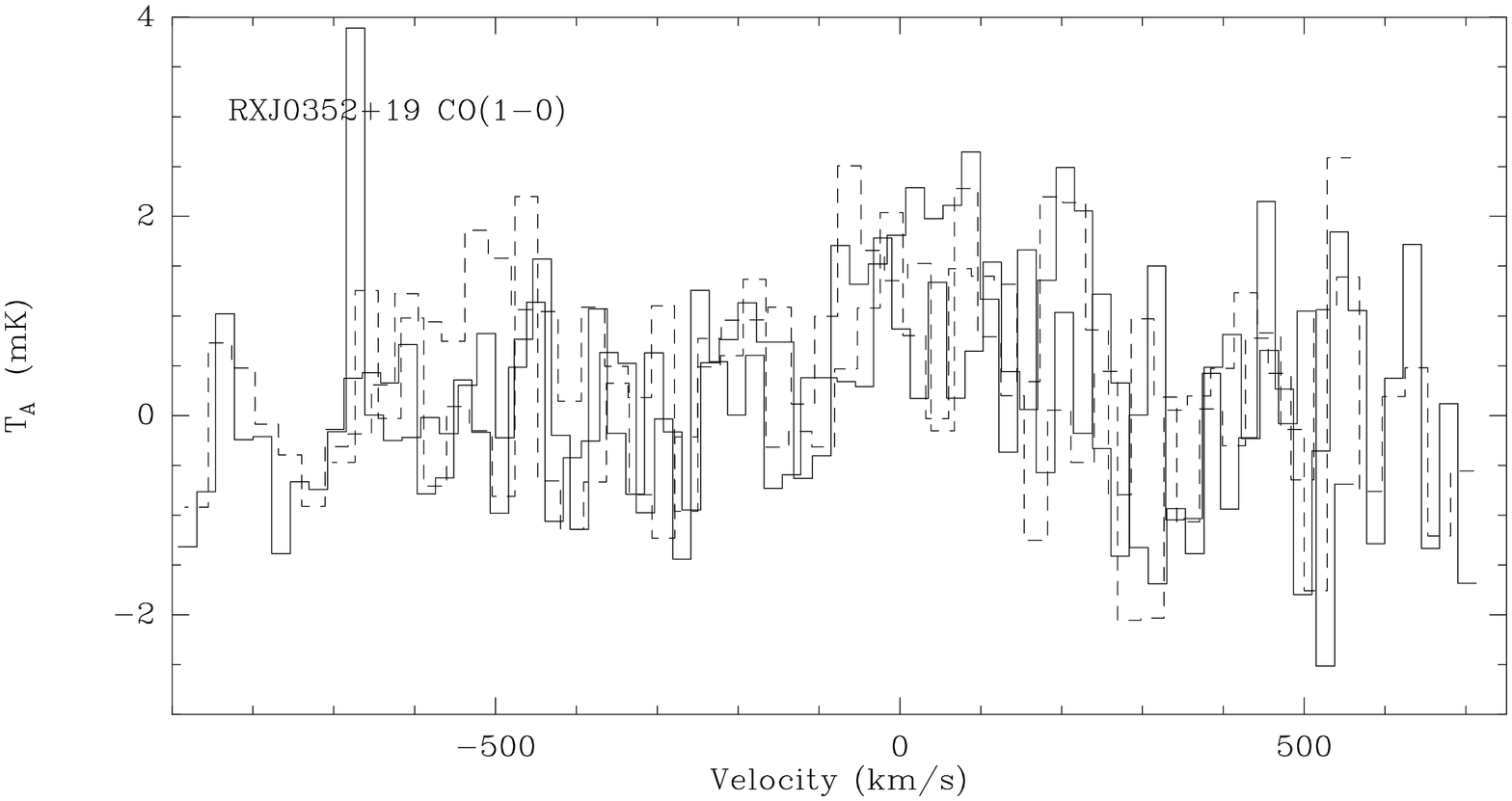,angle=270,width=8cm}
\psfig{file=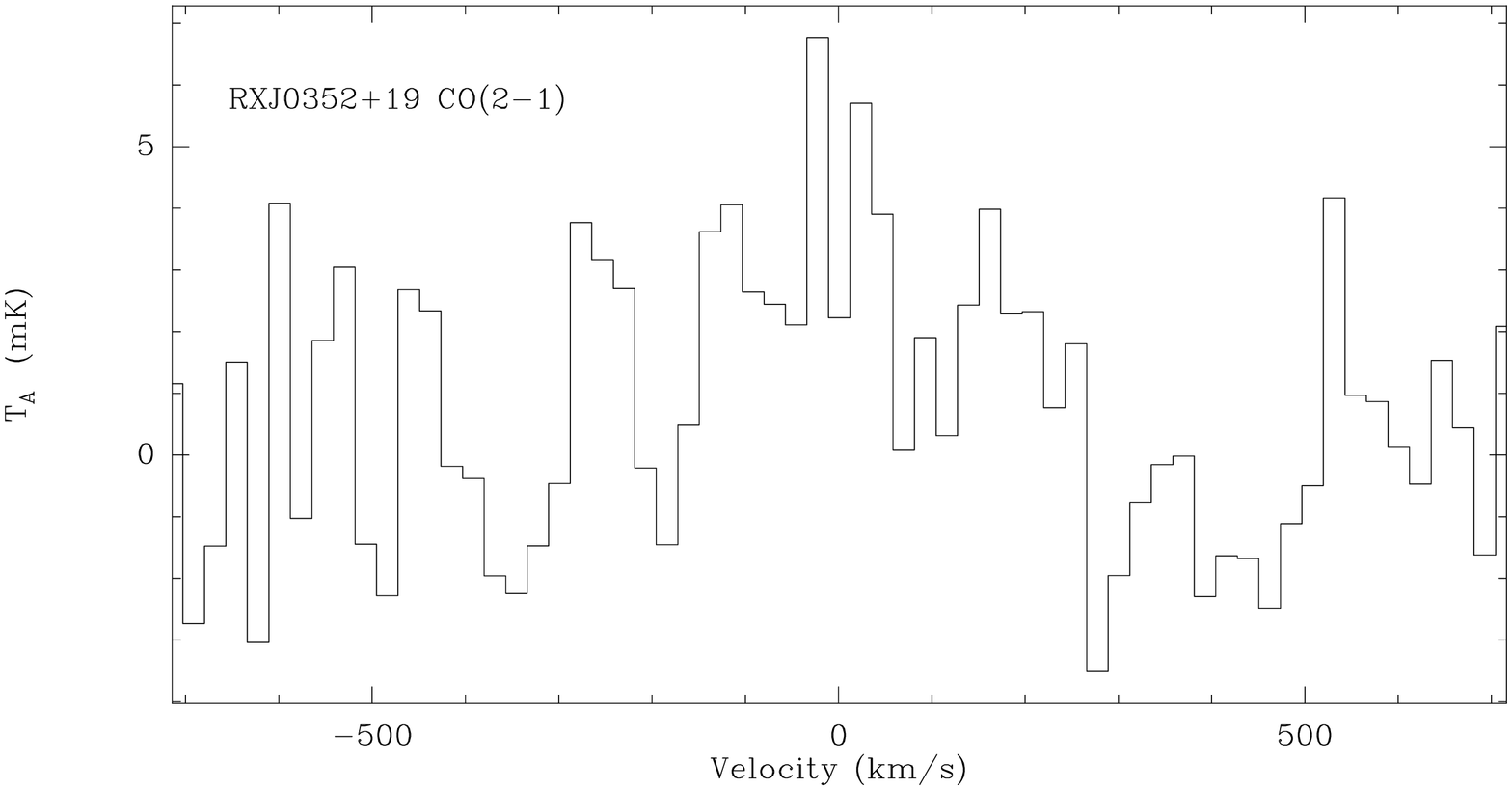,angle=270,width=8cm}}
\centerline{
\psfig{file=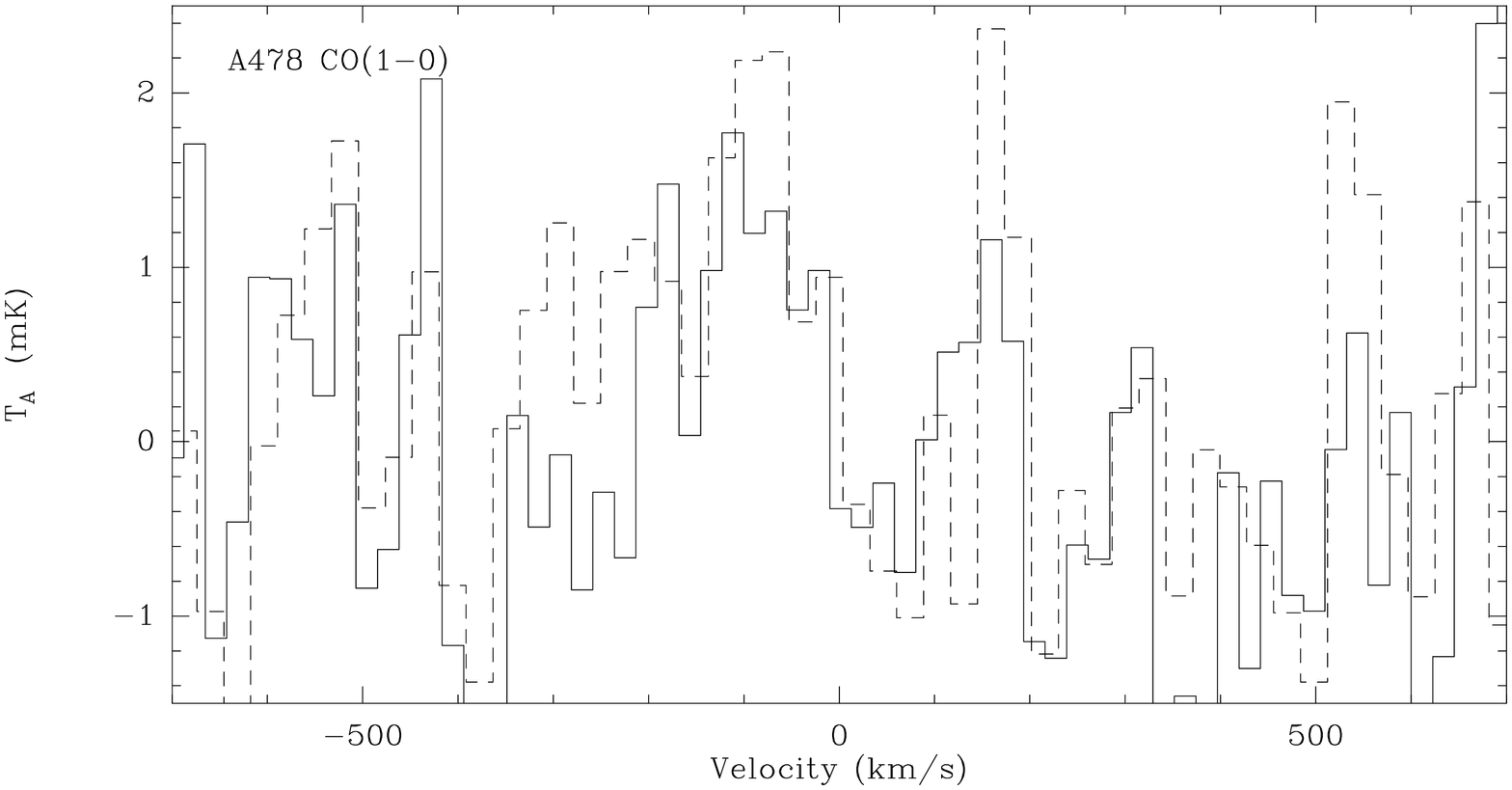,angle=270,width=8cm}
\psfig{file=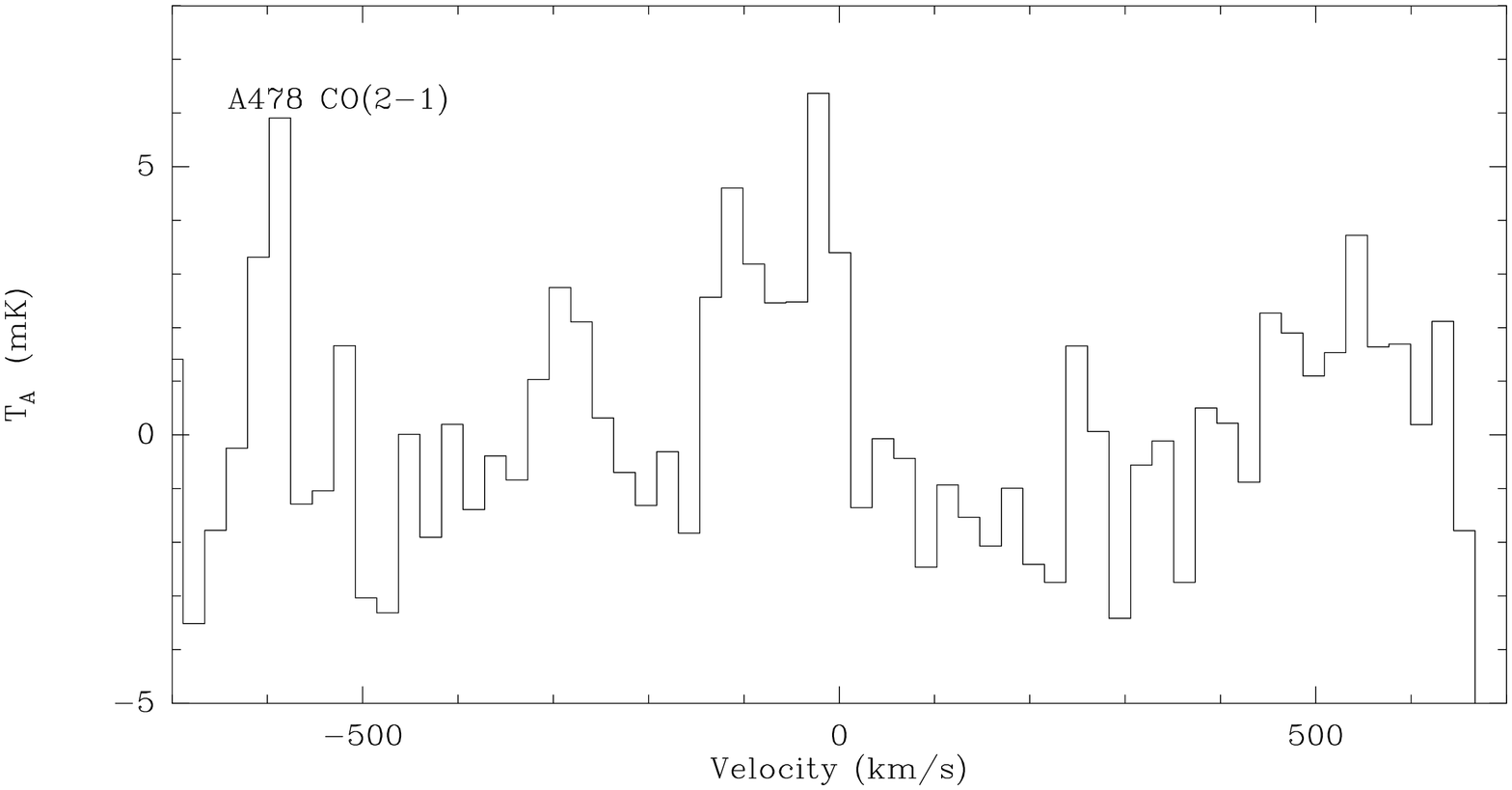,angle=270,width=8cm}}
\centerline{
\psfig{file=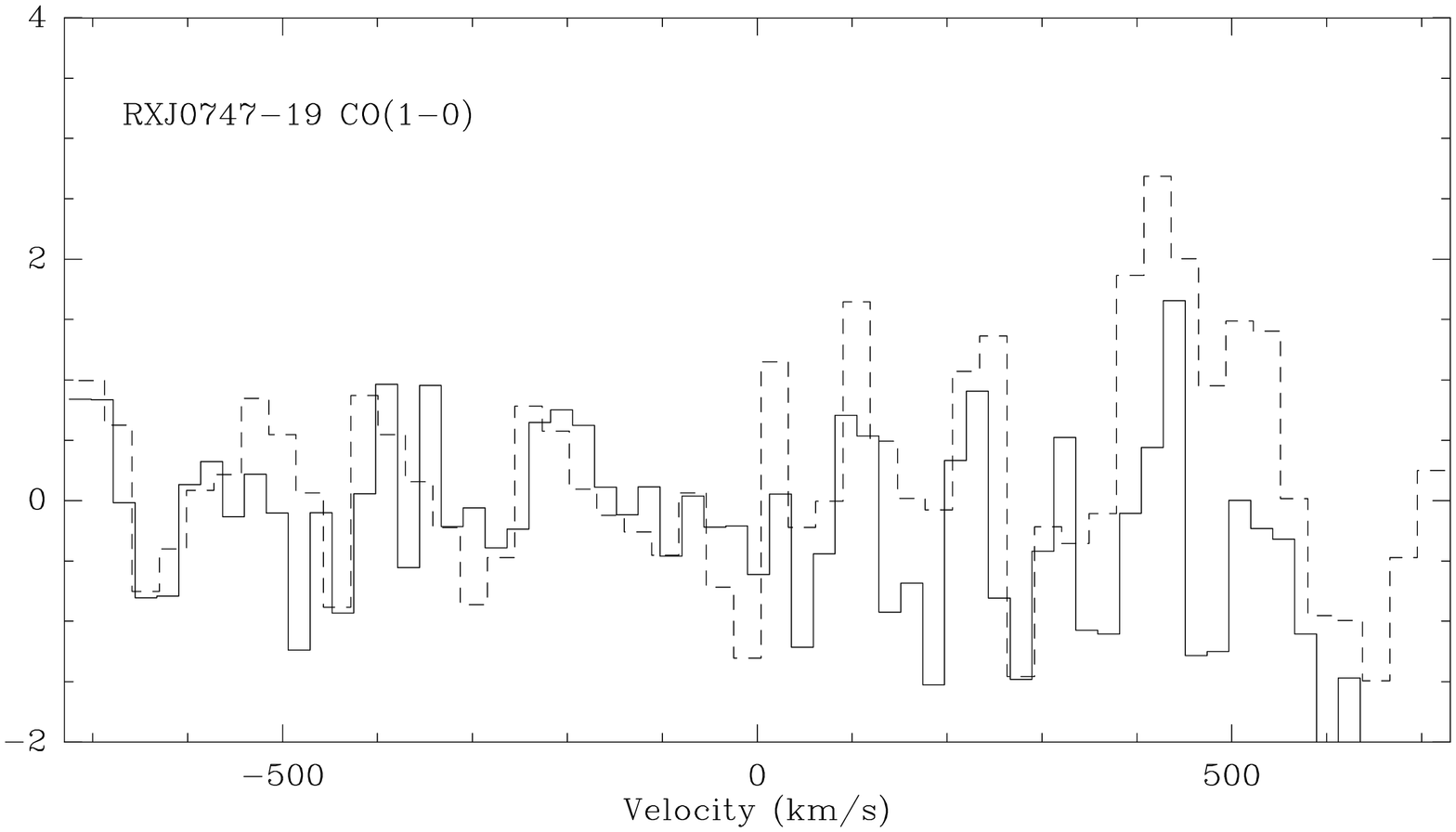,angle=270,width=8cm}
\psfig{file=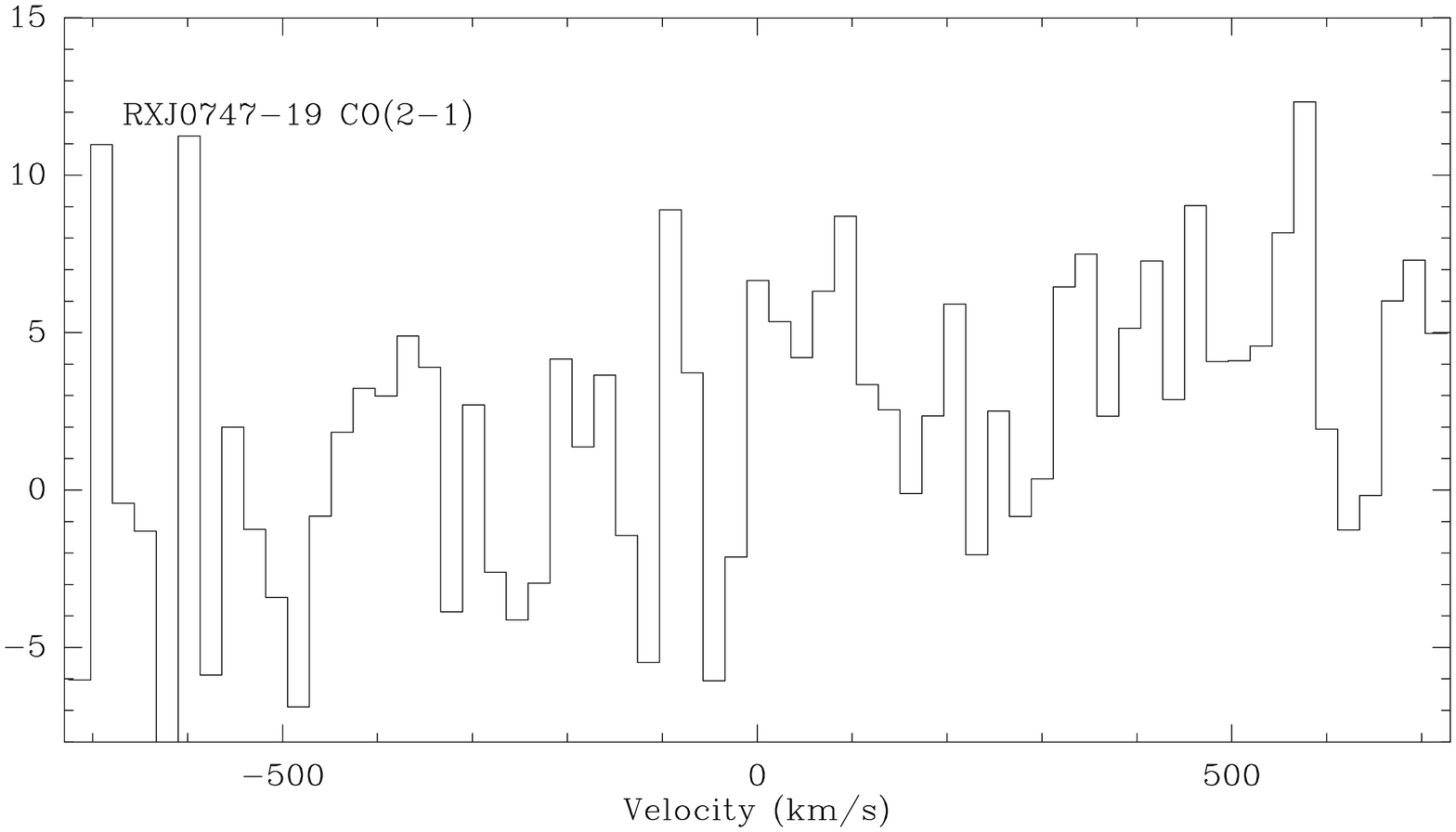,angle=270,width=8cm}}
\centerline{
\psfig{file=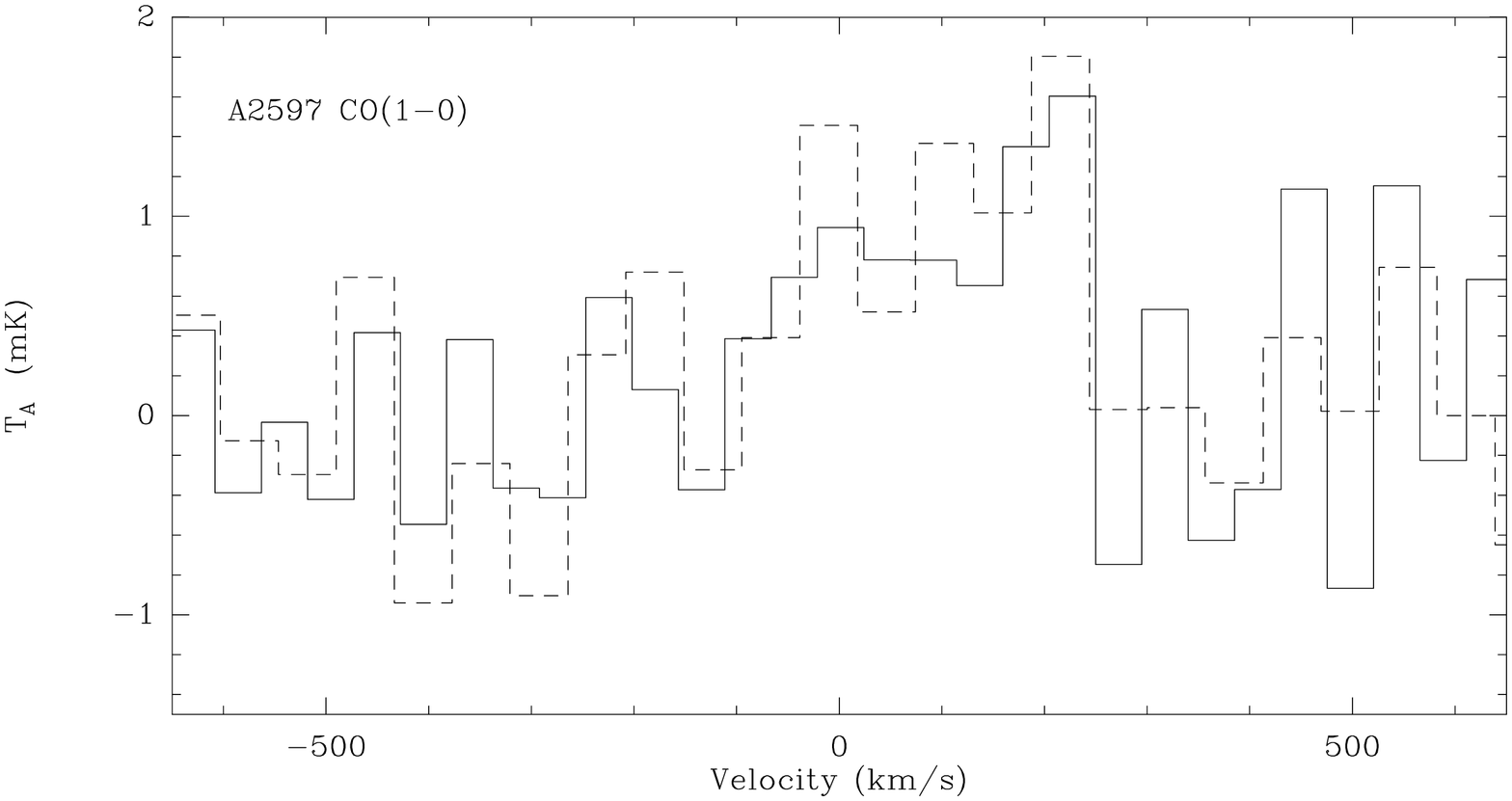,angle=270,width=8cm}
\psfig{file=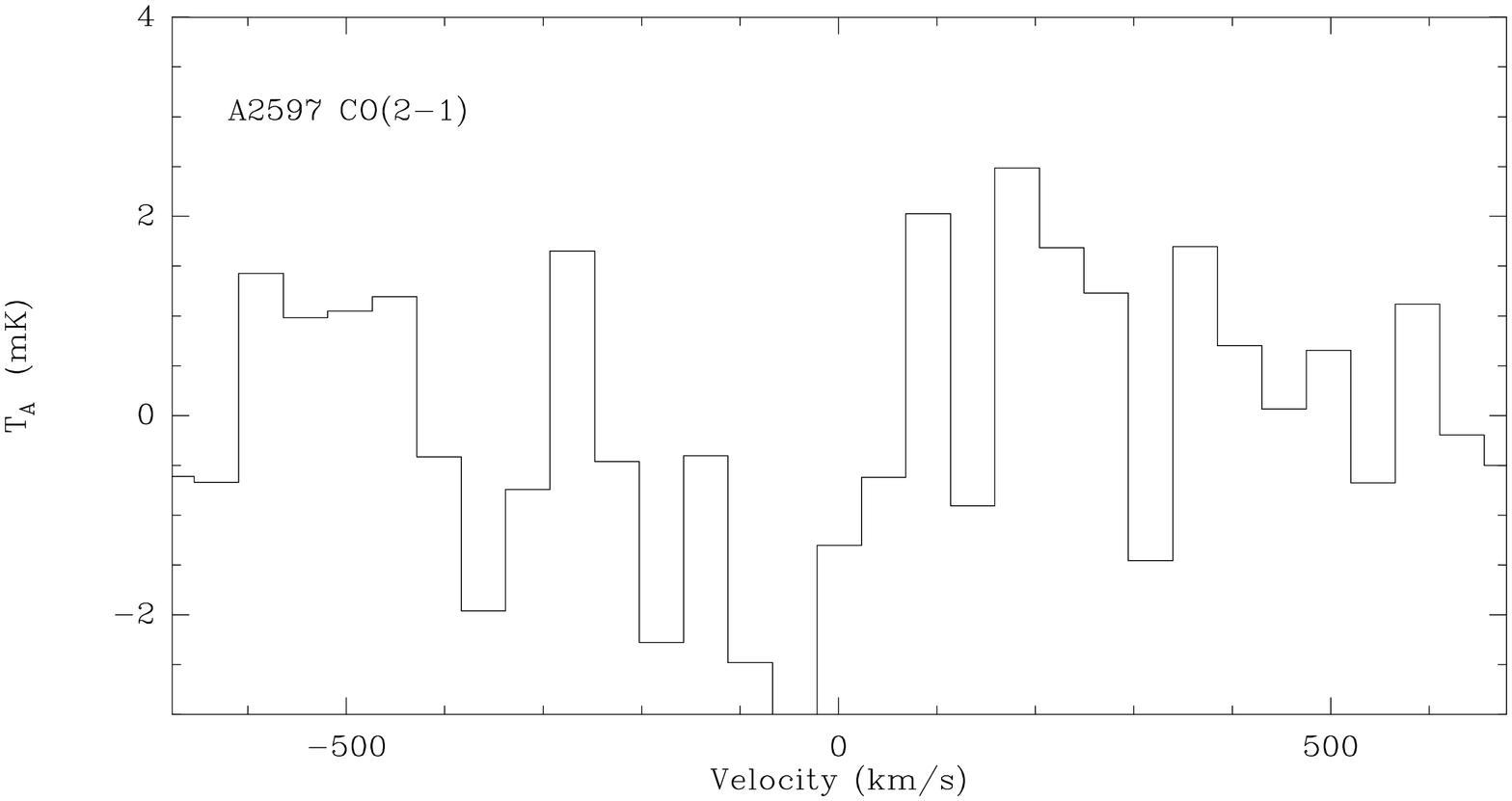,angle=270,width=8cm}}

\caption{{IRAM} 30m spectra for CO(1-0) and CO(2-1) for RXJ0352+19, A478,
RXJ0747-19 and A2597. The solid line is the
coadded 500~MHz data from A100 and B100 and the dashed line is the coadded
Autocorrelator data from A100 and B100. The plot for RXJ0352+19 
contains data from two, frequency-shifted observations.
The data for A2597 are smoothed more 
heavily than the other spectra to show the weak feature in both spectra.}

\end{figure*}

\begin{figure*}

\centerline{
\psfig{file=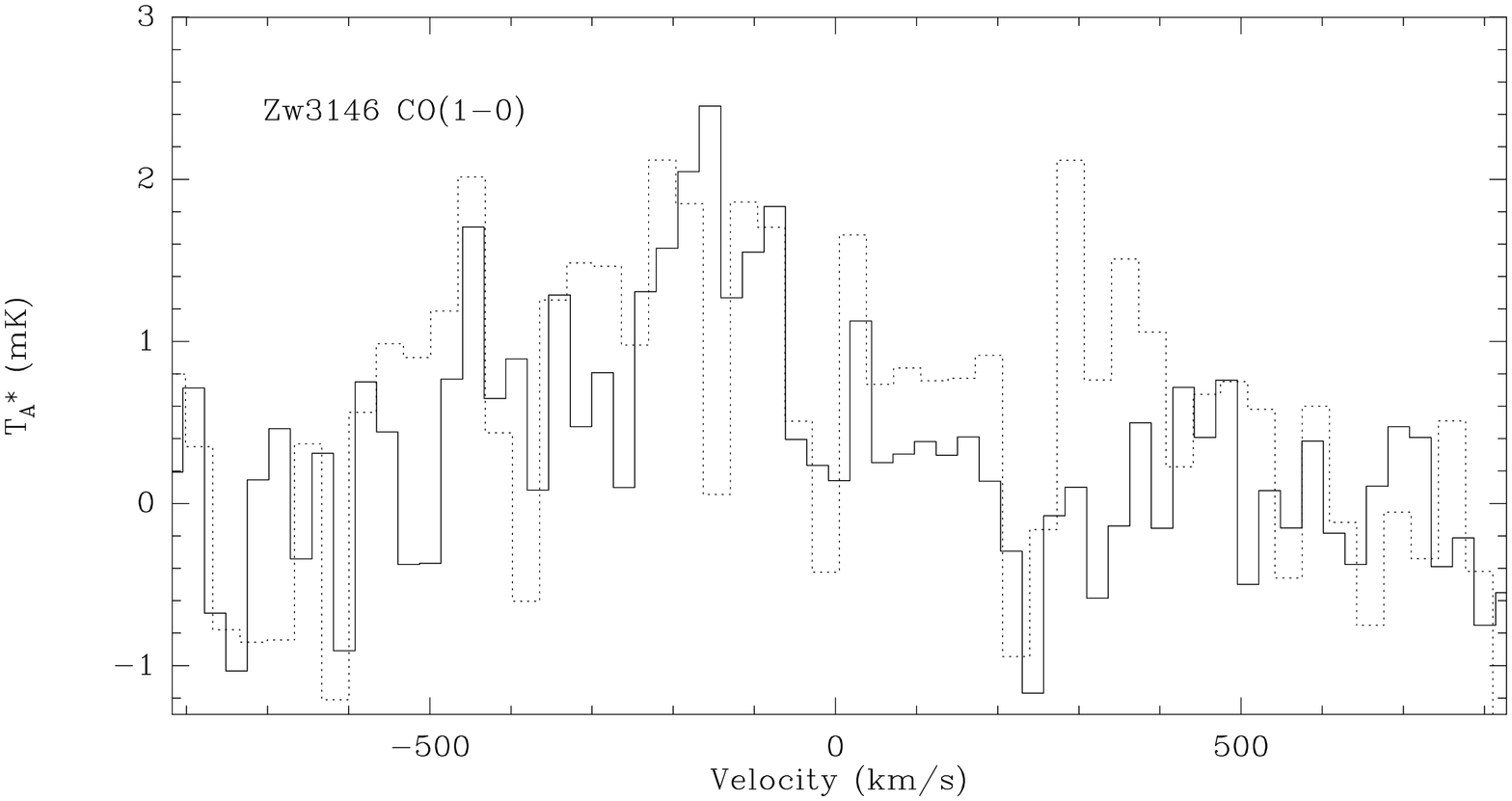,angle=270,width=8cm}
\psfig{file=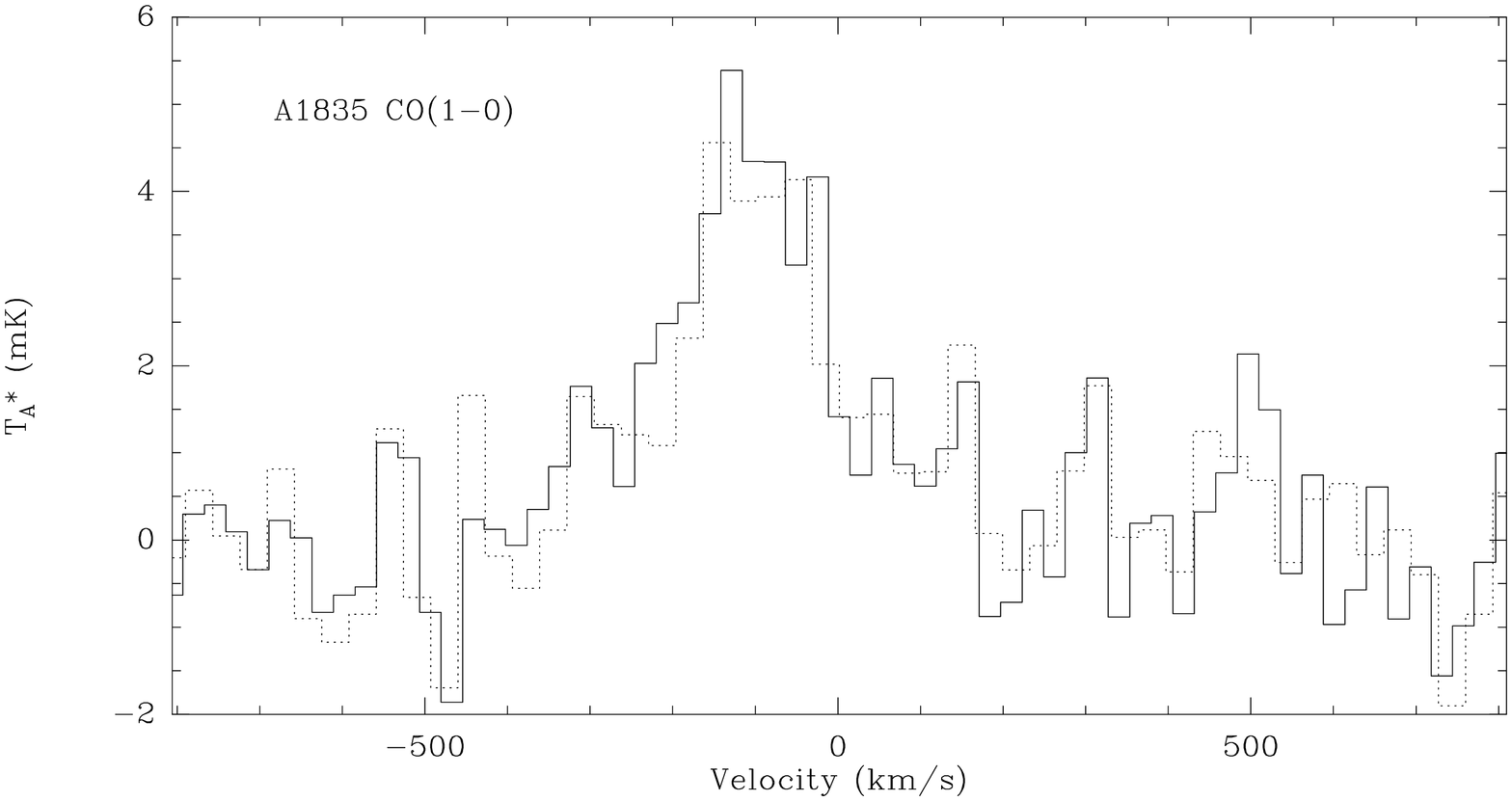,angle=270,width=8cm}}
\centerline{
\psfig{file=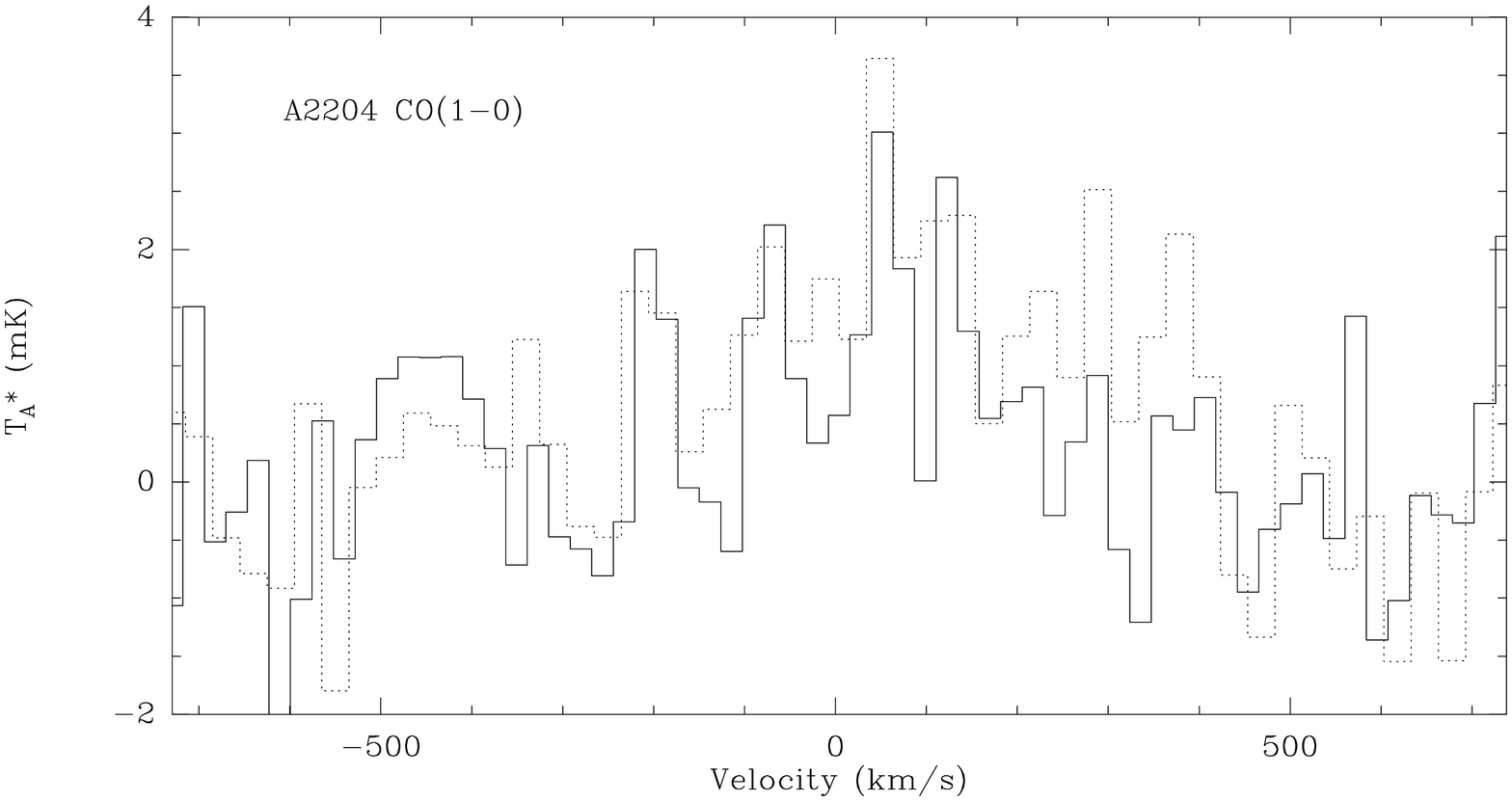,angle=270,width=8cm}
\psfig{file=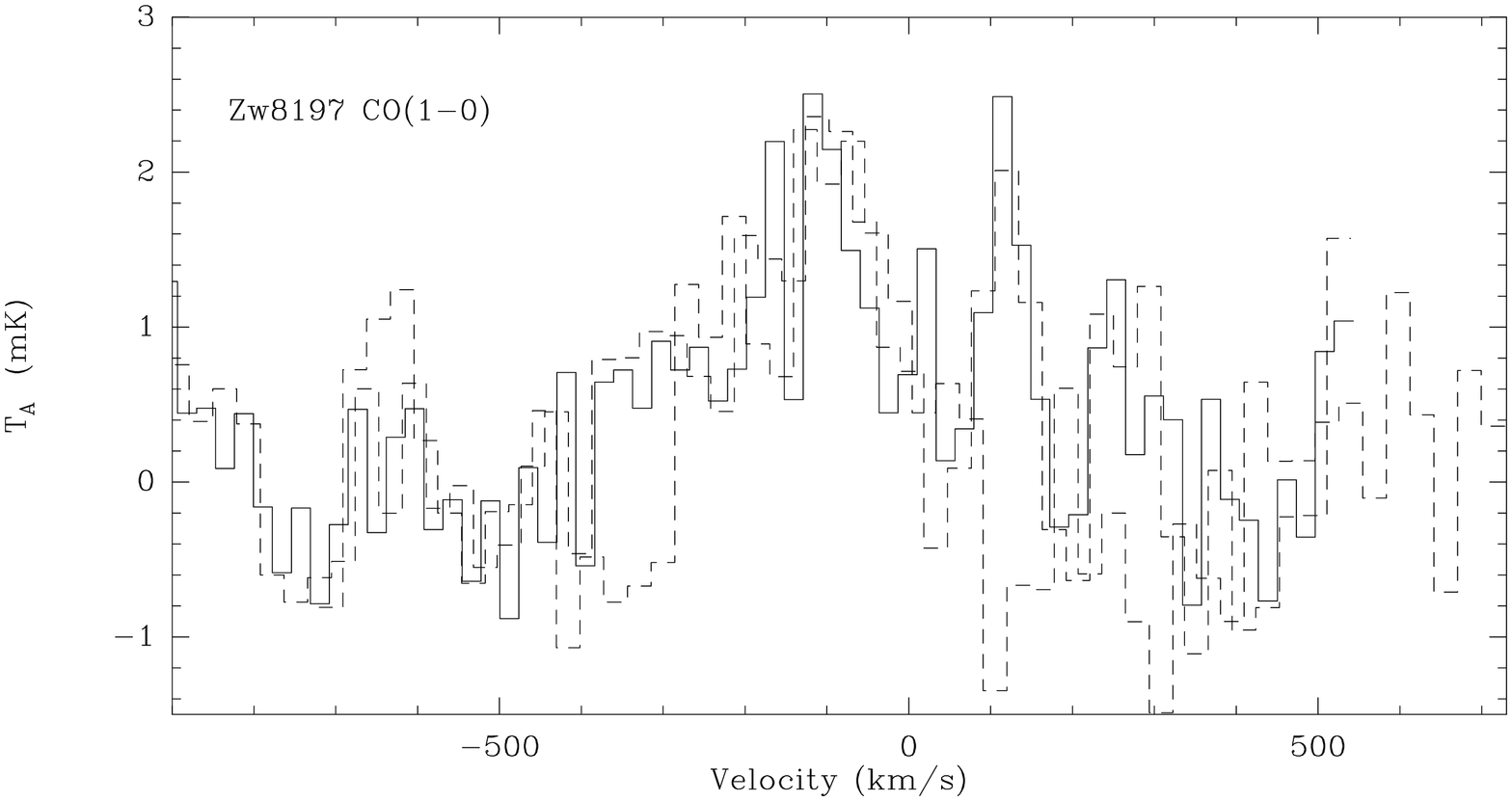,angle=270,width=8cm}}
\centerline{
\psfig{file=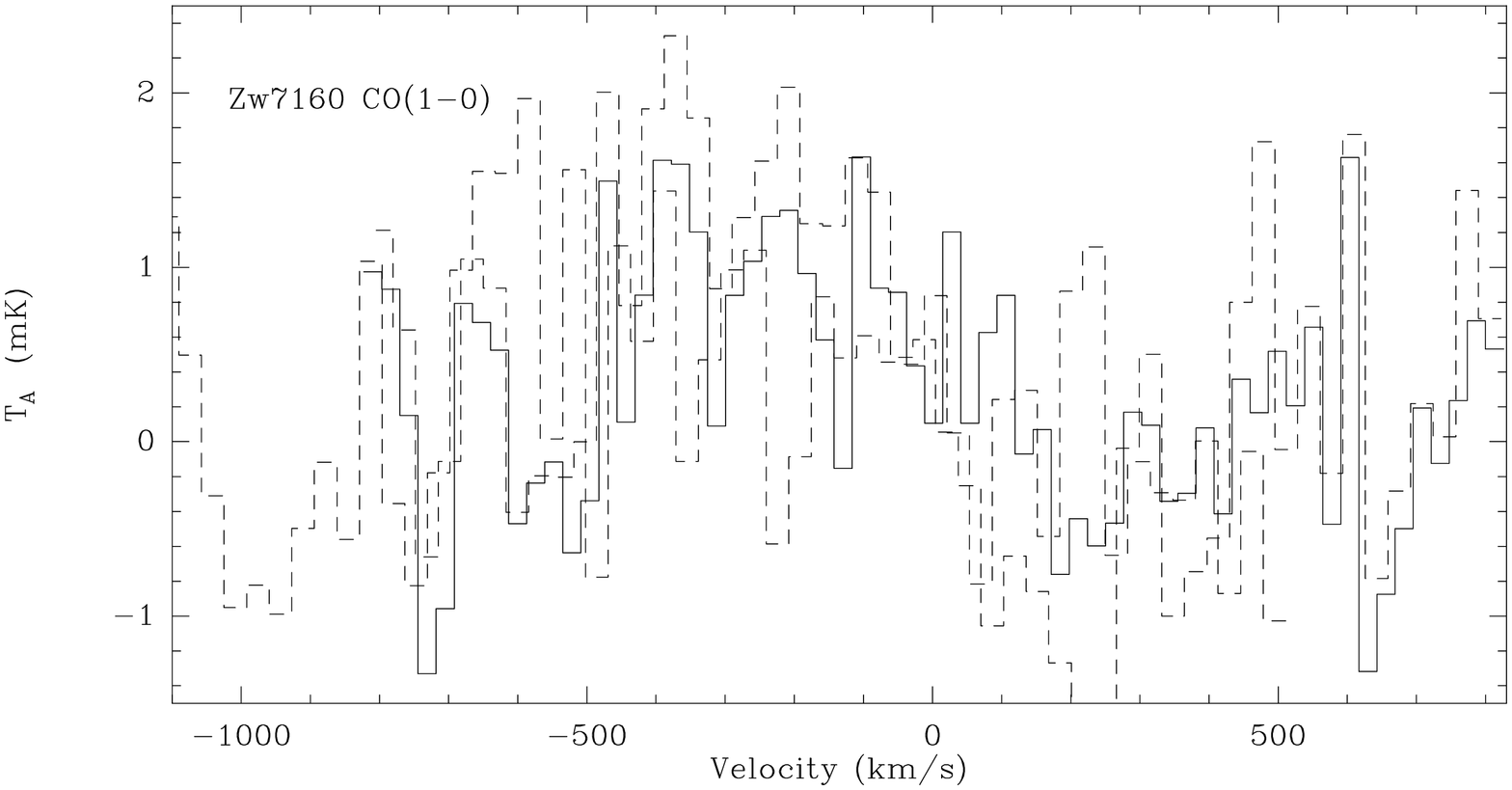,angle=270,width=8cm}
\psfig{file=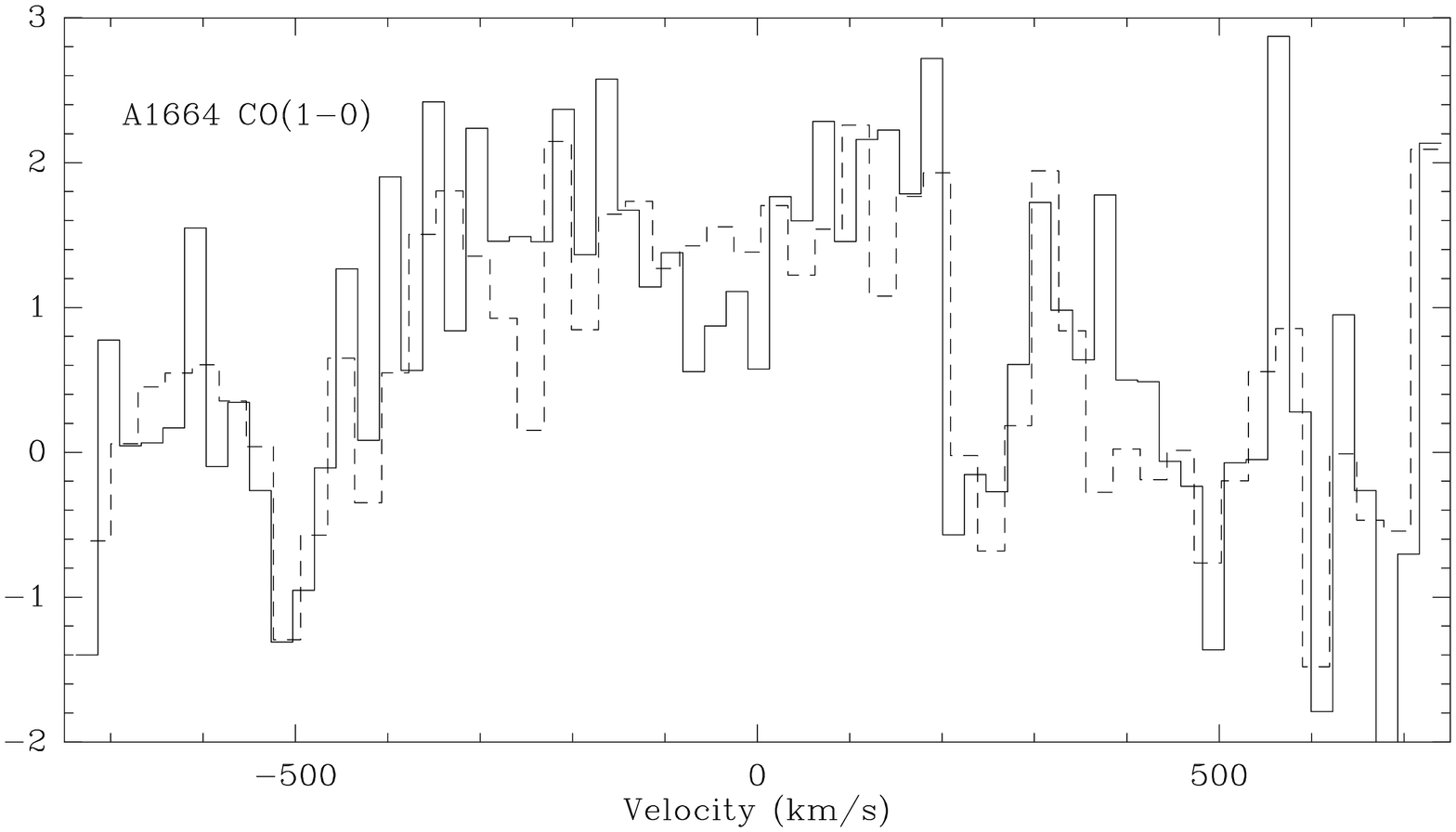,angle=270,width=8cm}}
\centerline{
\psfig{file=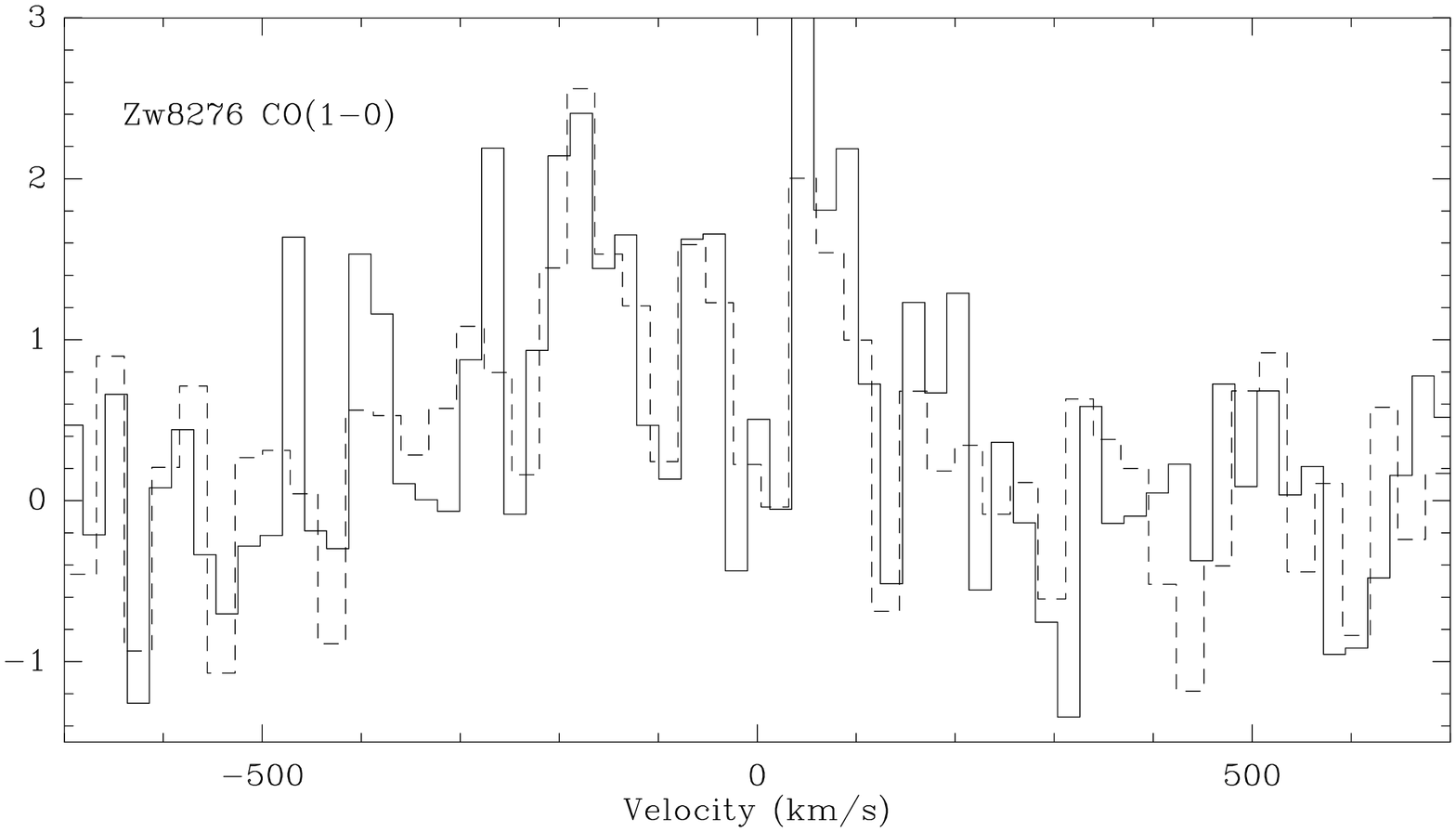,angle=270,width=8cm}
\psfig{file=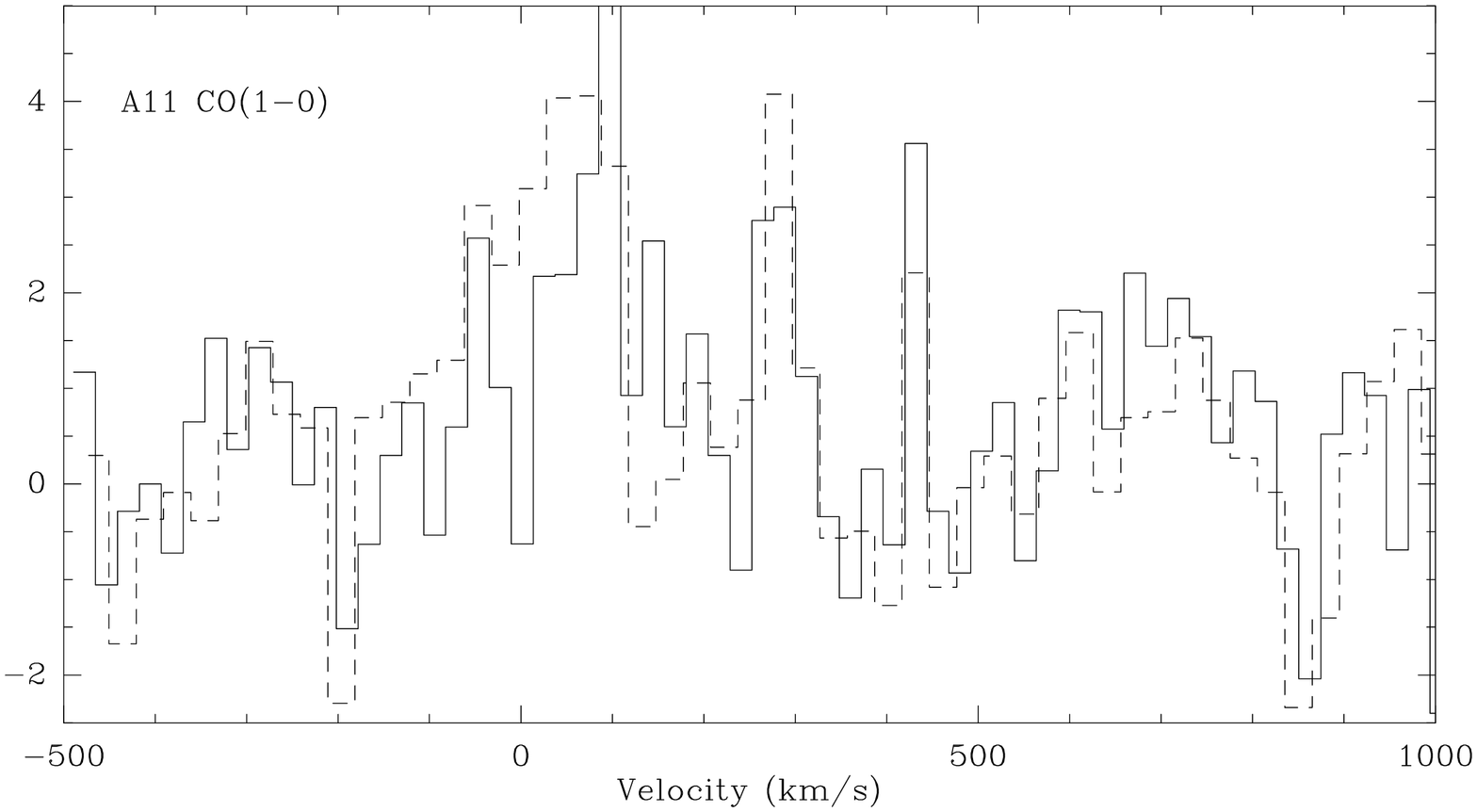,angle=270,width=8cm}}

\caption{{IRAM} 30m spectra for Zw3146, A1835, A2204, Zw8197, Zw7160, A1664, Zw8276 and
A11  where only CO(1-0) was observed due to redshift or weather
constraints.}

\end{figure*}

\begin{figure*}

\centerline{
\psfig{file=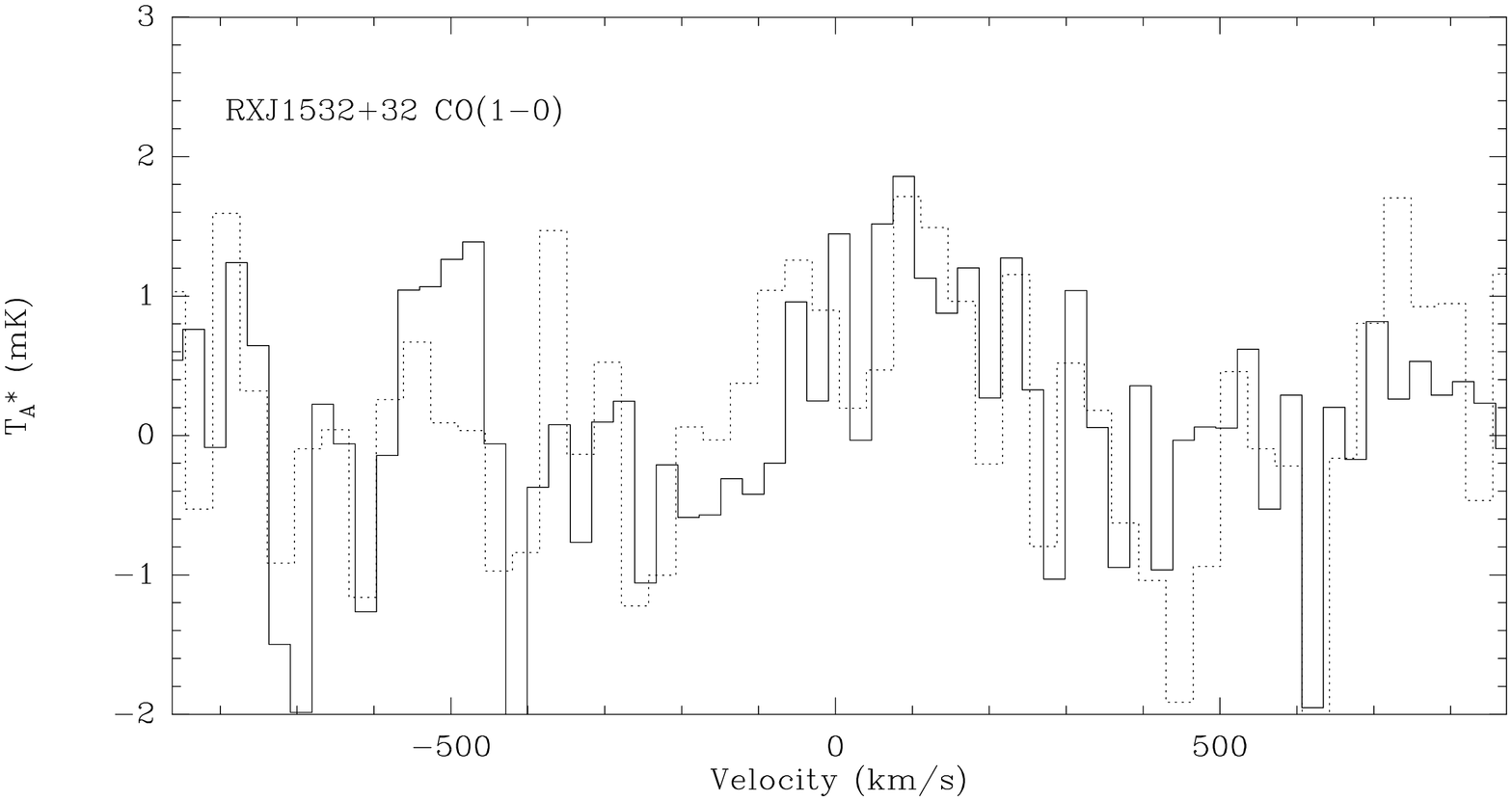,angle=270,width=5cm}
\psfig{file=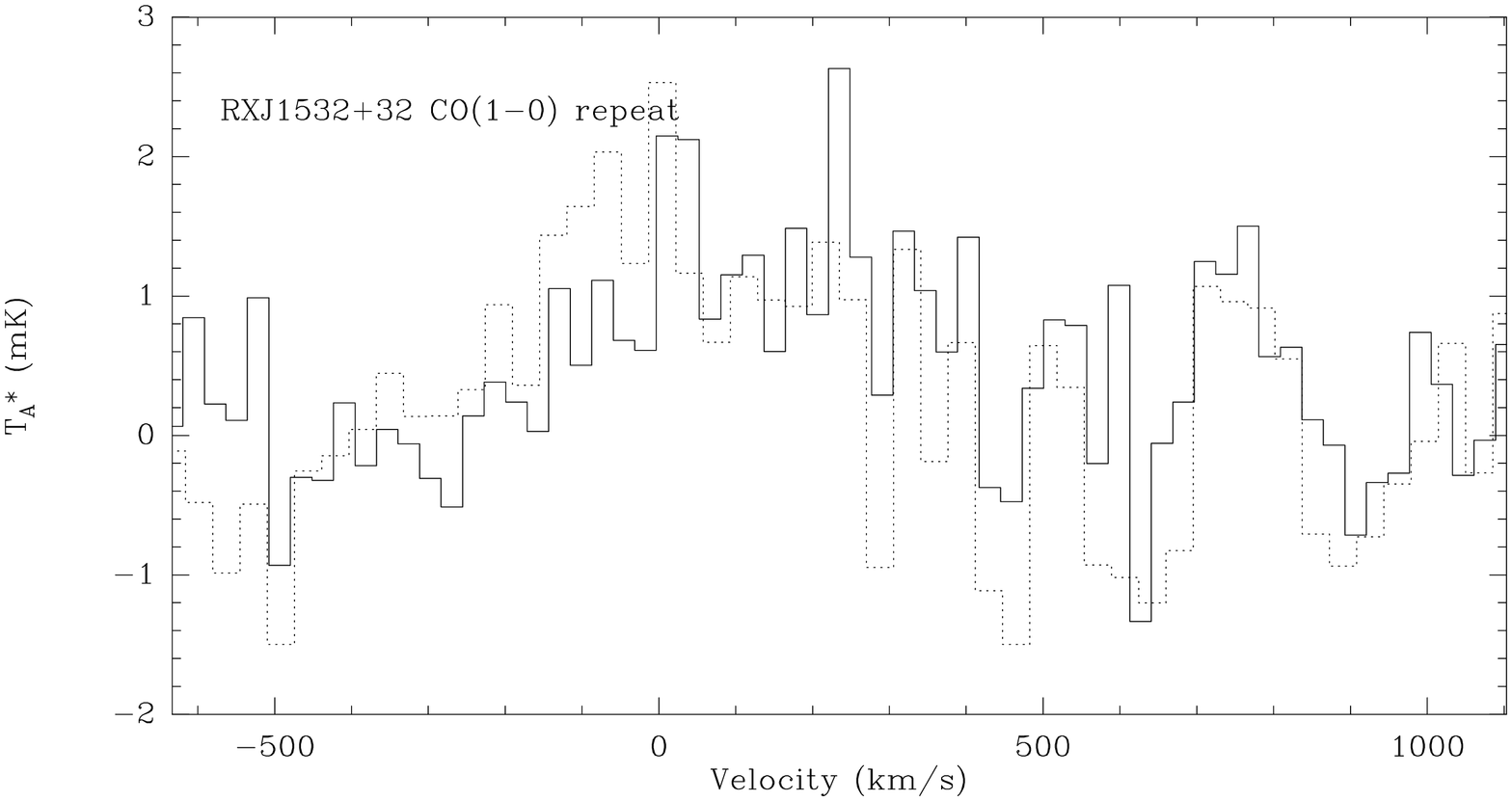,angle=270,width=5cm}
\psfig{file=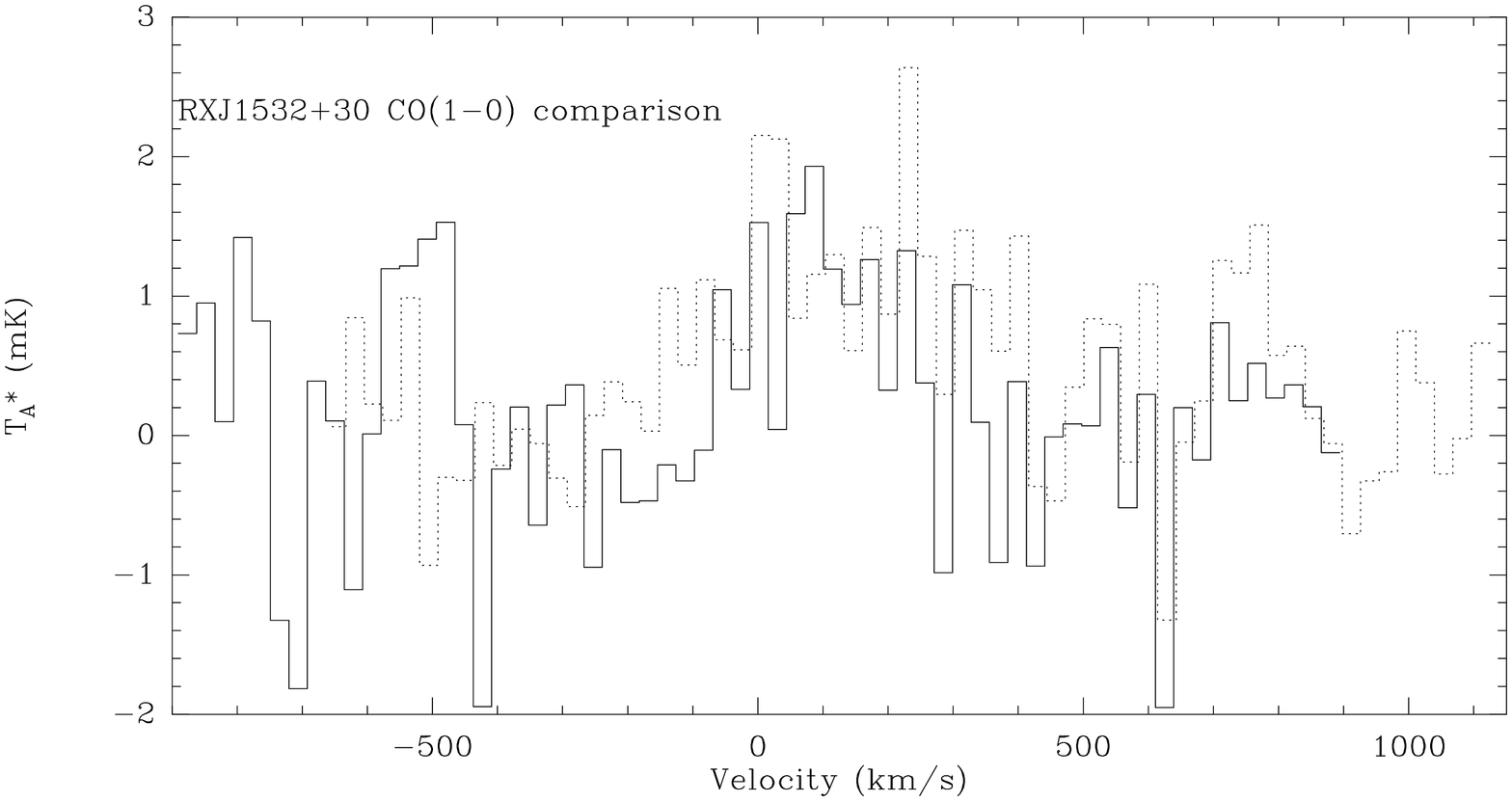,angle=270,width=5cm}}
\centerline{
\psfig{file=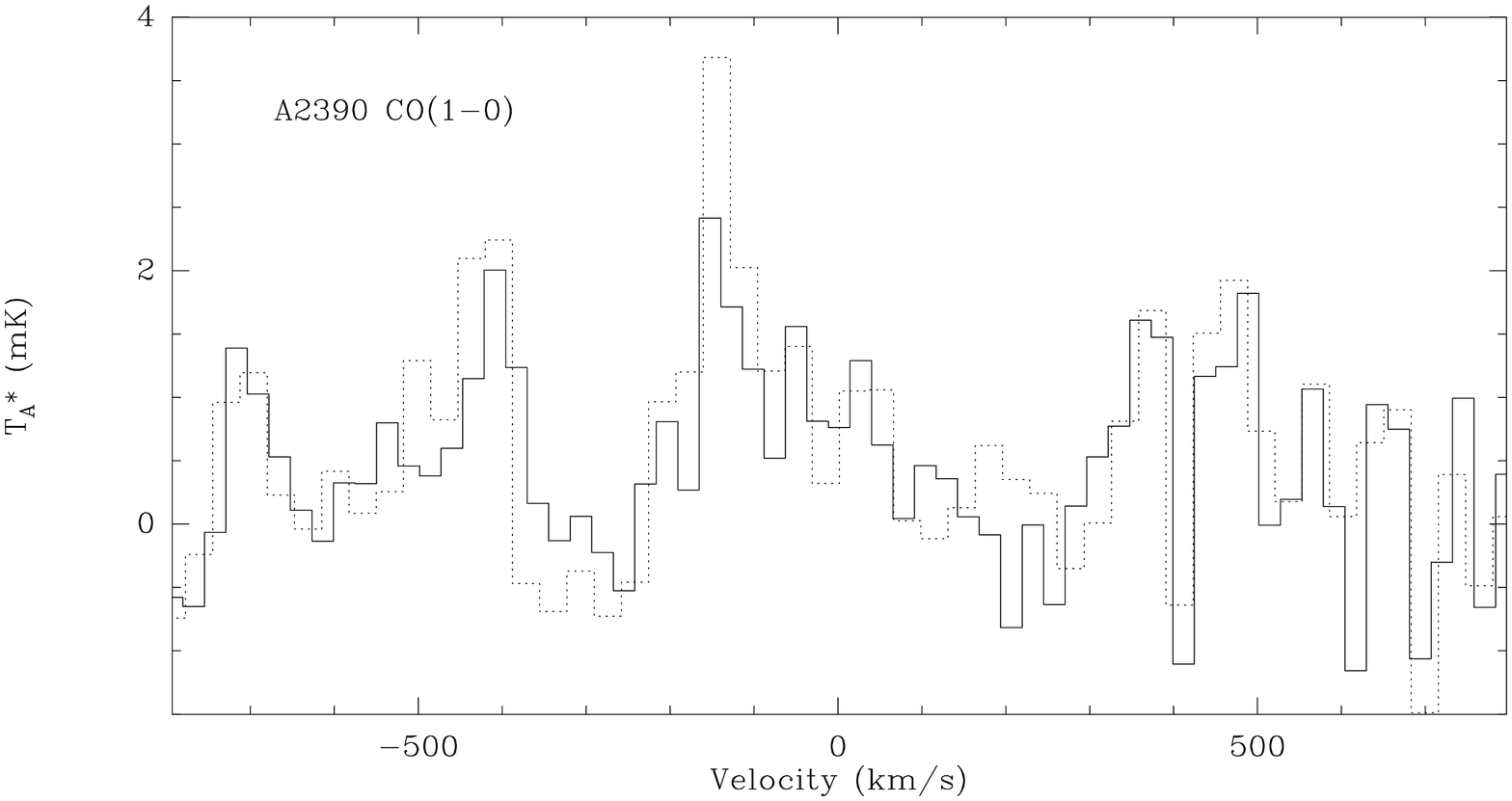,angle=270,width=5cm}
\psfig{file=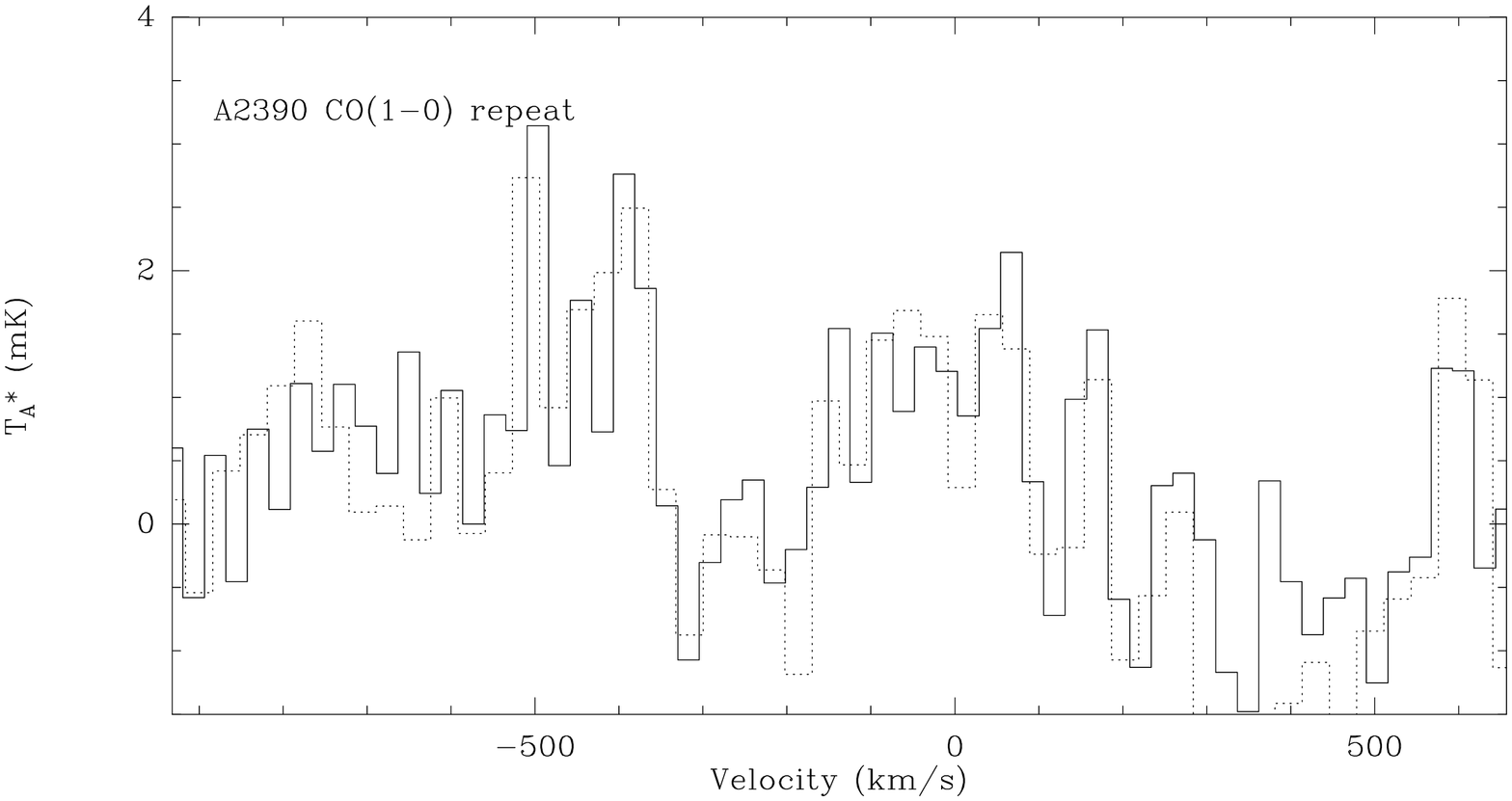,angle=270,width=5cm}
\psfig{file=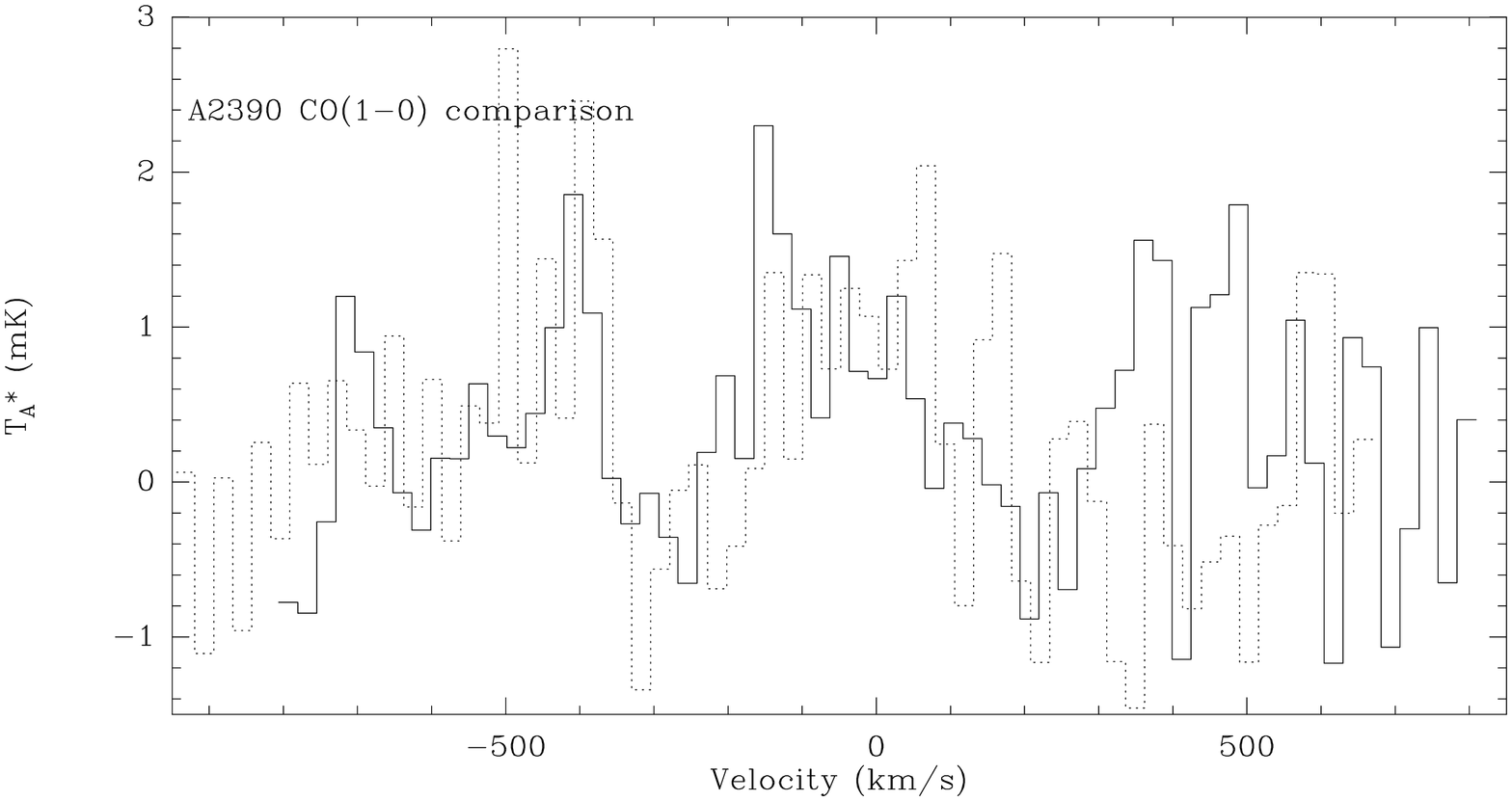,angle=270,width=5cm}}

\centerline{
\psfig{file=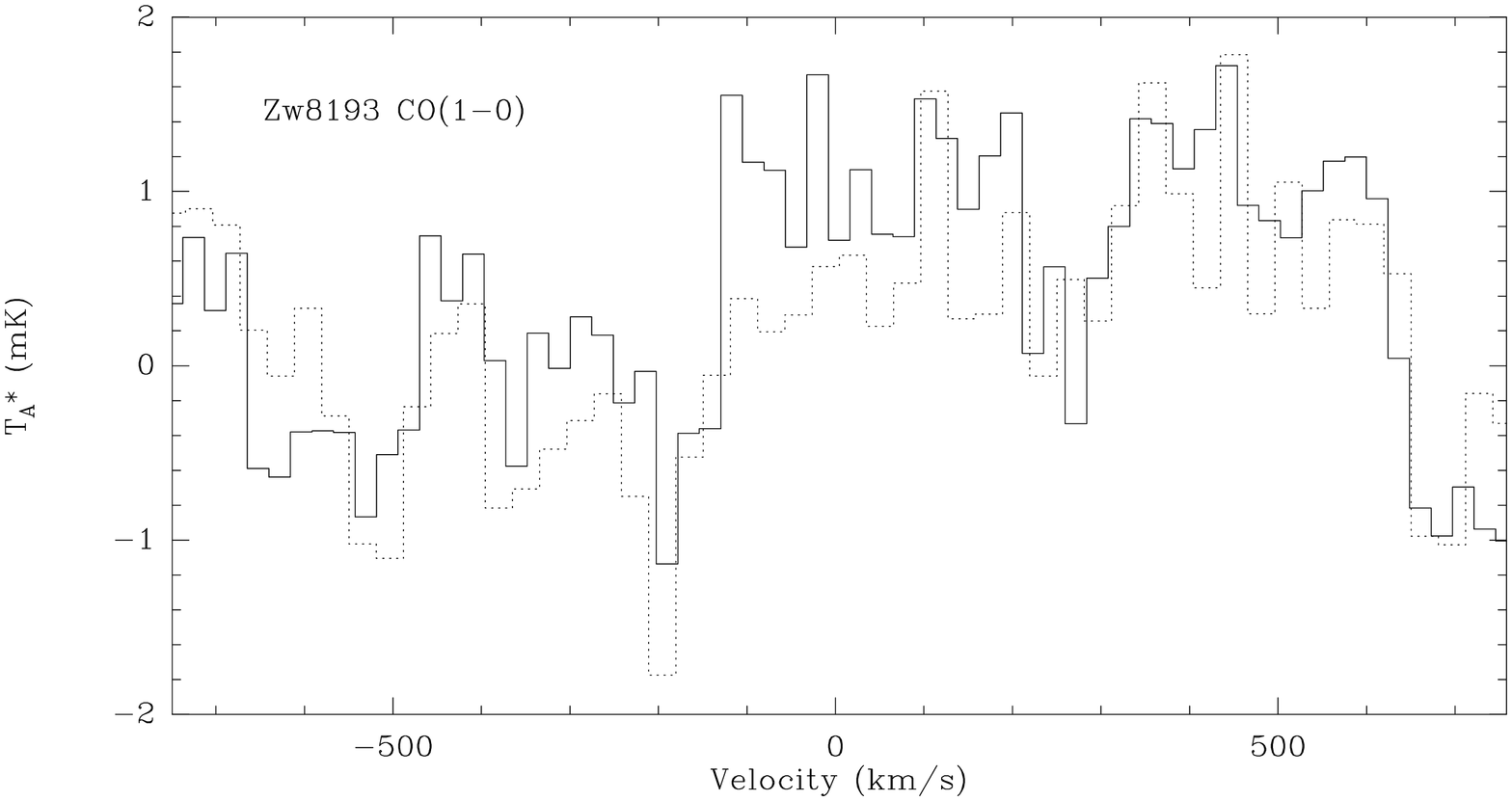,angle=270,width=5cm}
\psfig{file=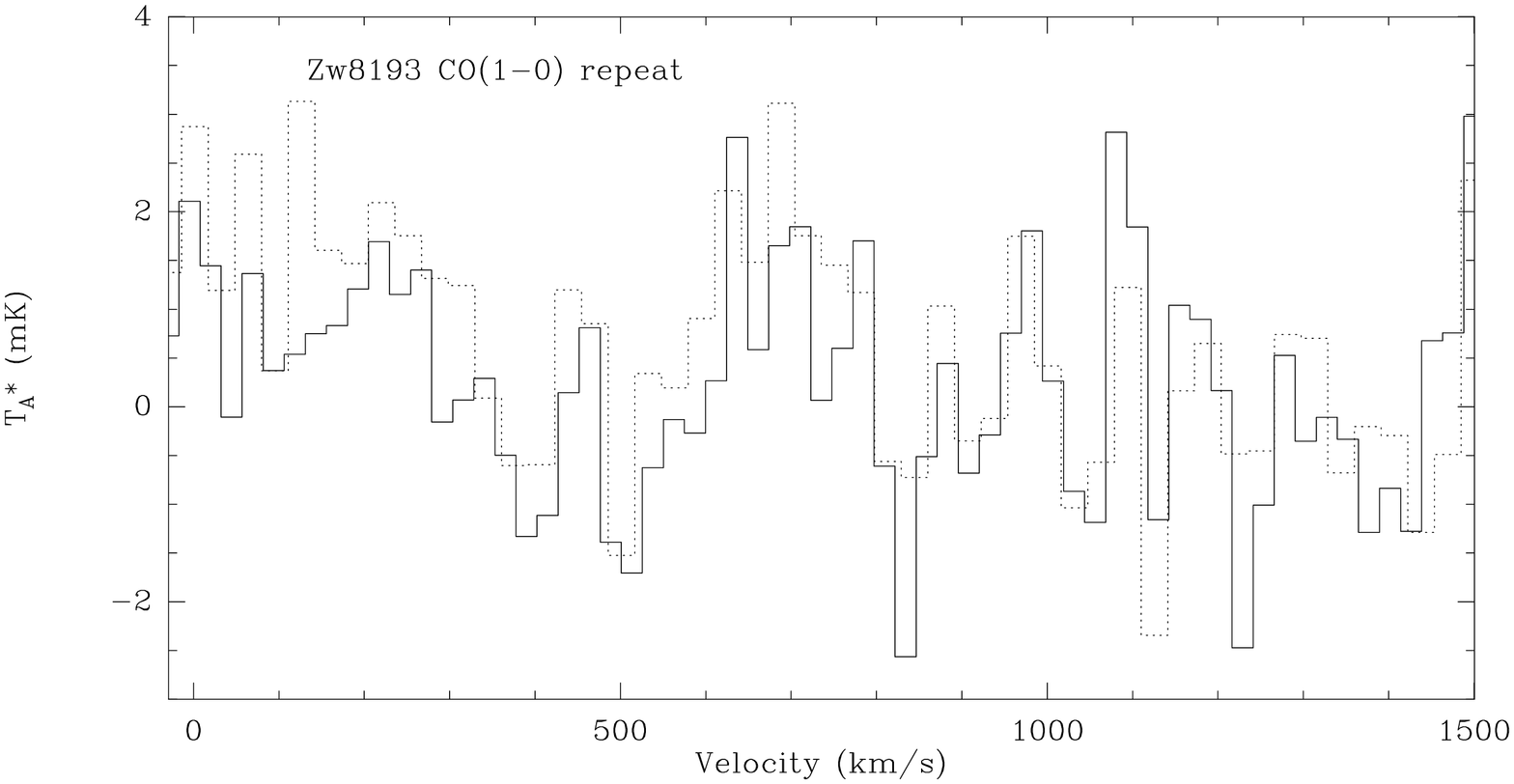,angle=270,width=5cm}
\psfig{file=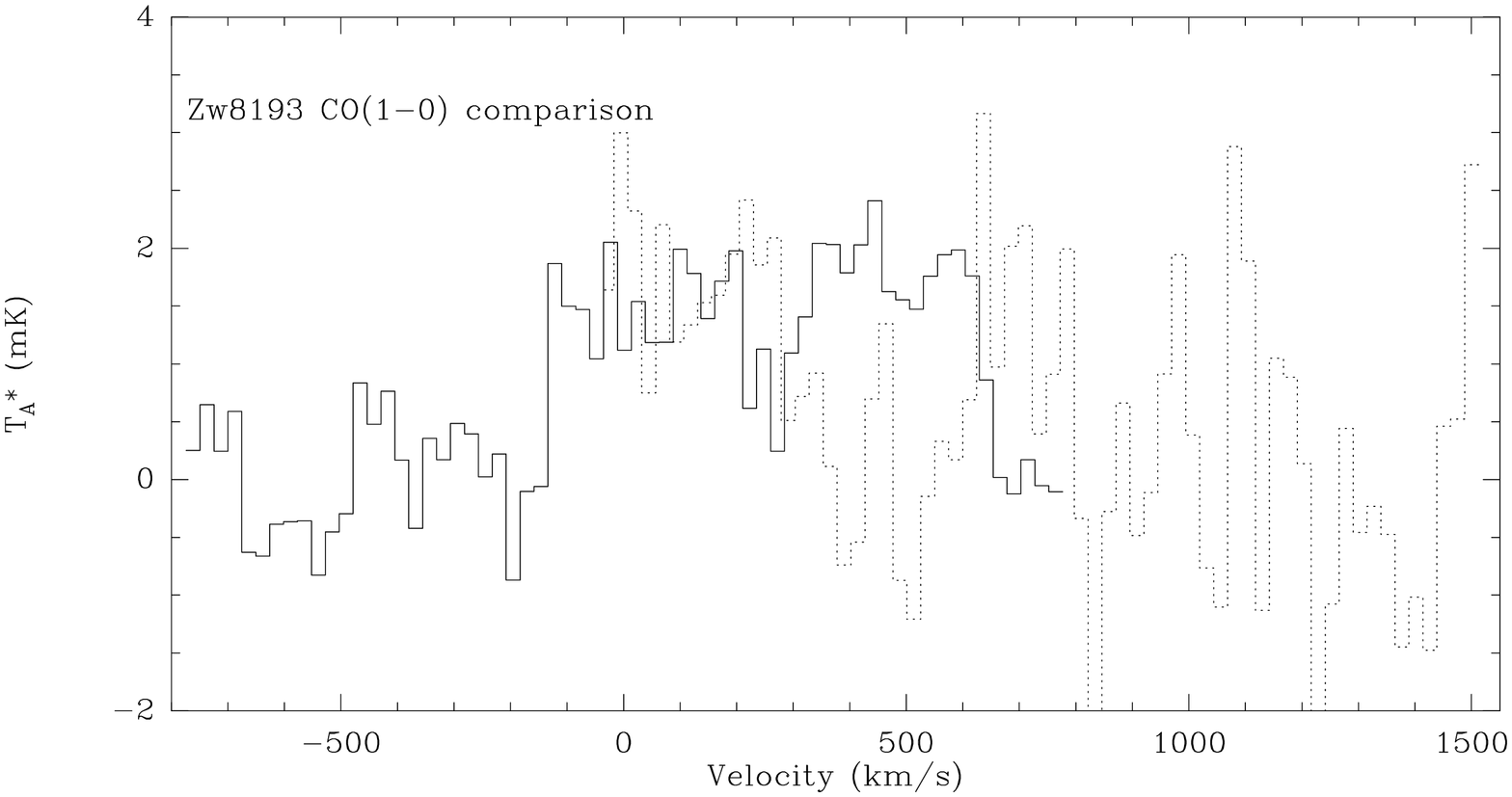,angle=270,width=5cm}}

\caption{{IRAM} 30m spectra for 
RXJ1532+30, A2390, and Zw8193  which were observed several times.
The left and central plots of each line are as in Figure 4. The right-hand
plot the solid line is the 500~MHz data for the first observation and the dashed
line is the 500~MHz data for the second observation. Note the change in 
velocity range between observations.}
\end{figure*}

\begin{figure*}

\centerline{
\psfig{file=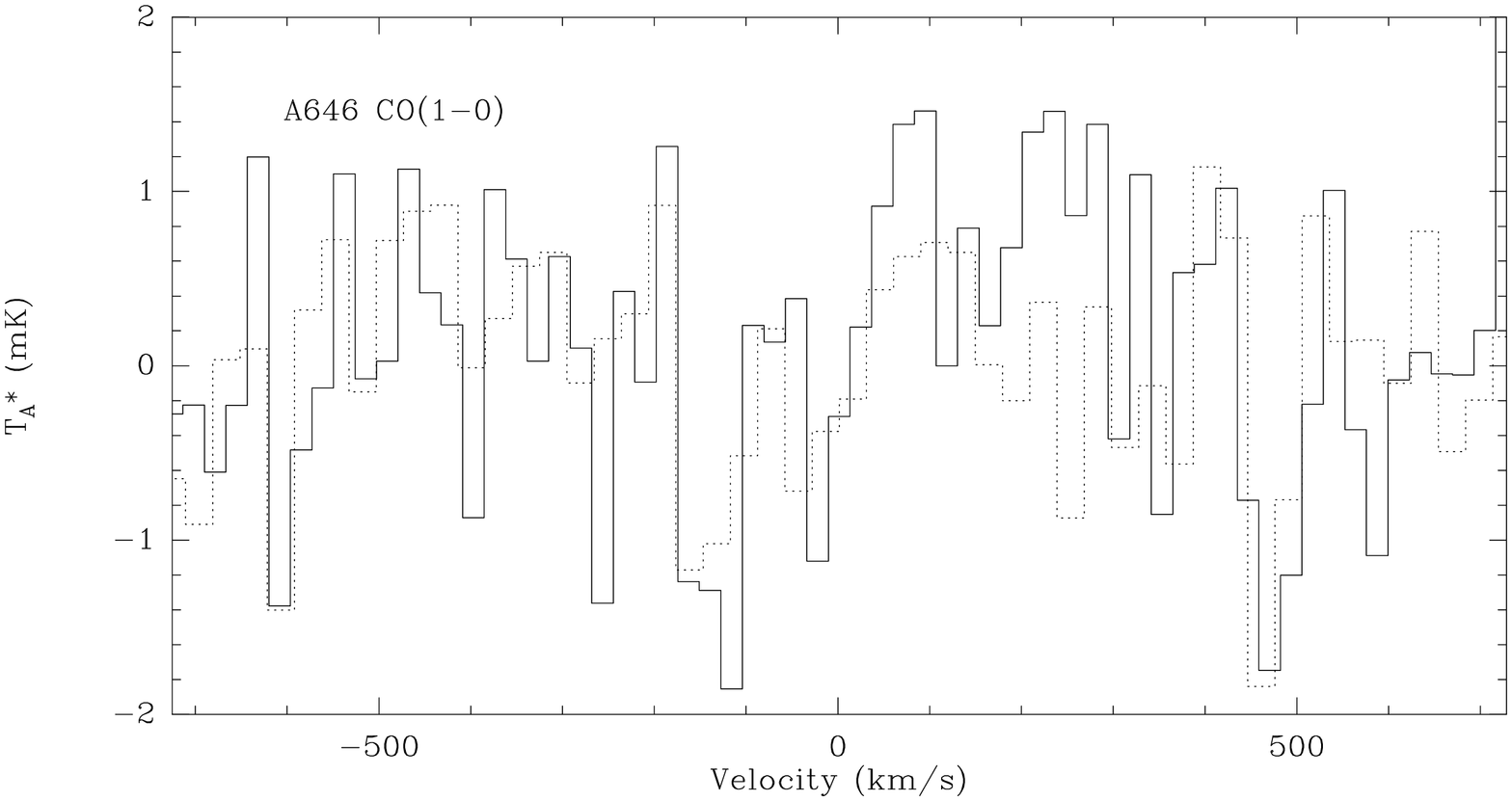,angle=270,width=8cm}
\psfig{file=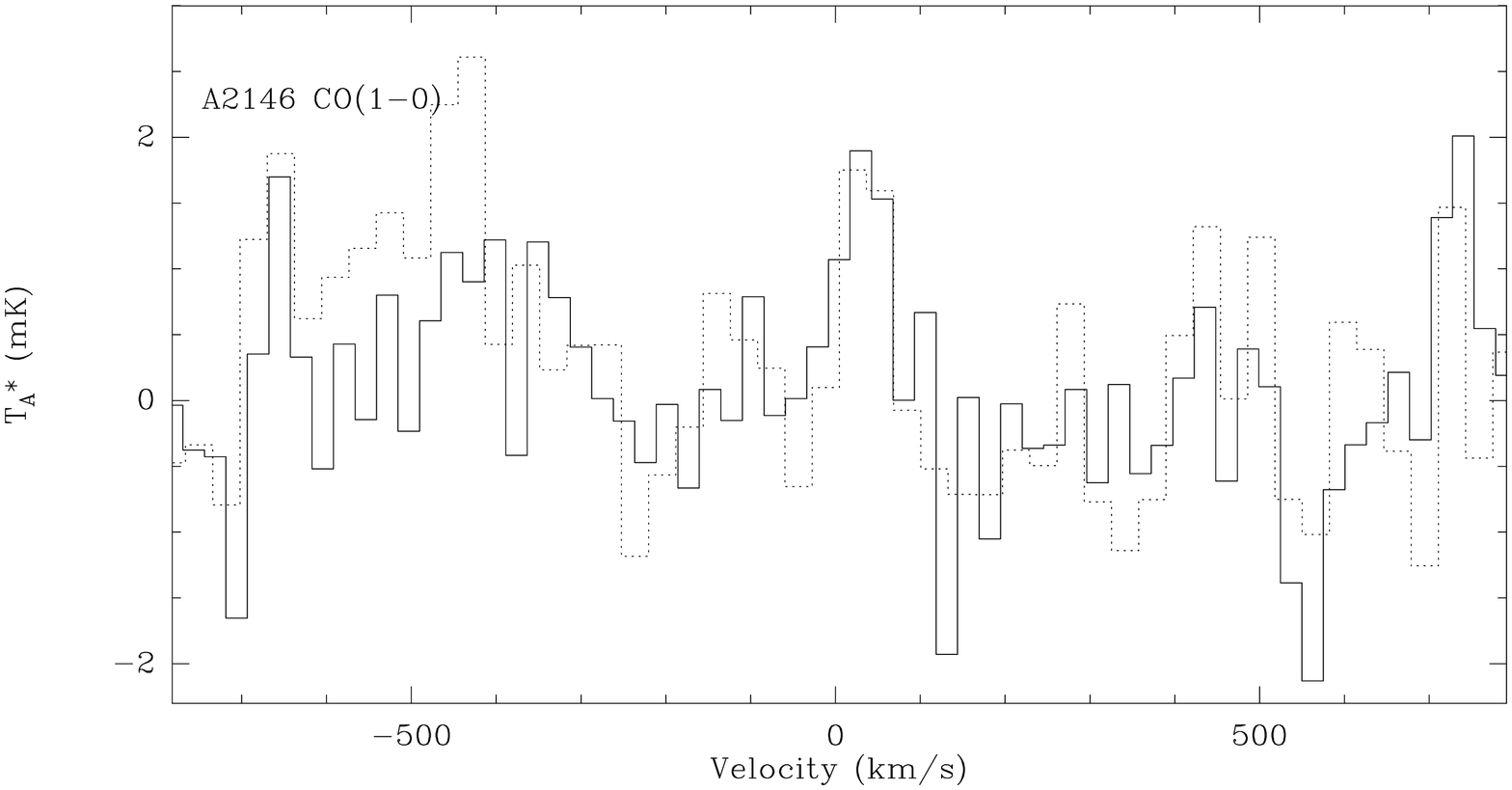,angle=270,width=8cm}}
\centerline{
\psfig{file=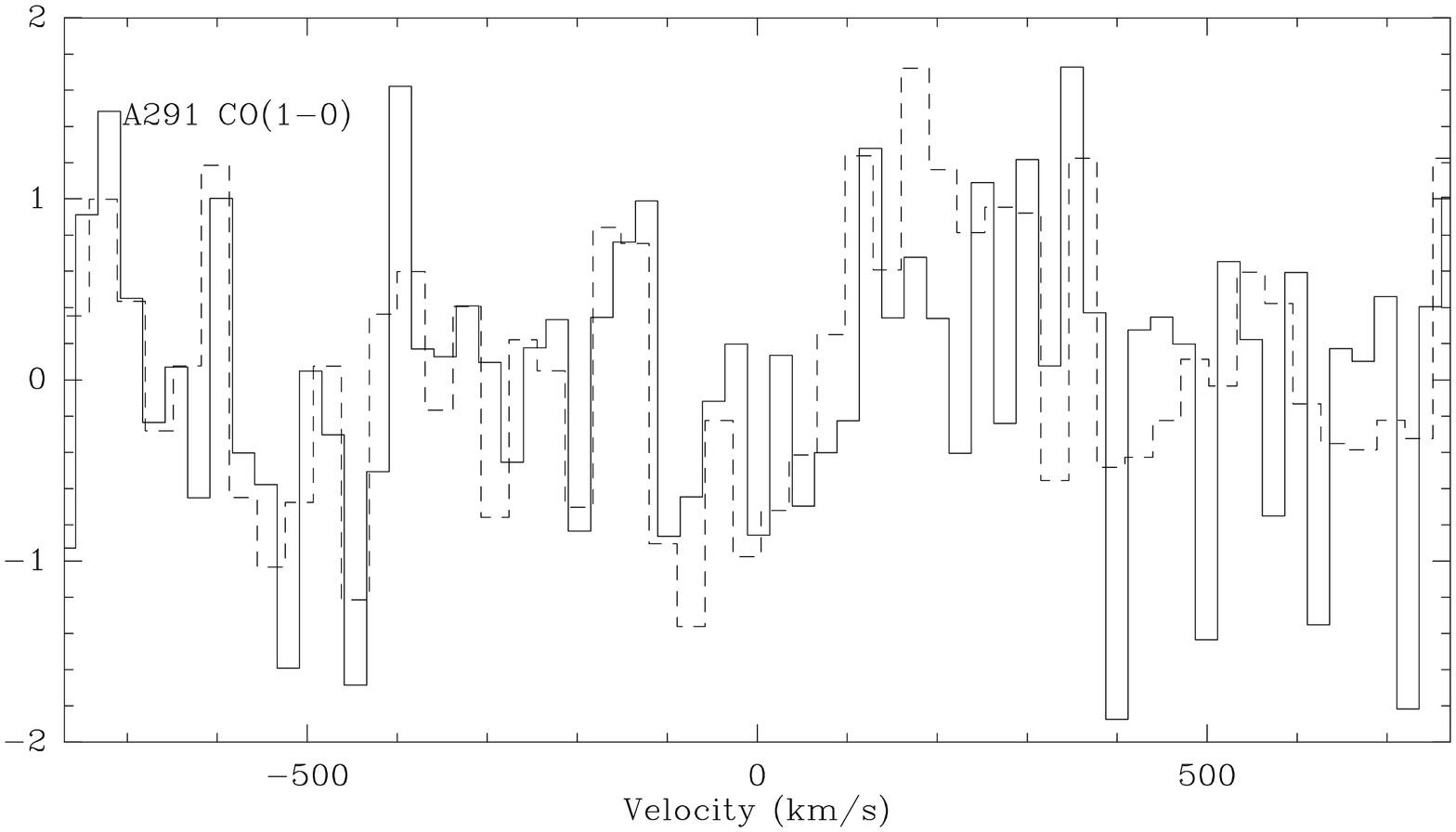,angle=270,width=8cm}
\psfig{file=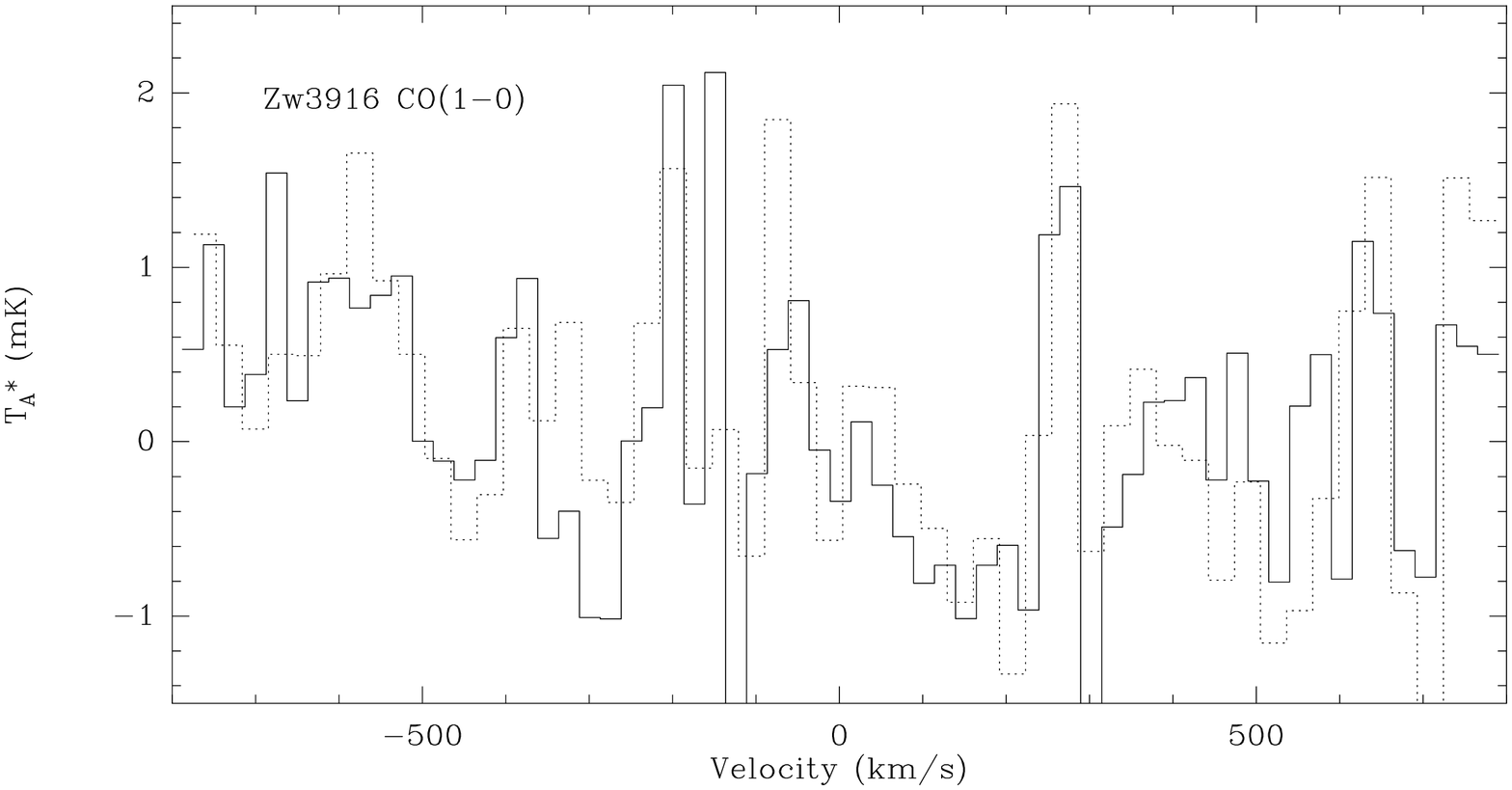,angle=270,width=8cm}}
\centerline{
\psfig{file=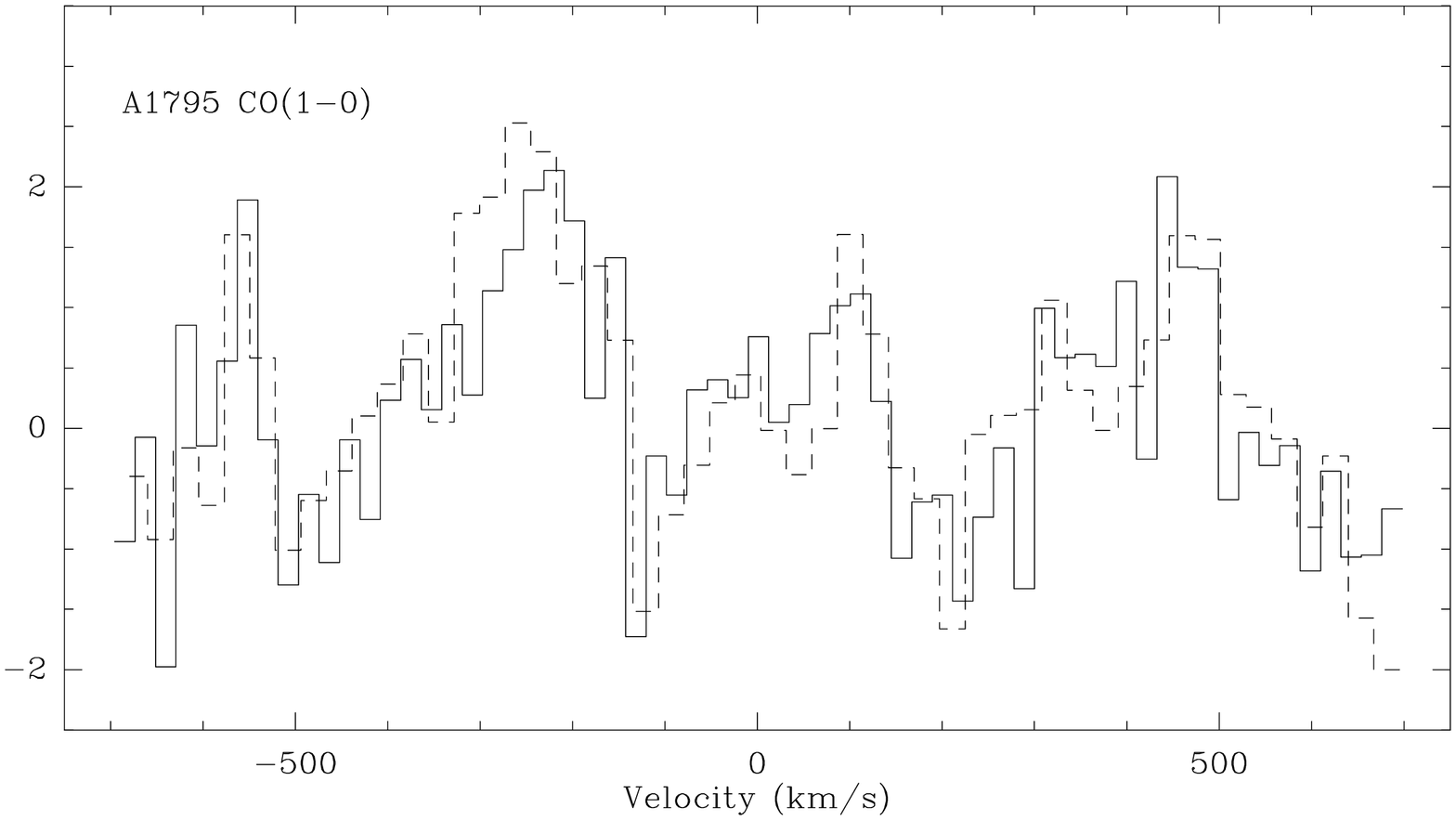,angle=270,width=8cm}
\psfig{file=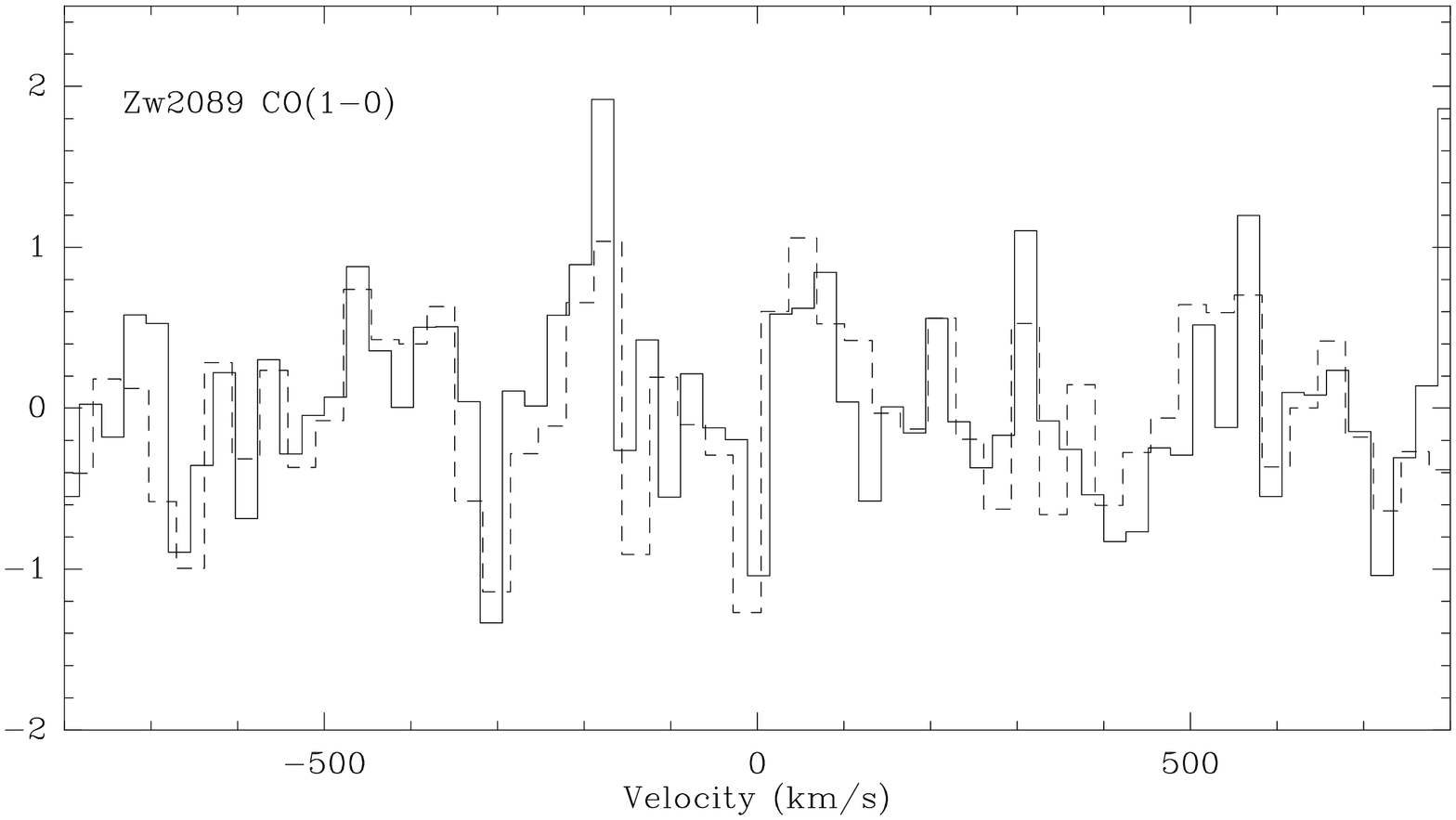,angle=270,width=8cm}}
\centerline{
\psfig{file=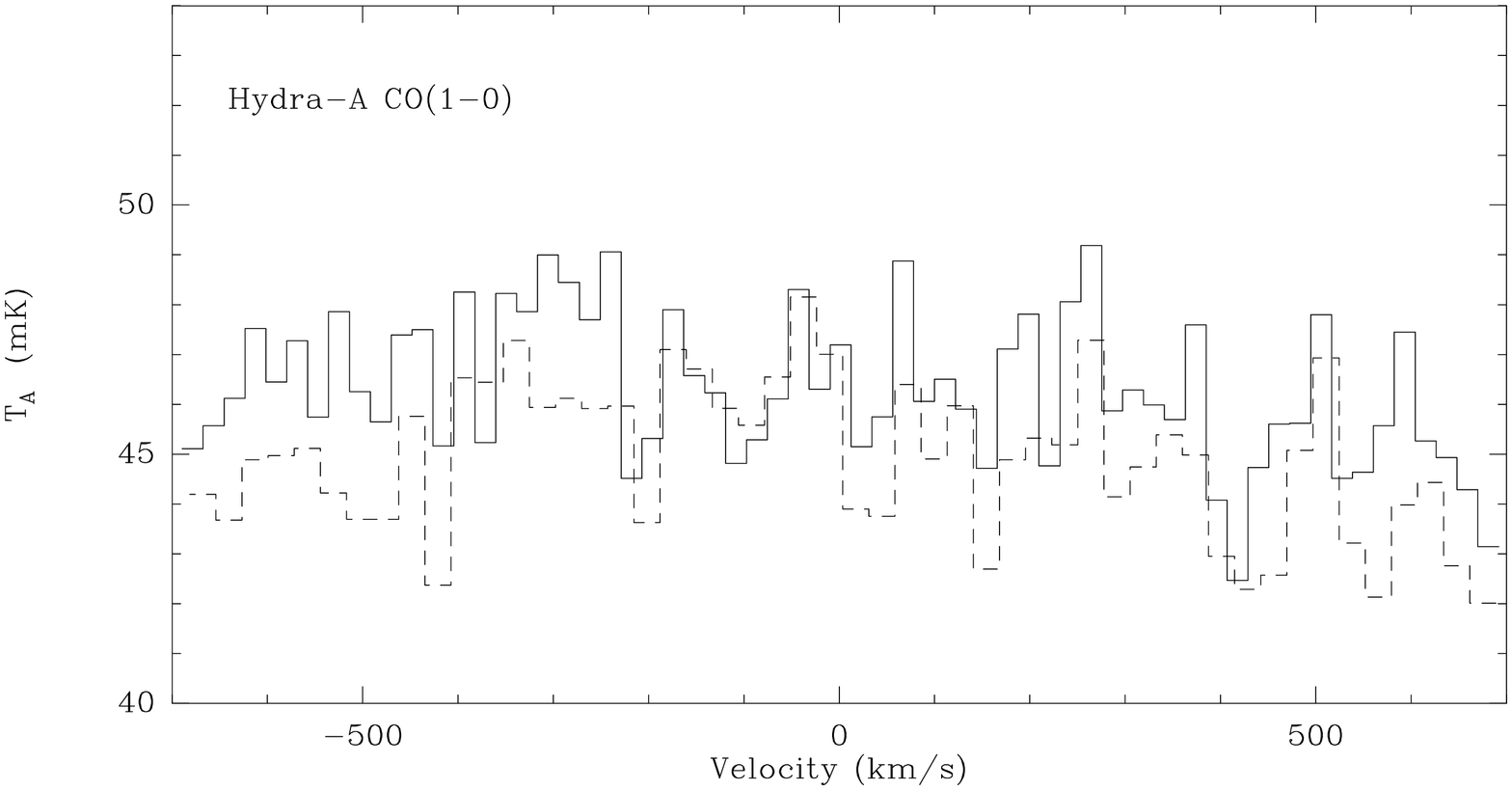,angle=270,width=8cm}
\psfig{file=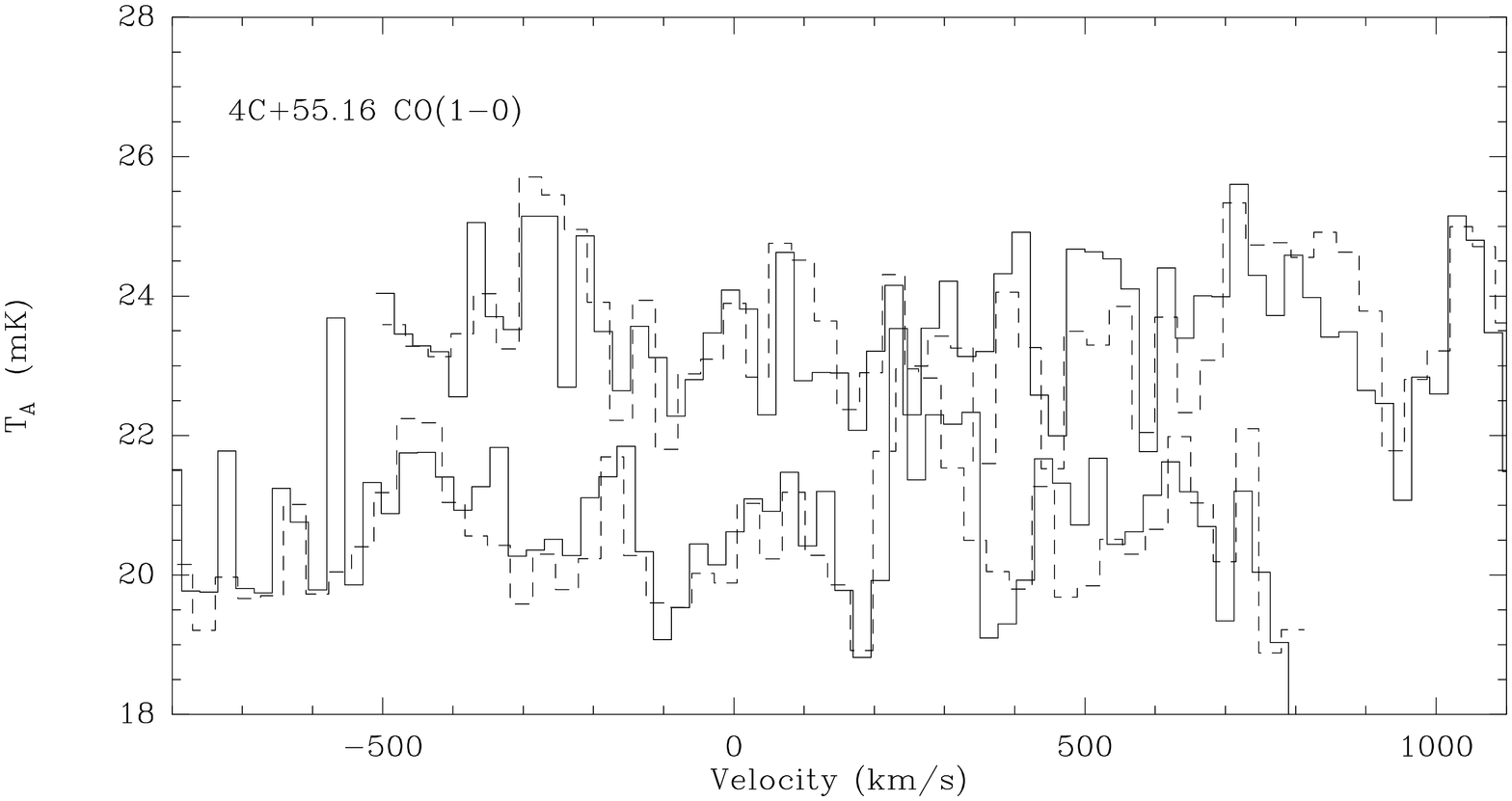,angle=270,width=8cm}}

\caption{{IRAM} 30m spectra for 
A646, A2146, A291, Zw3916, A1795, Zw2089, Hydra-A and 4C+55.16 which were only observed once
and no significant detection made. Note change in the continuum level between observations
of 4C+55.16 which is probably due to small pointing differences.}
\end{figure*}


%
\begin{figure*}
\centerline{\psfig{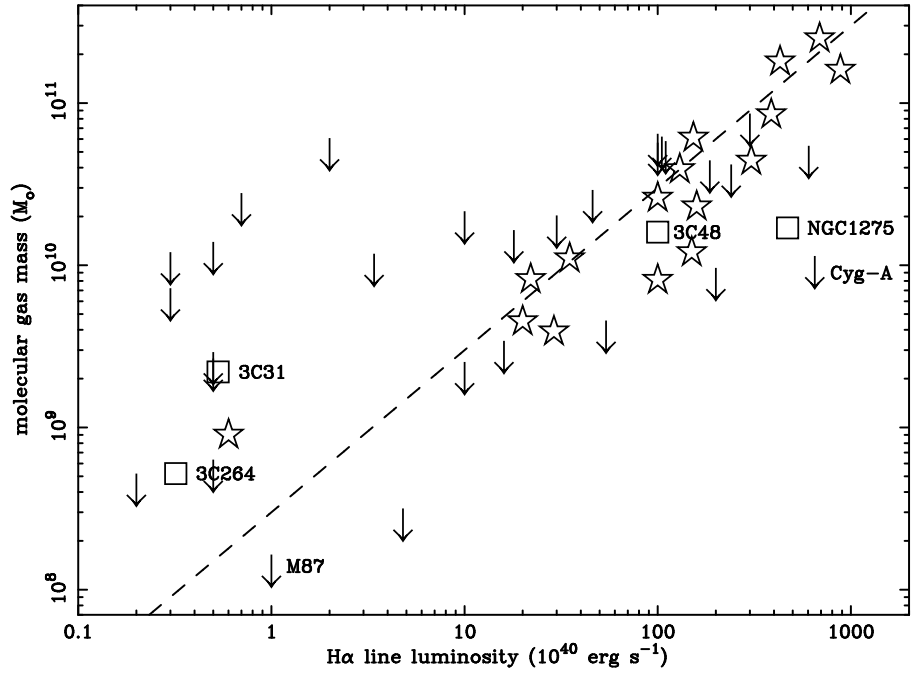}}
\caption{Molecular gas estimate plotted against optical line luminosity. The stars
are new detections presented here, the squares are detections of other central
cluster galaxies and the upper limits are from this work and the literature.
The dashed line marks a line with a constant ratio of molecular gas mass to
optical line emission and is not a fit to the data.}
\end{figure*}

\begin{figure*}
\centerline{\psfig{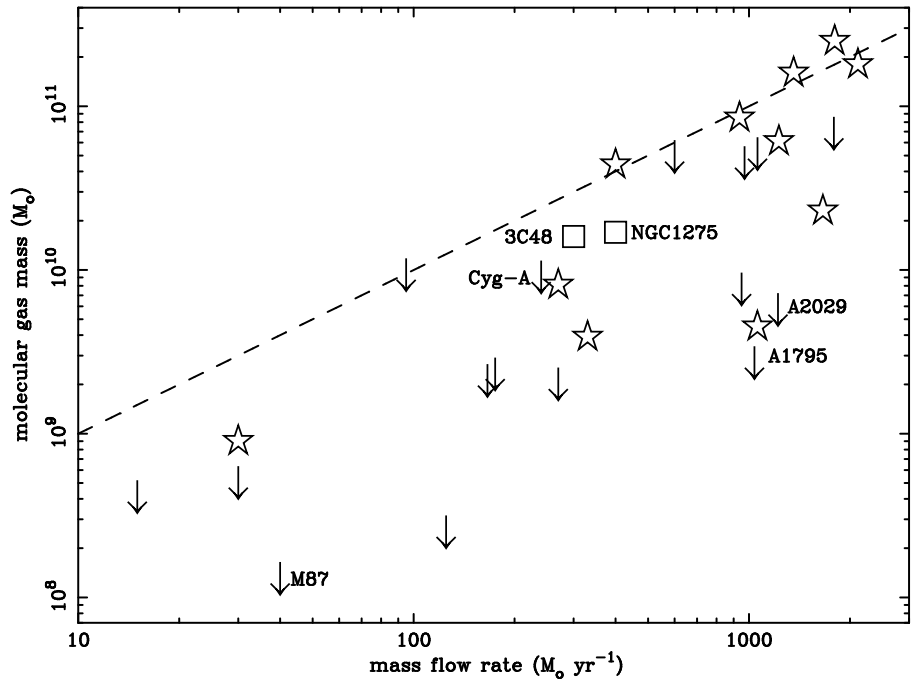}}
\caption{Molecular gas estimate plotted against global mass flow rate. The symbols and line are as
in Figure 9. Not all points are plotted as mass flow rates are not known for all
detections and upper limits.}
\end{figure*}

\end{document}